\documentclass[letter,11pt]{article}
\pdfoutput=1 

\usepackage{jcappub} 
\usepackage{xcolor} 
\usepackage[T1]{fontenc} 
\usepackage{subcaption}
\usepackage{enumitem}

\newcommand{\Neff}{N_{\textrm{eff}}}
\newcommand{\vp}{|\vec{p}|}

\usepackage{mathrsfs}

\def\({\left(}
\def\){\right)}
\def\[{\left[}
\def\]{\right]}

\title{\boldmath Dark radiation constraints on portal interactions with hidden sectors}

\author{Peter Adshead,}
\author{Pranjal Ralegankar,}
\author{and Jessie Shelton,}
\affiliation{Illinois Center for the Advanced Study of the Universe, Department of Physics, University of Illinois at Urbana-Champaign, Urbana, IL 61801, USA }

\emailAdd{adshead@illinois.edu}
\emailAdd{pranjal6@illinois.edu}
\emailAdd{sheltonj@illinois.edu}

\abstract{We update dark radiation constraints on millicharged particle (MCP) and gauged baryon-number-minus-lepton-number ($B-L$) extensions of the Standard Model (SM). In these models, a massive SM gauge singlet mediator couples the SM plasma to additional SM-singlet light degrees of freedom. In the early Universe, these new light particles are populated via the interaction of the SM with the MCP, or the new $B-L$ gauge boson, and act as dark radiation. The presence of dark radiation in the early Universe is tightly constrained by current and upcoming cosmic microwave background (CMB) measurements. We update bounds on MCPs from current measurements of $\Neff$ and show that future CMB experiments will be able to rule out or discover the extended MCP model invoked to explain the EDGES anomaly. Our analysis of the gauged $B-L$ model goes beyond previous studies by including quantum-statistical and out-of-equilibrium effects. Further, we account for the finite lifetime of the $B-L$ gauge boson, which boosts the subsequent right-handed neutrino energy density. We also develop a number of approximations and techniques for simplifying and solving the relevant Boltzmann equations. We use our approximations to develop a  lower bound on the radiation density in a generic hidden sector with a light relic that is insensitive to the details of the hidden sector, provided the mediator interacts more strongly with the hidden sector than with the SM.
}

\begin{document}
	\maketitle
	\flushbottom
	\section{Introduction}\label{sec:intro}
	
	Next-generation cosmic microwave background (CMB) experiments, such as CMB-Stage 4 \cite{Abazajian:2019eic}, will measure the details of the acoustic peaks in the microwave background with unprecedented accuracy. These measurements will result in subpercent-level determinations of the contents and geometry of the Universe. In particular, the fidelity with which the locations of the acoustic peaks are forecast to be determined will improve the measurement of the energy density in free-streaming radiation, parametrized by the effective number of neutrino species, $\Neff$, by almost an order of magnitude. Future CMB experiments, beyond Stage-4,  aim to reach a threshold of $\sigma_{\Neff} < 0.027$, where any new relativistic beyond-the-Standard Model (BSM) particle must be always out of equilibrium with the Standard Model (SM) in the early Universe  \cite{Green:2019glg} if the measured central value agrees with the SM prediction of $\Neff^{\rm SM}=3.044$ \cite{Mangano_2005, Grohs_2016, deSalas:2016ztq, Akita:2020szl, Abenza_2020}.  A measurement of $\Neff$ that deviates from $\Neff^{\rm SM}$ would be compelling evidence of physics beyond the standard model. Conversely, models that require additional light states must be coupled to the SM in such a way that does not violate bounds on $\Neff$.  
	
	Constraints on new relativistic degrees of freedom through $\Neff$ are often restated as a constraint on the decoupling temperature at which any BSM relativistic particle must lose thermal contact with the SM plasma in the early universe (see, e.g., Refs.~\cite{Olive:1980wz, Brust:2013ova, Weinberg:2013kea,Knapen:2017xzo}).  For an out-of-equilibrium relativistic particle, measurements of  $\Neff$ can be used to constrain the total energy transferred between BSM relativistic particles and the SM plasma in the early Universe, and can provide a powerful probe of the interactions of the SM with light, feebly-interacting particles. 
	
	Our primary interest in this work is the case where a SM singlet mediator particle has renormalizable couplings to both the SM and the dark radiation species.  This scenario is ultraviolet (UV) insensitive insofar as it yields interaction rates that grow more rapidly than the Hubble rate as the universe expands, provided that the SM temperature remains larger than the mediator mass. This UV insensitivity means that the asymptotic dark radiation density predicted in these models does not depend on the unknown early thermal history of our universe provided the reheating temperature is above the mediator mass. In this work, we focus on mediator masses $m > 0.1$ MeV where thermal production in the early universe provides one of the leading avenues to test these models.  Constraints from stellar cooling are typically stronger than cosmological constraints for masses $m < 0.1$ MeV \cite{An:2013yfc,Knapen:2017xzo}.  
	
	Similar UV-insensitive and out-of-equilibrium dark radiation production has been explored earlier in the context of specific models. For instance, in the case of axions, freeze-in production can receive important contributions from both heavy states in the UV completion \cite{DEramo:2021lgb} and fermion annihilation, which proceeds through infrared-dominated processes below the scale of electroweak symmetry breaking \cite{Ferreira:2018vjj,DEramo:2018vss,AriasAragon:2020shv,Green:2021hjh}. BSM neutrino model-building can also yield sizeable out-of-equilibrium dark radiation production \cite{Luo:2020fdt,Biswas:2022vkq}.  Meanwhile LHC searches  can provide a complementary window onto the freeze-in of dark radiation in scenarios where a weak-scale mediator carries SM charge \cite{Bernreuther:2022bdw}.  	
	
	In this paper, we study the production of dark radiation in minimal BSM models that consist of a massive ($m>0.1$ MeV)  SM gauge singlet mediator coupled to new light degrees of freedom. 
	We begin by considering two well-motivated extensions to the SM:  a millicharged particle (MCP) model \cite{Holdom:1985ag}, and a  model where the SM baryon-number-minus-lepton number ($B-L$) symmetry is gauged \cite{Carlson:1986cu,Feldman:2011ms}. In the MCP model, a dark photon that kinetically mixes with SM hypercharge is the dark radiation and the MCP is the mediator. In the gauged $B-L$ model the three right-handed neutrinos required to cancel gauge anomalies are the dark radiation, while the new $B-L$ gauge boson is the massive mediator. By developing and solving the relevant Boltzmann equations, we use the production of dark radiation  in these models to place constraints on the strength of their interactions with the SM. We update constraints on the minimal MCP model given in Refs.~\cite{Vogel:2013raa,Foot:2014uba} and present forecasts for future CMB observatories. We further demonstrate that future CMB experiments will be able to rule out (or discover evidence for) the extended model proposed by Ref.~\cite{Liu:2019knx} to explain the EDGES anomaly.  For the $B-L$ model, we improve on the analysis of Ref.~\cite{Abazajian:2019oqj} by incorporating two further effects that lead to more stringent constraints in the unequilibriated regime. In particular, we take into account the out-of-equilibrium production of right-handed neutrinos, and further show that the out-of-equilibrium decays of the $B-L$ gauge bosons lead to a more powerful constraint on the $B-L$ coupling in the relevant regions of parameter space. 
		
	In the process of deriving these results, we develop a number of  approximations which allow us to analytically solve the Boltzmann equations in the regions of parameter space where the new light degrees of freedom are out of equilibrium with the SM. We use these solutions to argue, on general grounds, that a conservative lower bound on the dark radiation density can be quickly obtained for a generic class of hidden sectors containing light degrees of freedom that interact with the SM via a heavier SM gauge singlet mediator. The lower bound is governed by the properties of the mediator and is insensitive to the details of the hidden sector, such as the number of degrees of freedom and their internal interactions, and relies solely on the assumption that the mediator preferentially transfers its energy into the HS rather than the SM. This amounts to assuming that the mediator interacts more strongly with the HS than the SM.
	
	This paper is organized as follows. In sections \ref{sec:MCP} and \ref{sec:RH_nu}, we study dark radiation production in the MCP and gauged $B-L$ models, respectively. We develop and solve the relevant Boltzmann equations to find the allowed regions of parameter space given current and projected CMB constraints on $\Neff$.  In both models, we develop approximations that allow us to analytically solve the Boltzmann equation in various regimes. 	In section \ref{sec:wide_application}, we consider the applicability of dark radiation constraints to generic classes of hidden sectors containing relativistic particles. We conclude in section \ref{sec:summary}. The details of many of our computations are relegated to appendices.  In appendix~\ref{sec:mcp_other_channels} we describe various processes transferring energy between the SM and the dark photons in the MCP model, and similarly in appendix~\ref{sec:BL_coll} we describe processes transferring energy from the SM into right-handed neutrinos in the gauged $B-L$ model. Finally, in appendix~\ref{sec:schannel_quant}, we simplify the phase space integral of the energy transfer collision terms for generic annihilations, decays, and elastic scatterings, while taking into account the quantum statistical distributions of relevant particles.

	\section{Millicharged particle model}\label{sec:MCP}

	In this section we derive constraints on the allowed parameter space of a MCP model from CMB measurements of $N_{\rm eff}$. In this model, a massless dark photon kinetically mixes with the SM hypercharge gauge boson, while the MCP is a massive Dirac fermion charged under the dark $U(1)$.  
	
	MCP models have recently been explored in detail as potential explanations of the anomalously small spin temperature of the hydrogen atoms inferred from the 21 cm signal measured by the EDGES experiment~\cite{Munoz:2018pzp,Bowman:2018yin,Barkana:2018qrx, Liu:2019knx,Mathur:2021gej}. This anomaly can be resolved if the baryons were cooled by scattering with DM particles. In the scenario where the MCP comprises some of the dark matter, the millicharge interactions can cool the baryons to explain the EDGES anomaly. However, the required values of the millicharge, $Q$, are ruled out by a combination of bounds from the CMB and $e^+e^-$ colliders \cite{Barkana:2018qrx}.\footnote{If the baryons are cooled by a millicharged dark fermion that is not coupled to dark radiation, then one can explain the EDGES result if the dark fermions compose a 0.4\% fraction of dark matter \cite{Berlin:2018sjs,Kovetz:2018zan,Slatyer:2018aqg, dePutter:2018xte}. However, Ref.~\cite{Creque-Sarbinowski:2019mcm} found that this solution is incompatible with the constraints on the millicharge and dark fermion mass imposed by its production history in the early universe.} 
	Recently, an extension of the minimal MCP model was proposed with multiple millicharged fermions that could resolve the EDGES anomaly while evading current constraints \cite{Liu:2019knx}. In this section we both update the current CMB constraints on the minimal MCP model and show that measurements of $N_{\rm eff}$ from future CMB experiments will provide a stringent test of these extended MCP models.
	
	This section is organized as follows. We begin by describing the MCP model in section \ref{sec:themodel}. In section~\ref{mcp:boltz}, we describe the relevant Boltzmann equations and solve them to find the region of parameter space that saturates the $\Neff$ bounds from current and upcoming CMB experiments, updating the results of \cite{Vogel:2013raa}. Next, in section~\ref{mcp:physics} we go into more detail about the physics responsible for the production of dark radiation, and the relevant features of the resulting parameter space constraints from $\Neff$ measurements. Finally, in section~\ref{mcp:constraint}, we show how these constraints can be extended to models with multiple MCPs in a detail-insensitive way.  We then apply these constraints to the MCP model proposed by Ref.~\cite{Liu:2019knx} and show that measurements of $N_{\rm eff}$ at the level of accuracy forecast by CMB-S4 can potentially rule out this explanation of the EDGES anomaly.

	\subsection{The millicharged particle model}\label{sec:themodel}

	The MCP model is an extension of the SM that contains a massless dark photon, $A_\mu'$, and an additional  Dirac fermion, $\psi$, with mass $m$. The dark photon kinetically mixes with the SM hypercharge gauge boson, $A_\mu$, and the Dirac fermion has charge $e'$ under the dark $U(1)$. The relevant interactions for our study are
	\begin{align}
		\mathcal{L}_{\rm int}=-\frac{\epsilon}{2}B^{\mu\nu}F'_{\mu\nu}+eJ^{\mu}_{\rm EM}A_{\mu}+eJ^{\mu}_{Z}Z_{\mu}+e'\bar{\psi}\gamma^{\mu}\psi A_{\mu}',
	\end{align}
	where $B_{\mu\nu}$ is the hypercharge field strength, $Z_{\mu}$ is the $Z$ boson, $J^{\mu}_{\rm EM}$ is the electromagnetic current, and $J^{\mu}_{Z}$ is the weak neutral current.
	
	We work in the basis where the gauge boson kinetic terms are diagonal and where $J^{\mu}_{\rm EM}$ and $J^{\mu}_{Z}$ do not couple to the dark photon. Thus the dark photon remains `dark'. After performing the relevant redefinitions of the $A$ and $A'$ fields and considering the limit of weak kinetic mixing, $\epsilon\ll 1$, the interaction Lagrangian is 
	\begin{align}\label{eq:L_mcp}
		\mathcal{L}_{\rm int} \approx e\left(J^{\mu}_{\rm EM}-Q\bar{\psi}\gamma^{\mu}\psi\right)A_{\mu}+e'\bar{\psi}\gamma^{\mu}\psi A'_{\mu}+Qe\tan\theta_W\bar{\psi}\gamma^{\mu}\psi Z_{\mu},
	\end{align}
	where $\theta_W$ is the weak mixing angle, and the dark fermion has obtained a millicharge, $Q$, given by
	\begin{align}
		Q\equiv \epsilon\frac{e'}{e}\cos\theta_W.
	\end{align}
	While the dark photon does not directly couple to SM degrees of freedom, dark photons are produced by annihilations of millicharged fermions, which themselves are produced by interactions with the SM plasma in the early Universe. In this work, we consider the regime where the fermion mass is $m>0.1$ MeV; stellar cooling observations provide the dominant constraint for smaller masses  \cite{Vogel:2013raa}.

	\subsection{Evaluation of the dark radiation density and the constraints on the model}\label{mcp:boltz}
	
	Dark photons contribute to the energy budget of the Universe as radiation, and their presence in the early Universe is constrained by measurements of the effective number of (free-streaming) relativistic species, $N_{\rm eff}$. Specifically, dark photons shift the value of $N_{\rm eff}$ away from its SM value of $\Neff^{\rm SM}=3.044$, by
	\begin{align}\label{eq:Dneff_mcp}
		\Delta\Neff \equiv N_{\rm eff} -\Neff^{\rm SM} = \frac{8}{7}\left(\frac{11}{4}\right)^{4/3}\frac{\rho_{A'}}{\rho_{\gamma}},
	\end{align}
	where $\rho_{A'}$ and $\rho_{\gamma}$ are the energy densities of the dark photon and the SM photon, respectively.  The dark photon energy density $\rho_{A'}$ during recombination is controlled by $Q$, $m$, and $e'$, and thus measurements of $N_{\rm eff}$ can be translated into constraints on the parameter space of the model.
	
	We demonstrate below that, for the regions of parameter space that lead to dark radiation densities that  saturate the bounds on $\Delta N_{\rm eff}$ from  upcoming experiments, the dark charge $e'$ must be large enough to enable almost all the MCPs to efficiently annihilate. In this limit,  the final  dark photon abundance is insensitive to the value of $e'$. Moreover, due to the tight coupling of the MCPs to the dark photons, the hidden sector (HS) thermal bath comprising the MCP and the dark photon is well-approximated by a fluid in chemical equilibrium.  Thus, instead of solving for the individual MCP and dark photon abundances, we can solve for the combined HS energy density through the Boltzmann equations
	\begin{align}\label{eq:Boltz_eps_full_mcp}
		\frac{d\rho_{\rm SM}}{dt}+3H(1+w_{{\rm SM}})\rho_{\rm SM}=-\mathcal{C}\nonumber\\
		\frac{d\rho_{\rm HS}}{dt}+3H(1+w_{{\rm HS}})\rho_{\rm HS}=\mathcal{C},
	\end{align}
	where $\mathcal{C}$ is the energy transfer collision term due to millicharge interactions, $\rho$ is the energy density, $H=\sqrt{\rho_{\rm HS}+\rho_{\rm SM}}/[\sqrt{3}M_{\rm Pl}]$, $w=\mathcal{P}/\rho$ is the equation of state,  $\mathcal{P}$ is the pressure, and $M_{\rm Pl}=2.435\times 10^{18}$ GeV is the reduced Planck mass. After the MCPs become non-relativistic and annihilate into dark photons, $\rho_{\rm HS}\approx \rho_{A'}$.
	
	Both $\rho_{\rm HS}$ and $w_{{\rm HS}}$ are determined in terms of $T_{\rm HS}$ by 
	\begin{align}\label{eq:def_whs}
		\rho_{\rm HS}=\frac{\pi^2}{30}g_{\rm HS}T_{\rm HS}^4,\quad \quad	w_{{\rm HS}}=\frac{g_{{\rm HS},p}}{3g_{\rm HS}},
	\end{align}
	where
	\begin{align}\label{eq:g_HS}
		g_{\rm HS}=2+\frac{30}{\pi^2T_{\rm HS}^4}\times4\int_{0}^{\infty}\frac{d^3p}{(2\pi)^3} E\frac{1}{\exp(E/T_{\rm HS})+1},\\
		g_{{\rm HS},p}=2+\frac{90}{\pi^2T_{\rm HS}^4}\times4\int_{0}^{\infty}\frac{d^3p}{(2\pi)^3} \frac{p^2}{3E}\frac{1}{\exp(E/T_{\rm HS})+1}.
	\end{align}
	Similarly, $w_{{\rm SM}}$ is related to $T_{\rm SM}$ via
	\begin{align}
		w_{{\rm SM}}=\frac{g_{*p}(T_{\rm SM})}{3g_{*}(T_{\rm SM})},
	\end{align}
	where $g_*$ and $g_{*p}$ count the effective degrees of freedom in the SM energy density and the SM pressure, respectively. We model the QCD phase transition using the $g_*$ tables from Ref.~\cite{Husdal:2016haj} for $T_{\rm SM}>100$ MeV.
	
	The collision term in eq.~\eqref{eq:Boltz_eps_full_mcp} includes all processes that transfer energy from the SM plasma into the HS bath due to the millicharge interactions. There are four important processes contributing to energy transfer: (1) SM fermion annihilation into MCPs; (2) $Z$-boson decays into MCPs; (3) plasmon decays into MCPs; and (4) Coulomb scattering of SM fermions with MCPs.\footnote{Energy transfer from Compton-like scattering, $A+\psi\rightarrow \psi+A'$, can be more important than the processes mentioned here for large values of the dark coupling constant, $e'>0.9$. We neglect this process for simplicity and genericity.} We include the quantum statistical distributions of SM particles while deriving the collision term for each of these processes, relegating the details to  appendix~\ref{sec:mcp_other_channels}. The use of quantum statistics instead of Maxwell-Boltzmann distributions provides a $\sim 20\%$ correction to the net energy transfer. Among the three $s$-channel processes (numbers 1-3 above), we find that energy transfer via fermion annihilation dominates over the other two in the bulk of parameter space. For instance, the energy transferred by fermion annihilations  dominates over that from $Z$-boson decays except for the region of parameter space where $1{\rm\ GeV}\lesssim m\lesssim 40$ GeV. The energy transferred via plasmon decays is typically around $\sim 20\%$ of that transferred by fermion annihilations.
	
	Finally, we find that the energy transferred by Coulomb scattering dominates over that from fermion annihilations for the values of $Q$ and $m$ that saturate the bounds on $\Delta N_{\rm eff}$ from both current and upcoming experiments.
	Naively, one might expect energy transfer via Coulomb scattering to be subdominant in the out-of-equilibrium regime because these processes are suppressed by the small MCP abundance in the initial state. However, due to the forward-scattering singularity, the energy transfer via Coulomb scattering dominates over that via SM fermion annihilations for $T_{\rm HS}$ as low as $0.35T_{\rm SM}$ (for more detail, see appendix~\ref{sec:mcp_other_channels}).  Temperature ratios of $T_{\rm HS}/T_{\rm SM}>0.35$ during recombination produce enough dark radiation to shift $\Delta\Neff>0.06$, which can be detected in the upcoming CMB-S4 experiments \cite{Abazajian:2019eic}. Hence, Coulomb scattering processes are key for evaluating the dark radiation densities relevant for the values of $\Delta\Neff$ that can be tested in upcoming as well as current experiments. 
	\begin{figure}
		\centering
		\includegraphics[width=\textwidth]{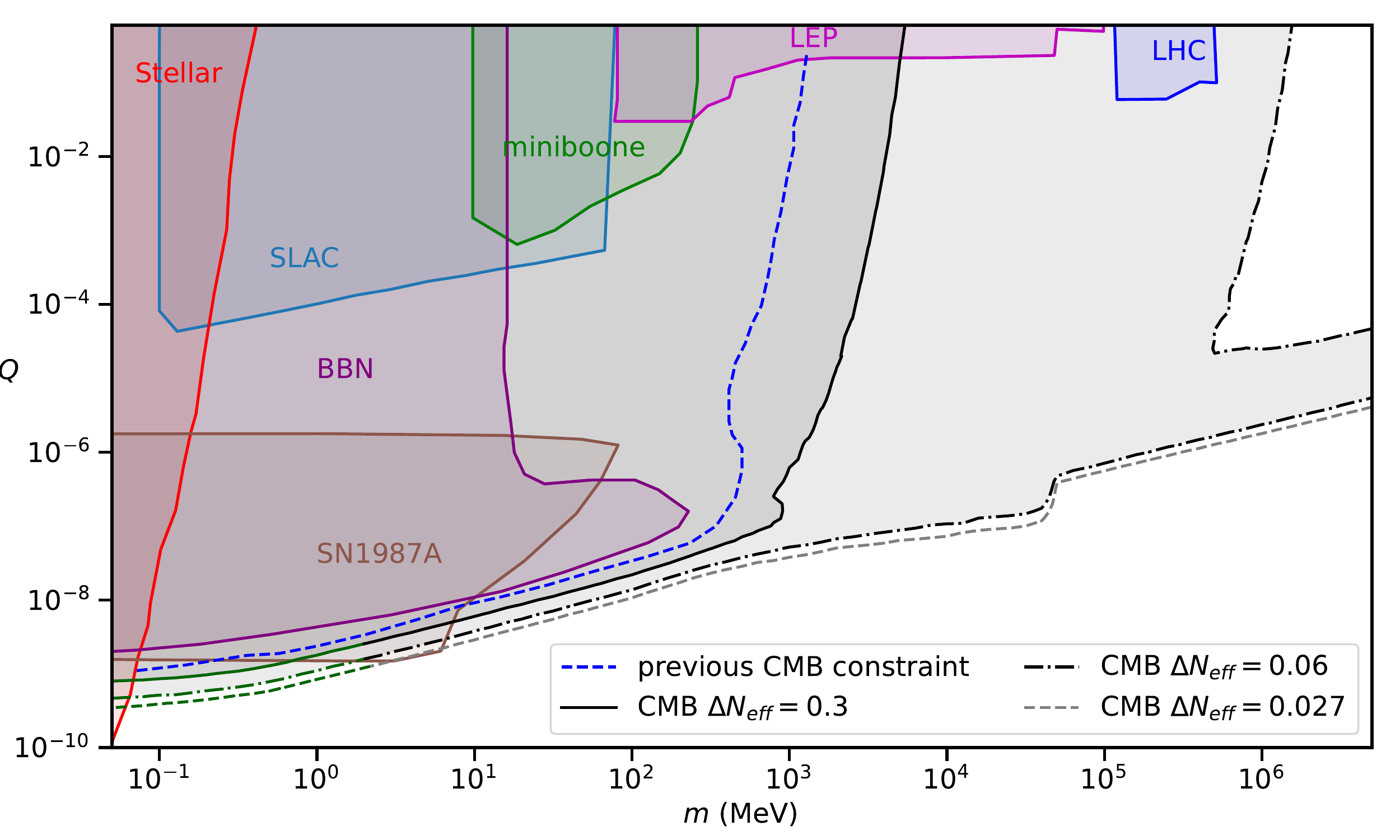}
		\caption{Constraints on the mass and millicharge of the millicharged particle. The black solid, gray solid, and gray dashed contours indicate the parameter space that yields $\Delta N_{\rm eff}= 0.3$, $\Delta N_{\rm eff}= 0.06$, and $\Delta N_{\rm eff}= 0.027$, respectively. The blue dashed contour is the CMB constraint derived in Ref.~\cite{Vogel:2013raa} for $\Delta N_{\rm eff}= 0.8$.  The green color on our $\Delta\Neff$ contours marks the region where we expect the millicharge interactions to cool the electron-photon bath relative to neutrinos and strengthen our constraints by an $\mathcal{O}(1)$ factor. Also shown are constraints from SLAC \cite{Prinz:1998ua}, MiniBooNE \cite{Magill:2018tbb}, LEP \cite{Davidson:2000hf}, LHC \cite{Jaeckel:2012yz}, BBN \cite{Vogel:2013raa}, supernova 1987A \cite{Chang:2018rso} and stellar observations \cite{Vogel:2013raa}. }
		\label{fig:mcp_constraint}
	\end{figure}
	
	To determine the relic dark radiation density, and the resulting  parameter space allowed by  $\Neff$ measurements, we solve the Boltzmann equations in eq.~\eqref{eq:Boltz_eps_full_mcp} from  an initial SM temperature $T_{\rm SM}\gg m$ until the energy injection into the HS ends, $T_{\rm SM}\ll m$. We assume the HS is initially empty, which provides a conservative constraint; any significant initial HS density  only increases the final dark radiation density and thus $\Delta\Neff$. 
	
	The various shaded regions in figure~\ref{fig:mcp_constraint} show the regions of parameter space where the resulting energy density in dark radiation exceeds various current and future experimental sensitivities to shifts in $N_{\rm eff}$. The constraint contours saturate the current one-sided 2$\sigma$ upper bound from Planck \cite{Aghanim:2018eyx} $\Delta\Neff= 0.3$ (black solid), the projected 2$\sigma$ sensitivity for CMB-S4 \cite{Abazajian:2019eic} $\Delta\Neff = 0.06$ (gray solid), and the threshold goal for future CMB experiments $\Delta\Neff = 0.027$ (gray dashed). The threshold of $\Delta \Neff = 0.027$ physically corresponds to the shift in $N_{\rm eff}$ due to the energy density at recombination in a relativistic spin-zero particle that was in thermal equilibrium with the SM in the early Universe and decoupled while all SM species were relativistic. 
	
	For comparison, in figure~\ref{fig:mcp_constraint} we also display the results of Ref.~\cite{Vogel:2013raa} as the blue dashed line, which shows the parameter points that lead to $\Delta\Neff = 0.8$. We have verified that our results agree with Ref.~\cite{Vogel:2013raa} within $\mathcal{O}(1)$ when we assume Maxwell-Boltzmann statistics for all particles. The use of Maxwell-Boltzmann statistics as opposed to Fermi-Dirac statistics overestimates the dark radiation density by around 20\%. As the energy density in dark radiation depends on $Q^2$, using Fermi-Dirac statistics for SM fermions weakens the $\Neff$ constraint on $Q$ by around $~10\%$. 
	
	In figure~\ref{fig:mcp_constraint} we also show the constraints on the MCP from collider experiments, stellar evolution, and supernova observations (see Ref.~\cite{Fabbrichesi:2020wbt} for a review). We omit limits from direct detection experiments because those constraints are dependent on the interaction of MCPs with the magnetic fields in the galaxy \cite{Chuzhoy:2008zy,Stebbins:2019xjr}. Among the displayed constraints, the current Planck limit is already the dominant bound in a substantial portion of parameter space, while upcoming CMB observations will provide the strongest constraint for the entire region with $m\gtrsim 0.1$ MeV, assuming no deviation is observed from the SM value of $\Neff^{\rm SM}=3.044$.
	
	The curves of constant $\Delta\Neff$ in the MCP model parameter space shown in figure~\ref{fig:mcp_constraint} have four key features. First, at low masses, the contours of constant dark radiation density at recombination (and therefore constant $\Delta \Neff$) relate the millicharge, $Q$, to the MCP mass via $Q \propto  \sqrt{m\Delta\Neff}$. In this region the HS is out-of-equilibrium with the SM.  Second, as one moves along the contour of constant $\Delta \Neff$ toward increasing $m$, one reaches a threshold mass $m_{\rm th}$ where the millicharge $Q$  becomes large enough that the HS thermalizes with the SM. When the HS is thermalized with the SM plasma, the net energy transfer between sectors becomes insensitive to the specific value of $Q$ since forward and backward processes balance each other.  In this regime the asymptotic dark radiation density, and therefore the constraint from  $\Delta N_{\rm eff}$, depends primarily on $m$ and only logarithmically on $Q$, as seen in the figure. 
	
	Third, the contour corresponding to $\Delta\Neff = 0.06$ has a narrow exclusion region (where $\Delta\Neff > 0.06$) extending from $m\approx m_{\rm th}$ up to arbitrarily large $m$, while no such excluded strip exists for either the $\Delta\Neff = 0.3$ or the $\Delta\Neff = 0.027$ contours. The existence (or non-existence) of this strip beyond the threshold mass is related to the fact that $\Delta\Neff<0.06$ still allows the dark photon itself to have been in equilibrium with the SM plasma for temperatures above the TeV scale, but is not compatible with the MCP also having entered equilibrium, which would increase the hidden sector relativistic degrees of freedom to an unacceptably large value at early times. 
	Finally, the $\Delta\Neff<0.06$ and $\Delta\Neff<0.027$ constraints have a bump below $m\sim m_Z/2$ which is due to energy injection from on-shell $Z$-boson decays. 	
	In the following subsection we elaborate on this discussion by analytically solving the Boltzmann equations in the relevant regimes.
	
	For $m\lesssim 2$ MeV, energy transfer into the HS occurs predominantly after neutrino decoupling. In this part of parameter space, the production of dark photons as well as the relative cooling of the electron-photon bath compared to neutrinos contributes to $\Delta\Neff$ during recombination, while our analysis only considers the contribution from dark photons. Taking into account the relative cooling of photons should further strengthen the $\Neff$ constraints calculated in this study by an $\mathcal{O}(1)$ factor. We indicate this region in figure~\ref{fig:mcp_constraint} by coloring the $\Neff$ contours green. A full treatment of early universe constraints on the MCP model below $m\lesssim 2$ MeV requires a detailed treatment of neutrino decoupling as well as light element formation during BBN, and is beyond the scope of this work.
	\subsection{Dark radiation production in different regimes}\label{mcp:physics}

	The parameter space that saturates the bounds on $\Neff$ can be  separated into two distinct regions: a region where the HS remains out of equilibrium with the SM plasma and a region where the HS thermalizes with the SM. In this subsection we focus on the evolution of the HS energy density, $\rho_{\rm HS}$, in these two regions of parameter space. By studying the Boltzmann equations, we develop approximate analytic descriptions that enable a deeper understanding of the shapes of the curves in figure~\ref{fig:mcp_constraint}.

	\paragraph{Collisions, redshifting, and the evolution of $\rho_{\rm HS}$:}  The evolution of the energy density in dark radiation is controlled by two factors. The first is the (net) rate at which energy is injected into the HS, $\mathcal{C} = \mathcal{C}_{\rm f}- \mathcal{C}_{\rm b}$, where $\mathcal{C}_{\rm f}$, and $\mathcal{C}_{\rm b}$ are the forward and backward collision terms describing energy transfer from the SM into the HS. The second factor is the rate at which the energy density is redshifting,  $H\rho_{\rm HS} $. The ratio $\mathcal{C}_{\rm f}/H$, then, indicates the energy density transferred to the HS within a Hubble time.  When  $\rho_{\rm HS}$ is out of equilibrium with the SM, $\mathcal{C}_{\rm f}/H$ serves as a useful indicator of whether energy injection is important ($\mathcal{C}_{\rm f}/H > \rho_{\rm HS}$) or not ($\mathcal{C}_{\rm f}/H < \rho_{\rm HS}$) in governing its evolution.  When  $\rho_{\rm HS}$ is in equilibrium with the SM, $\rho_{\rm HS,eq}=[\pi^2g_{\rm HS}(T_{\rm SM})/30](T_{\rm SM})^4$ and $\mathcal{C}=\mathcal{C}_{\rm f}- \mathcal{C}_{\rm b} = 0$, as both forward and backward rates become large.  The HS remains in equilibrium with the SM plasma as long as the fractional energy injection rate, $\Gamma_E\equiv \mathcal{C}_{\rm f}/\rho_{\rm HS,eq}$, is larger than $H$.
	
\begin{figure}
		\begin{subfigure}{.5\textwidth}
			\includegraphics[width=1.00\textwidth]{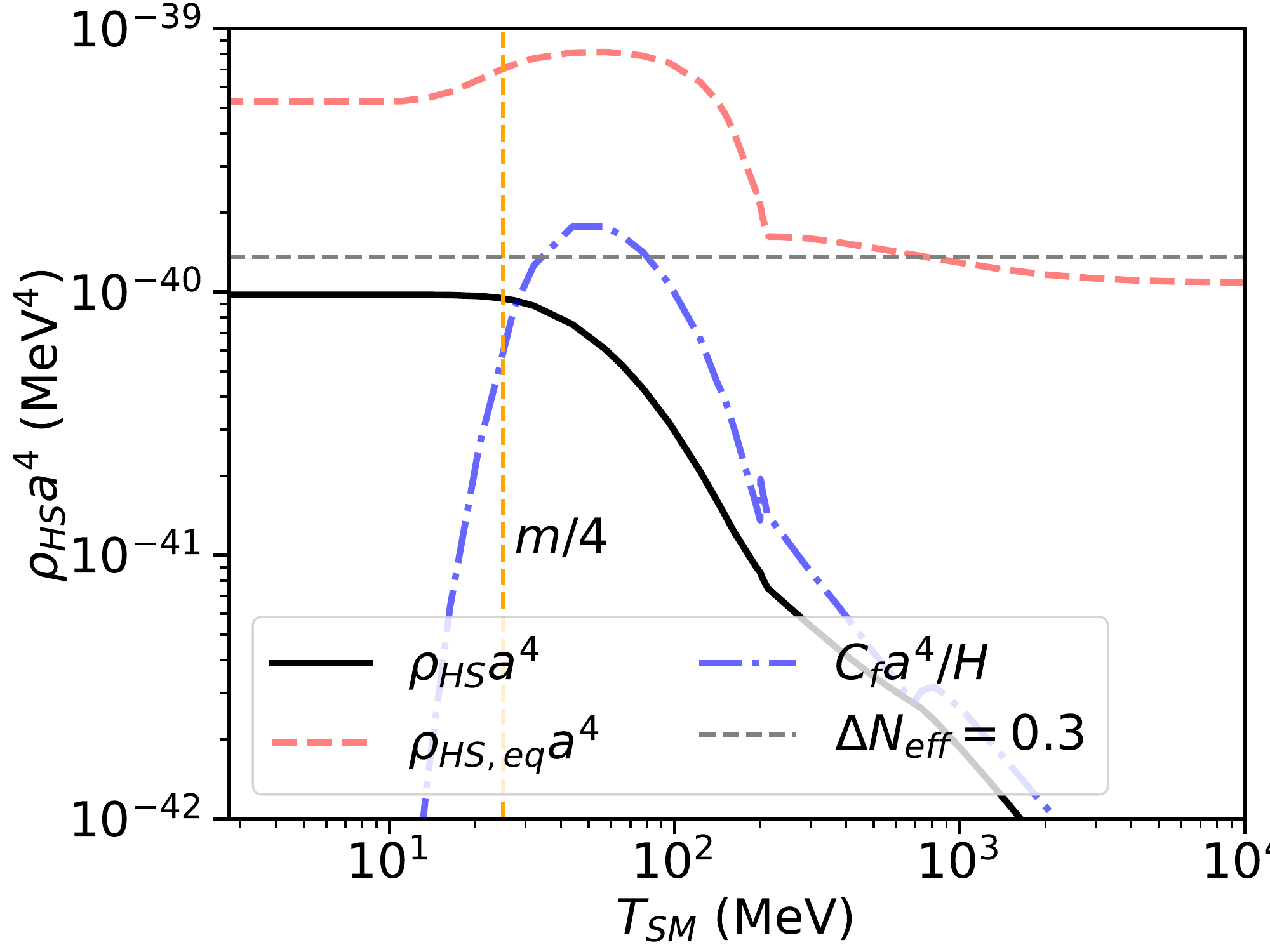}
		\end{subfigure}
		\begin{subfigure}{.5\textwidth}
			\includegraphics[width=1.00\textwidth]{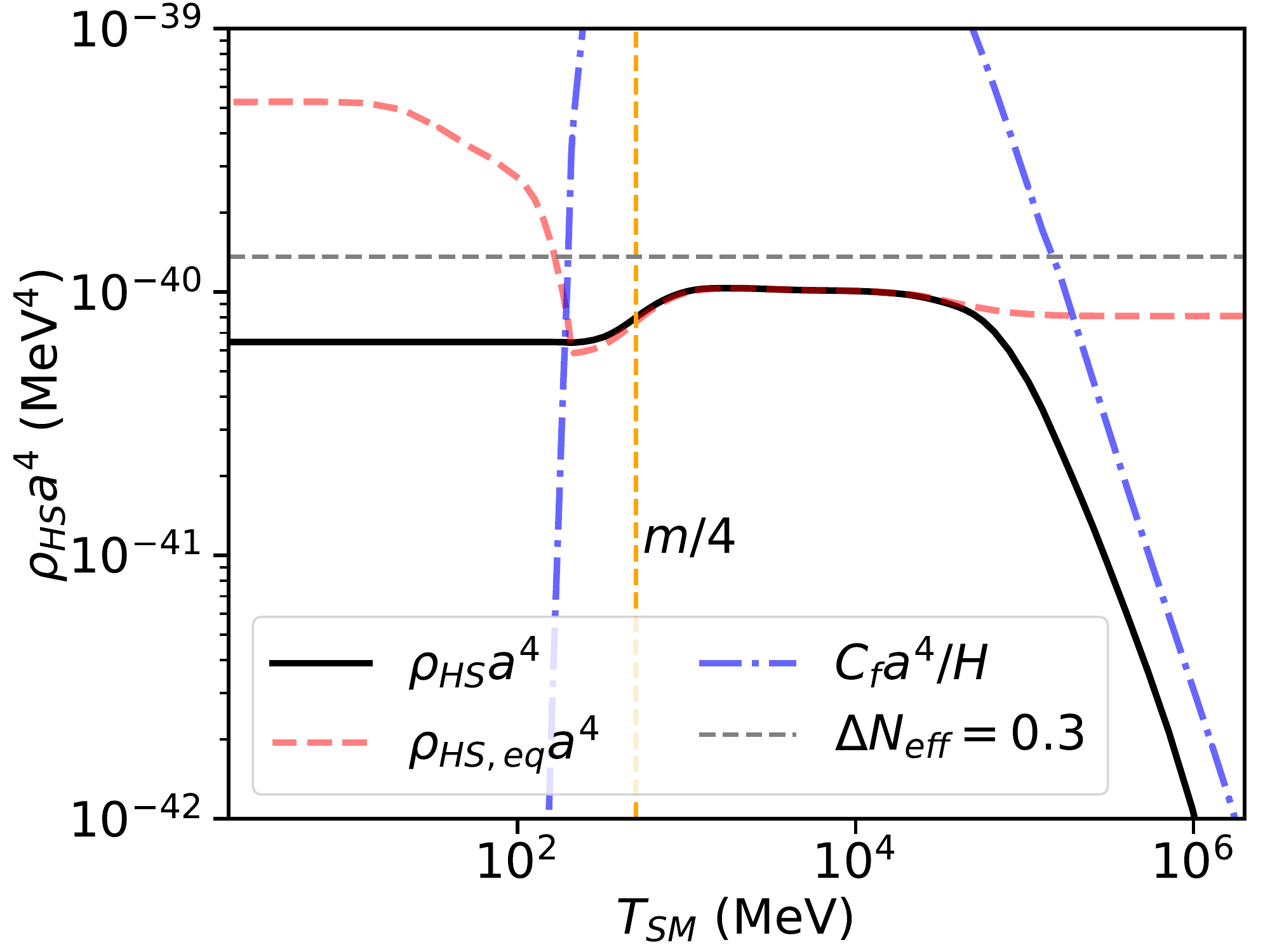}
		\end{subfigure} 
		\caption{Evolution of the comoving HS energy density (solid black) as a function of SM temperature for MCP mass and charges $\{m,Q\}=\{10^2\ \textrm{MeV}, 2\times 10^{-8}\}$ (left) and $\{m,Q\}=\{10^4\ \textrm{MeV}, 8\times 10^{-5}\}$ (right). The red dashed line shows $(\pi^2/30)g_{\rm HS}(T_{\rm SM})T_{\rm SM}^4a^4$, and the blue dot-dashed line is $\mathcal{C}_{\rm f}a^4/H$. The vertical orange dashed line marks $T_{\rm SM}=m/4$. The horizontal black dashed line marks the dark radiation density that produces $\Delta N_{\rm eff}= 0.3$. For an out-of-equilibrium HS, the  final dark radiation density depends on the maximum value of $\mathcal{C}_{\rm f}a^4/H$, while for a HS that thermalizes with the SM plasma the  dark radiation density depends on the decoupling temperature.}
		\label{fig:mcp_densityevolve}
	\end{figure}

	To develop some intuition about the evolution of these rates, and their impact on the resulting dark radiation density, in figure~\ref{fig:mcp_densityevolve} we show the evolution of $\rho_{\rm HS}a^4$ (black line) along with $\mathcal{C}_{\rm f}a^4/H$ (blue dot-dashed line) after numerically solving the Boltzmann equations given in  eq.~\eqref{eq:Boltz_eps_full_mcp}. The red-dashed line shows the evolution of $\rho_{\rm HS,eq}a^4=[\pi^2g_{\rm HS}(T_{\rm SM})/30](aT_{\rm SM})^4$.\footnote{We use the same scale factor for both the red dashed and black solid lines, which is obtained after numerically solving for $\rho_{\rm HS}$ indicated by the black line. The red line should not be confused with the solution for the comoving energy density for a HS always in thermal equilibrium.  The bump in the red line near $T_{\rm SM}=200$ MeV is due to the sudden decrease in $g_{*}$ below the QCD phase transition. The red line decreases for $T_{\rm SM}\lesssim m/4$ because the degrees of freedom in the HS decreases when MCPs become non-relativistic.} The ratio between the black and red lines is proportional to $(T_{\rm HS}/T_{\rm SM})^4$ and thus indicates how far away the HS is from equilibrating with the SM plasma. Two parameter choices are shown to illustrate the two different regimes for computing the resulting dark radiation density. The left panel shows a parameter point where $\mathcal{C}_{\rm f}/H$ is always smaller than $\rho_{\rm HS,eq}$, and consequently the HS remains out-of-equilibrium with the SM plasma. The right panel shows a second choice of parameters where the HS comes into thermal equilibrium with the SM for some period of time, indicated by the overlapping red and black lines. In both panels, the initial hidden sector energy density is small compared to the energy injection from the SM, $\rho_{\rm HS} < \mathcal{C}_{\rm f}/H$, and the evolution of $\rho_{\rm HS}$ is driven by the energy injection, giving the initial increase in $\rho_{\rm HS}a^4$. 
	
	In the left panel of figure \ref{fig:mcp_densityevolve}, energy injection into the HS ceases to be important after $\mathcal{C}_{\rm f}$ becomes Boltzmann-suppressed and $\mathcal{C}_{\rm f}/H$ falls below $\rho_{\rm HS}$. In particular, $\mathcal{C}_{\rm f}a^4/H$ attains its maximum around $T_{\rm SM}=m/2$, but it is not until $T_{\rm SM}=m/4$ (yellow dashed line) that energy injection into the HS effectively ends. For this choice of parameters, the HS does not come into thermal equilibrium with the SM, and consequently the final value of $\rho_{\rm HS}a^4$ can be estimated from the maximum value of $\mathcal{C}_{\rm f}a^4/H$. As $\Delta\Neff$ parametrizes the energy density of dark photons, it constrains the maximum value of $\mathcal{C}_{\rm f}a^4/H$, which is proportional to $Q^2M_{\rm Pl}/m$.
	
	In the right panel, $\mathcal{C}_{\rm f}/H$ grows until it exceeds $\rho_{\rm HS,eq}$ and subsequently the HS thermalizes with the SM plasma.  The two sectors remain in equilibrium until $\mathcal{C}_{\rm f}/H$ falls below $\rho_{\rm HS,eq}$. The final value of $\rho_{\rm HS}a^4$ is given by $\rho_{\rm HS,eq}a^4$ evaluated at $T_d$, where $T_d$ is the temperature below which HS thermally decouples from the SM plasma, $H(T_d)=\Gamma_E$. Consequently, if the HS thermalizes with the SM plasma, measurements of $\Neff$  probe $T_d$, which is only logarithmically sensitive to $Q$.
	
	We now separately study the regimes where the HS remains out of equilibrium with the SM plasma and where it equilibrates.
	
	\paragraph{Dark radiation production in the out-of-equilibrium regime:}

	For scenarios where the HS remains out of equilibrium with the SM plasma, the Boltzmann equations simplify because for $T_{\rm HS}\ll T_{\rm SM}$, the cooling of the SM plasma due to millicharge interactions is negligible, and the Hubble rate is dominated by $\rho_{\rm SM}$. Consequently, the SM plasma evolves adiabatically  and we only need to integrate the Boltzmann equation for $\rho_{\rm HS}$.
	
	Integrating the remaining Boltzmann equation for $\rho_{\rm HS}$ is non-trivial because the collision term depends on both $T_{\rm SM}$ and $T_{\rm HS}$. In particular, for $T_{\rm HS}\ll T_{\rm SM}$, $\mathcal{C}\approx\mathcal{C}_{\rm f}$,  and while $\mathcal{C}_{\rm f} = \mathcal{C}_{\rm f}(T_{\rm SM})$ for $s$-channel processes, for Coulomb scattering processes $\mathcal{C}_{\rm f} = \mathcal{C}_{\rm f}(T_{\rm HS}, T_{\rm SM})$. The energy transfer from Coulomb scattering process dominates over that from $s$-channel processes for the regions of parameters space that saturate $\Delta\Neff = 0.3$ (see section~\ref{mcp:boltz}). However, if $\Delta\Neff$ is constrained to smaller values by future experiments, the HS will be constrained to regions of parameter space with lower temperatures, and consequently, the contribution from Coulomb scattering processes will become less important compared to the contribution from $s$-channel processes. 
	
	In order to obtain a simple expression for a conservative lower bound on the asymptotic dark radiation density, we neglect the Coulomb scattering processes.  This allows us to take
	$\mathcal{C}_{\rm f} = \mathcal{C}_{\rm f}(T_{\rm SM})$. Then with the additional simplifying assumption that $w_{\rm SM} = w_{\rm HS} = 1/3$, we can integrate the Boltzmann equation for $\rho_{\rm HS}$ to obtain
	\begin{align}\label{eq:C_integrate}
		\left(\frac{\rho_{\rm HS}}{\rho_{\rm SM}}\right)_F-\left(\frac{\rho_{\rm HS}}{\rho_{\rm SM}}\right)_I\approx\int_{T_{{\rm SM},F}}^{T_{{\rm SM},I}}\frac{dT_{\rm SM}}{T_{\rm SM}}\frac{\mathcal{C}_{\rm f}}{H\rho_{\rm SM}}=\frac{\sqrt{3}M_{\rm Pl}}{(g_{*}\pi^2/30)^{3/2}}\int_{T_{{\rm SM},F}}^{T_{{\rm SM},I}}\frac{dT_{\rm SM}}{T_{\rm SM}^7}\mathcal{C}_{\rm f}.
	\end{align}

	Because the annihilations of SM fermions into MCPs typically dominates the $s$-channel energy transfer processes, we  focus on its contribution to the production of dark photons. In appendix~\ref{sec:schannel_quant}, we compute the collision term describing the forward energy transfer for these annihilation processes. 	The corresponding collision term for the forward energy transfer in SM fermion annihilations into MCPs is given by\footnote{While deriving eq.~\eqref{eq:Cf_gondolo_mcp_e} we make two key approximations. First, we neglect the Pauli-blocking effect from MCPs; second, we assume $T_{\rm SM}\gg m_f$. The first approximation is valid in the parameter space where MCPs are produced out-of-equilibrium with $T_{\rm HS}\ll T_{\rm SM}$. The second approximation has negligible impact on the production of dark radiation for $m\gg m_f$ because MCP production is Boltzmann-suppressed by the time $T_{\rm SM}\sim m_f$, while for $m<m_f$ the energy injection is dominated by lighter fermions that are relativistic during $T_{\rm SM}\sim m$.}
	\begin{align}\label{eq:Cf_gondolo_mcp_e}
		\mathcal{C}_{\rm f}^{\rm an}=&\sum_f\frac{1}{32\pi^4}\int_{4\max(m_f,m)^2}^{\infty} ds (s-4m_{f}^2)s\sigma_{ff\rightarrow \psi\bar{\psi}}(s)T_{\rm SM}G_{\zeta_f}(\sqrt{s}/T_{\rm SM}),
	\end{align}
	where $\sigma_{ff\rightarrow \psi\bar{\psi}}$ is the spin-summed center-of-mass (CM) frame cross-section (see eq.~\ref{eq:cross_ff_ann_Z}) and the summation runs over all SM fermions. The dimensionless function  $G_{\zeta_f} (z)$, given by eq.~\eqref{eq:G_def}, is determined by the quantum statistical distribution $f(p)=[e^{-E/T}+\zeta]^{-1}$, where $\zeta=1$ for fermions and $\zeta=-1$ for bosons. In the limit when SM fermions can be approximated to have a Maxwell-Boltzmann distribution ($\zeta_f\rightarrow0$), $G_{\zeta_f} (z)$ asymptotes to the second-order modified Bessel function of the second kind, $K_2 (z)$, and eq.~\eqref{eq:Cf_gondolo_mcp_e} then matches with the well-known result of Ref.~\cite{Gondolo:1990dk}.
	
	The integral on the RHS of eq.~\eqref{eq:C_integrate} can be simplified for the collision term of eq.~\eqref{eq:Cf_gondolo_mcp_e} by first rewriting the integral as
	\begin{align}\label{eq:mcp_temp}
		\int_{T_{{\rm SM},F}}^{T_{{\rm SM},I}}\frac{dT_{\rm SM}}{T_{\rm SM}^7}\mathcal{C}_{\rm f}^{\rm an}=&\sum_f\frac{1}{32\pi^4}\int_{4\max(m,m_f)^2}^{\infty} ds \frac{(s-4m_f^2)}{s\sqrt{s}}\sigma_{ff\rightarrow \psi\psi}\int_{x_F}^{x_I}\frac{dx}{x^6}G_{\zeta_f}\left(\frac{1}{x}\right),
	\end{align}
	where $x=T_{\rm SM}/\sqrt{s}$. One can show, to an excellent approximation, that the integration limits for $x$ can be replaced by 0 and infinity for $T_{{\rm SM},F}\ll \max(m,m_e)\ll T_{{\rm SM},I}$.\footnote{This approximation is possible for two reasons. First, the terms outside of the $x$ integral peak at energy scale $\sqrt{s}\sim \max(m,m_e)$. Second, the integrand of the $x$ integral goes to 0 as $x\rightarrow\infty$ and as $x\rightarrow0$. Thus, as long as $T_{{\rm SM},F}\ll\max(m,m_e)\ll T_{{\rm SM},I}$, the total integral is insensitive to the initial and final temperatures.} With this approximation the integral over $x$ yields a factor of $15\pi\kappa_{\zeta_f}/2$, where $\kappa_{1}=0.80$, $\kappa_0=1$ and $\kappa_{-1}=1.5$. Taking into account quantum statistics in the phase-space distribution of fermions therefore leads to a correction of about $20\%$ to the final dark photon density. This result is indicative of the size of quantum-statistical effects in all energy transfer processes we considered (including Coulomb scattering).
	
	Numerically, we find that the integral over the collision term obtains its asymptotic value at $T_{\rm SM}\sim \max(m,m_e)/4$. The dependence on $m_e$ is a consequence of the fact that for $T_{\rm SM}\ll m_e$ the abundance of all electromagnetically-charged SM fermions is Boltzmann-suppressed.
	
	Using the simplified collision integral, we find the fraction of energy transferred from the SM plasma into the HS is given by
	\begin{align}\label{eq:density_ratio_an_mcp}
		\left(\frac{\rho_{\rm HS}}{\rho_{\rm SM}}\right)_{\rm leak}\equiv \left(\frac{\rho_{\rm HS}}{\rho_{\rm SM}}\right)_{T_{\rm SM}= \Lambda}-\left(\frac{\rho_{\rm HS}}{\rho_{\rm SM}}\right)_I\approx\frac{15\sqrt{3}}{64\pi^3[g_{*}(4\Lambda)\pi^2/30]^{3/2}}\frac{M_{\rm Pl}}{\Lambda}\times L,
	\end{align}
	where
	\begin{align}\label{eq:def_leak_mcp}
		\Lambda\equiv \frac{\max(m,m_e)}{4}, && L=\Lambda\kappa_{1}\sum_f\int_{4\max(m,m_f)^2}^{\infty} ds \frac{(s-4m_f^2)}{s\sqrt{s}}{\sigma}_{ff\rightarrow \psi\psi}(s).
	\end{align}
	The energy injection decoupling temperature, $\Lambda$, determines the SM temperature below which energy injection ceases to be important, and the leak factor, $L$, parameterizes the leakage of energy from the SM plasma due to BSM interactions. While the above approximation assumes a constant $g_*$ until $T_{\rm SM}<\Lambda$, numerically we find that $\left(\rho_{\rm HS}/\rho_{\rm SM}\right)_{T_{\rm SM}=\Lambda}$ is primarily sensitive to $g_*$ at $T_{\rm SM}=4\Lambda$. Finally, this result is only valid as long as the HS does not thermalize with the SM, or equivalently, if $\left(\rho_{\rm HS}/\rho_{\rm SM}\right)_{T_{\rm SM}=\Lambda}<(g_{\rm HS}/g_*)_{T_{\rm SM}=\Lambda}$.
	
	The leak factor $L$ has a simple analytical form when the MCP mass $m$ is not close to  any of the SM fermion masses. Using the cross-section for fermion annihilation to MCP pairs given in eq.~\eqref{eq:cross_ff_ann_Z}, and neglecting $Z$-mediated contributions, we obtain
	\begin{align}\label{eq:leak_mcp}
		L\approx \kappa_{1}\sum_f^{m_f<4\Lambda}\frac{3\pi^2\alpha^2 Q^2Q^2_fN_c(f)}{8},
	\end{align}
	where $\alpha$ is the fine structure constant, $Q_f$ and $N_c(f)$ are the charge and color factor of the SM fermions, respectively, and the sum over $f$ runs over all SM fermions that are relativistic at $T_{\rm SM}\sim 4\Lambda$. 
	
	To evaluate the  final dark photon energy density we adiabatically evolve $\rho_{\rm HS}$ from the end of energy injection until recombination, $(\rho_{A'}a^4)_{\rm CMB}=\left(\rho_{\rm HS}/\rho_{\rm SM}\right)_{\rm leak}(\rho_{\rm SM}a^4)_{T_{\rm SM}=\Lambda}$. The dark photons act as free-streaming dark radiation and cause a shift in $N_{\rm eff}$ given in eq.~\eqref{eq:Dneff_mcp}. Requiring that the resulting $\Delta N_{\rm eff}$ remains below the $(\Delta\Neff)_{\rm max}$ upper bound set by CMB measurements limits the value of the charge to
	\begin{multline}\label{eq:mcp_analytical_Neff}
		Q^2<\frac{10^{-14}}{\kappa_{1}}\!\left( \frac{1/137}{\alpha}\right)^2\!\left(\frac{g_*(\Lambda)}{10}\right)^{1/3}\!\!\left(\frac{g_{*}(4\Lambda)}{10}\right)^{3/2}\!\!\left( \frac{4}{\sum_fQ_f^2N_c(f)}\right)\! \left(\frac{(\Delta\Neff)_{\rm max}}{0.3}\right)\!\left(\frac{\Lambda}{\rm GeV}\right).
	\end{multline}
	In deriving eq.\ \eqref{eq:mcp_analytical_Neff}, we set $g_{*s}(T_{\rm CMB})=3.94$ and approximated $g_{*s}(\Lambda)= g_{*}(\Lambda)$, where $g_{*s}$ counts the effective entropic degrees of freedom in the SM and $T_{\rm CMB}=0.25$ eV is the temperature of photons near recombination. The constraint on $Q$ for $m>m_e$ is roughly proportional to $\sqrt{m}$, with the proportionality constant determined by $\sqrt{(\Delta\Neff)_{\rm max}}$. Note that taking into account the Fermi-Dirac statistics of SM fermions weakens the constraint on $Q$ by $\sim10\%$. 
	
	\begin{figure}
		\begin{subfigure}{.5\textwidth}
			\includegraphics[width=1.00\textwidth]{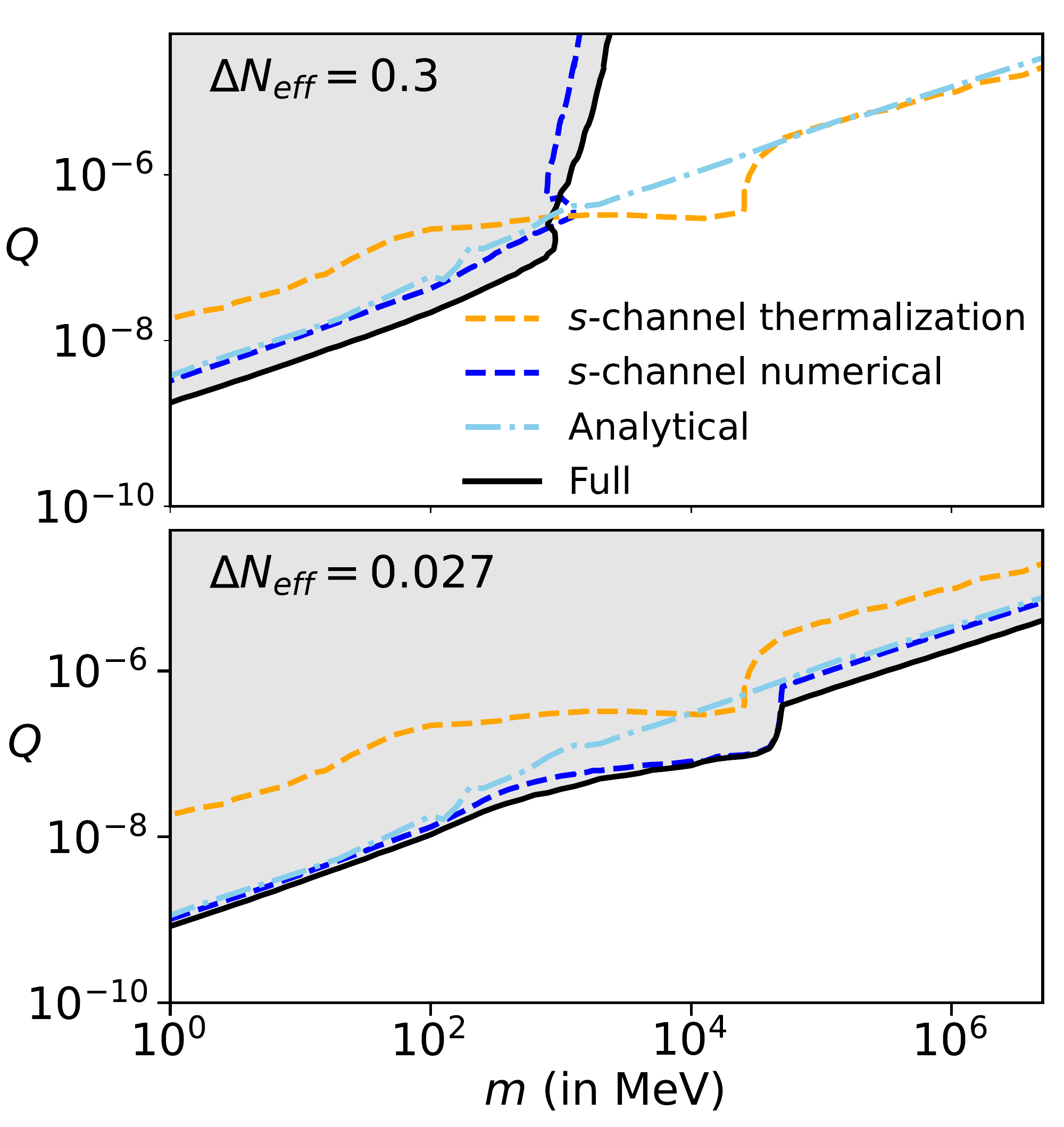}
		\end{subfigure}
		\begin{subfigure}{.5\textwidth}
			\includegraphics[width=1.00\textwidth]{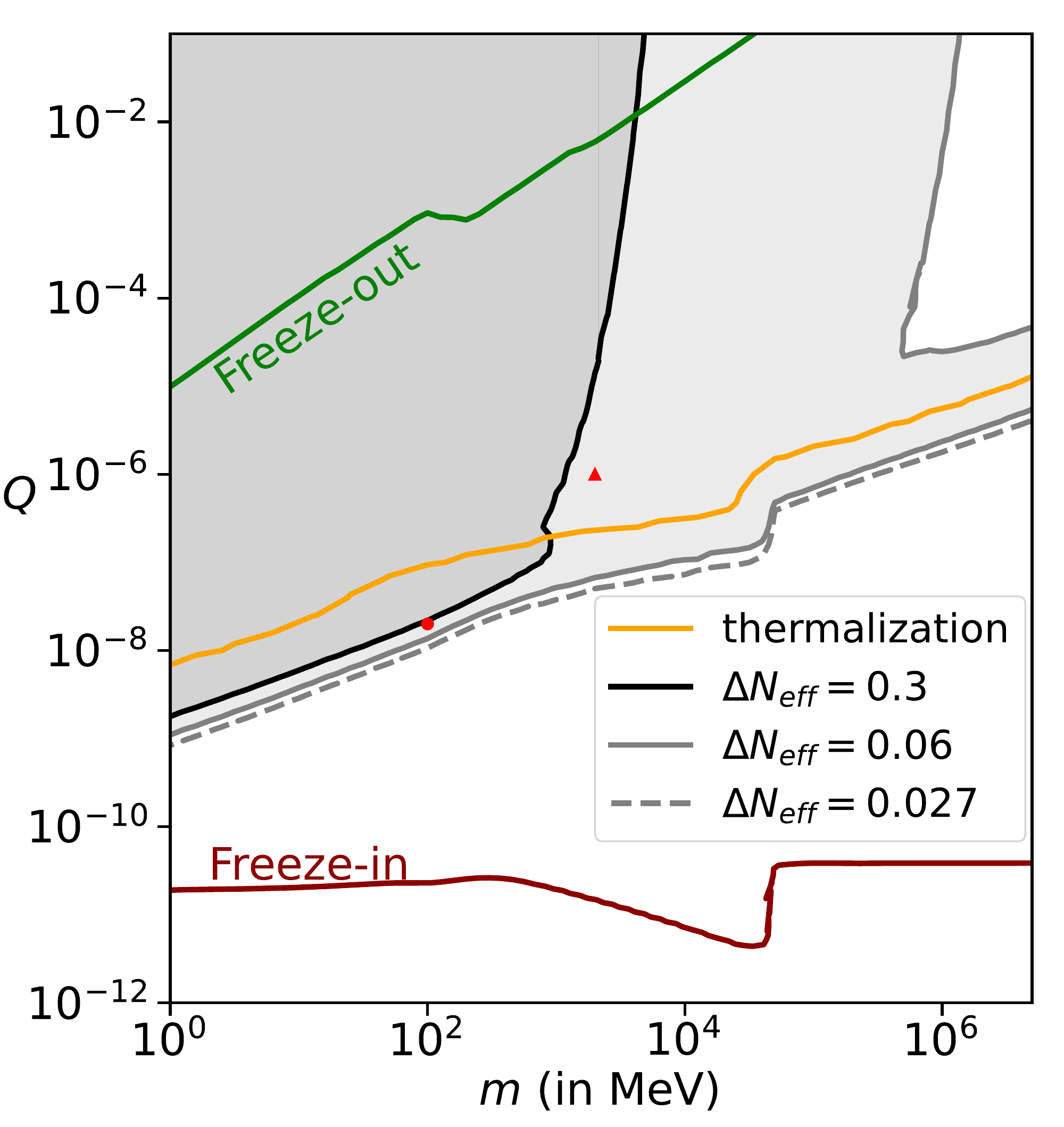}
		\end{subfigure} 
		\caption{\textbf{Left}: Solid black lines mark the parameter space for the MCP model that yields $\Delta\Neff =0.3$ (top) and $\Delta\Neff =0.027$ (bottom). The blue dashed lines show the numerical solution after neglecting contributions from Coulomb scattering processes. The light blue dot-dashed line is our analytical approximation to the blue dashed line, as given in eq.~\eqref{eq:mcp_analytical_Neff}. The orange dashed line marks the values of $Q$  at which the HS thermalizes with the SM plasma after neglecting Coulomb scattering processes. The exact $\Neff$ constraint is well described by the constraint calculated with only $s$-channel processes as $\Neff$ measurements are improved. \textbf{Right}: Solid black, solid gray and dashed gray lines mark the MCP parameter space that yields $\Delta\Neff=0.3$, $\Delta\Neff=0.06$, and $\Delta\Neff=0.027$, respectively, and are the same as those in figure~\ref{fig:mcp_constraint}. The orange line marks the parameter space above which the hidden sector thermalizes with the SM plasma. The green and maroon lines mark the parameter space where the MCP relic density matches the observed dark matter density via freeze-out \cite{Creque-Sarbinowski:2019mcm} and freeze-in \cite{Chu:2011be}, respectively. For values of $Q$ relevant for $\Neff$ constraints, almost all MCPs produced in the early universe must annihilate into dark photons to avoid overclosure of the universe. 
		}
		\label{fig:mcp_phasespace}
	\end{figure}
	
	In the left panels of figure~\ref{fig:mcp_phasespace}, the dark blue dashed lines show the values of the parameters that saturate  various $\Neff$ thresholds.  These points are evaluated by numerically solving the Boltzmann equations after including all $s$-channel energy transfer processes but not the $t$-channel Coulomb scattering processes. We compare this $s$-channel result with the full result, which includes Coulomb scattering processes, given by the black solid lines. Note that the agreement between the full and the $s$-channel-only results improves as $\Delta\Neff$ is restricted to smaller values. The light blue dot-dashed lines show the analytical result given by eq.~\eqref{eq:mcp_analytical_Neff}. Our analytical result does not include the contribution from $Z$-boson decays and hence underestimates the dark radiation density in the range 1 GeV$\lesssim m\lesssim$ 40 GeV in the bottom left panel of figure~\ref{fig:mcp_phasespace}.\footnote{One can straightforwardly incorporate $Z$-boson decays into the approximate analytical treatment by substituting the  corresponding collision term, given in eq.~\eqref{eq:Zdecay}, into eq.~\eqref{eq:C_integrate}. We omit this calculation for brevity.}

	\paragraph{Dark radiation production in the equilibrium regime:} 

	The analysis in the previous section is only valid when the HS remains out of equilibrium with the SM plasma. However, starting in the out-of-equilibrium regime, as one follows a contour of constant $\Delta N_{\rm eff}$ by increasing $m$, the value of $Q$ increases. At some point the coupling can become large enough that the HS  thermalizes with the SM. Once the sectors are thermalized, the dark radiation density is no longer sensitive to the maximum of the forward energy transfer $\mathcal{C}_{\rm f}a^4/H$.  Instead, the final dark radiation density is determined by the temperature $T_d$ at which the HS decouples from the SM. This decoupling temperature is determined by the Boltzmann suppression of the collision term, and is principally determined by the mass of the  MCP, while remaining only weakly dependent on the coupling $Q$. This is illustrated in the right panel of figure~\ref{fig:mcp_densityevolve}, which shows the evolution of the densities for a parameter point where the HS and SM thermalize. Here, decoupling occurs with the Boltzmann suppression of the collision term at $T \sim m/4$.
	
	The orange lines in figure~\ref{fig:mcp_phasespace} mark the values of $Q$  above which the HS thermalizes with the SM plasma for a given mass $m$. In the left panel, the orange lines are plotted after considering only $s$-channel energy transfer processes while in the right panel they are plotted after including all processes. In the left panel, the $s$-channel result that saturates $\Delta\Neff = 0.3$  becomes largely insensitive to the coupling $Q$ once the curve crosses above the $s$-channel thermalization contour; similar weakening occurs in the right panel for the full result. 
	
	We can determine the thermalization threshold, the mass scale beyond which we can no longer use the out-of-equilibrium result in eq.~\eqref{eq:mcp_analytical_Neff}, as follows. On the one hand, a given relic dark radiation density, or value of $\Delta \Neff$, can be translated to a decoupling temperature, $T_d$, by assuming that entropy is separately conserved in the HS and SM sectors after $T_d$. This leads to the implicit relation 
	\begin{align}\label{eq:def_Td}
		\frac{8}{7}\left(\frac{11}{4}\right)^{4/3}\left(\frac{g_{*s}(T_{\rm CMB})}{g_{*s}(T_d)}\right)^{4/3} \frac{g_{\rm HS}}{2} =  \Delta\Neff,
	\end{align}
	which can be solved to determine $T_{d}(\Delta\Neff ,g_{\rm HS})$.\footnote{ 
		Note that there is a many-to-one map from $T_d$ to $\Delta\Neff$ because $g_{*s}(T_d)$ is constant away from mass thresholds. For $\Delta\Neff$ values that exactly coincide with regions where $g_{*s}(T_d)$ is constant, we calculate $T_{d}(\Delta\Neff ,g_{\rm HS})$ by finding the minimum $T_d$ that satisfies eq.~\eqref{eq:def_Td}.}  This expression for $T_{d}(\Delta\Neff ,g_{\rm HS})$ is independent of the masses and couplings in the hidden sector, depending only on the effective number of degrees of freedom.  
	
	On the other hand, given a model, in this case the MCP model, we can compute the decoupling temperature directly from the collision term by 	setting the energy transfer rate $\Gamma_E(T)\equiv \mathcal{C}_{\rm f}(T)/\rho_{\rm HS,eq}(T)$ equal to the Hubble rate at $T_d$.  This condition determines the decoupling temperature in terms of the model parameters $Q$ and $m$, $T_d(Q,m)$. 
	Consequently, when the HS is thermalized with the SM plasma, the contour in MCP parameter space that yields a given value of $\Delta\Neff$ is found by setting
	\begin{align}
		T_d(Q,m)= T_{d}(\Delta\Neff ,g_{\rm HS}).
	\end{align}
	
	The energy transfer rate $\Gamma_E$ increases compared to the Hubble rate until  $T_{\rm SM}\sim m/2$, after which it starts decreasing. Consequently, the decoupling temperature has to be smaller than $~m/2$. Thus the lowest value of $m$ for which the HS can be in equilibrium with the SM plasma for a given $(\Delta\Neff)_{\rm max}$ is determined by $T_{d}(\Delta\Neff ,g_{\rm HS})$. Empirically we find that the Boltzmann suppression of $\Gamma_E$ becomes prohibitive for $T_{{\rm SM}}\lesssim m/4$, and thus the precise location of the decoupling temperature becomes logarithmically sensitive to the value of $Q$ for $T_d<m/4$. Therefore, the value of $m$ above which the dark radiation constraint on $Q$ become exponentially weak occurs at
	\begin{align}\label{eq:thresh_mcp}
		m_{\rm th}\equiv4T_{d}[(\Delta\Neff)_{\rm max} ,g_{\rm HS}].
	\end{align}
	Notice that the evaluation of $m_{\rm th}$ is independent of the strength of energy transfer processes and only depends on the sensitivity of the $\Neff$ measurement and the degrees of freedom in the HS. Consequently, eq.~\eqref{eq:thresh_mcp} does not depend on the detailed calculation of $\mathcal{C}$, and in particular whether we do or do not include contributions from Coulomb scattering.
	
	If future CMB missions continue to see an agreement with the SM value of $\Neff$, the thermalization threshold  $m_{\rm th}$ will be pushed to larger values. The gray solid and dashed lines in the right panel of figure~\ref{fig:mcp_phasespace} show the values of the parameters that lead to $\Delta\Neff = 0.06$ and $\Delta\Neff = 0.027$, respectively. The excluded regions extend to much larger values of $m$ because more of the parameter space is required to have the HS  remain out of equilibrium with the SM plasma. For $\Delta\Neff<0.027$, there is no allowed thermalization threshold.
	
	The exponential behavior of the constant $\Delta\Neff$ contours for $m>m_{\rm th}$  eventually stops at sufficiently large values of $Q$, when direct energy transfer from SM into dark photons through off-shell MCPs become larger than the Boltzmann-suppressed energy transfer into on-shell dark fermions. These off-shell processes depend on additional model parameters, in particular the dark gauge coupling constant, and are beyond the scope of the paper. 
	
	\paragraph{Requirement of chemical equilibrium:} 
	
	Our analysis assumes that the HS energy density can be treated as a whole, including both the MCP and the dark photon, instead of tracking their energy densities separately.  This assumption is strictly valid when the HS is in internal chemical equilibrium throughout the period of energy transfer, which is not necessarily true everywhere throughout our parameter space.   However, this assumption of internal chemical equilibrium is only critical to our final result for the dark radiation abundance in the regions near and above the thermalization threshold(s) for the MCP, where it does hold (as we discuss below). Below the thermalization threshold, where the MCPs remain out-of-equilibrium with the SM, the assumption of internal chemical equilibrium remains an excellent approximation as long as (i) the HS energy density is dominated by radiation throughout the period of energy transfer, 
	and (ii) we can treat all the entropy carried by the MCPs as deposited into dark radiation, rather than the SM, after it becomes non-relativistic. Given these two conditions, the detailed evolution of the MCP number density itself is unimportant to the final dark radiation abundance. In fact condition (ii) follows from condition (i) when the MCPs are out of equilibrium with the SM, as requiring the HS to be dominated by radiation means that almost all the produced MCPs must rapidly annihilate, and  if the MCP is out of equilibrium with the SM, then necessarily $n^2_{\psi}\langle\sigma v\rangle_{\psi\psi\rightarrow \gamma\gamma}<H$.  Thus the MCP must dominantly annihilate into dark photons.

	The condition that almost all the produced MCPs efficiently annihilate into dark radiation {\em is} met in the regions of our parameter space relevant for current and forecast out-of-equilibrium constraints, given the mild constraint on the dark gauge coupling $e'$ that follows from requiring that the relic MCP abundance does not overclose the universe, as we now argue.	The green line in  figure~\ref{fig:mcp_phasespace} indicates where the freezeout of SM annihilations into pairs of MCPs would produce the observed DM relic density in the absence of dark photons, i.e., if the MCP's only annihilation channel is to SM fermions \cite{Creque-Sarbinowski:2019mcm}. Meanwhile the maroon line indicates where the freezein production of MCPs from the SM produces the observed DM relic density, again turning off the MCP annihilations into dark photons \cite{Chu:2011be}.  As current and future $\Neff$ constraints lie between these two lines (except for a small region above the thermalization threshold in the case of current constraints), in the region of parameter space  relevant for evaluating these constraints, SM processes alone overproduce MCPs by multiple orders of magnitude.  Thus the dark gauge coupling constant must be large enough to enable the vast majority of MCPs to annihilate efficiently into dark photons.  If this condition is not met, the model is excluded simply by overclosure; our $\Neff$ analysis applies to the surviving model parameter space where $e'$ is large enough to avoid overclosure, and otherwise does not depend on the detailed value of $e'$.  Thus avoiding overclosure alone establishes the requirement for out-of-equilibrium case discussed above, which suffices as long as the final result for dark radiation density does not depend on the evolution of $g_{{\rm HS}}$ with temperature.
	
	To accurately determine the production of dark radiation when the hidden sector is close to the thermalization threshold, we  need to track the evolution of $g_{{\rm HS}}$ with temperature, and thus the evolution of the MCP number abundance with temperature, up until $T_{{\rm HS}} \lesssim m/3$. Once $T_{\rm HS} < m/3$, the hidden sector equation of state is given by $w_{\rm HS}=1/3$ to an excellent approximation.  In particular this is necessary to accurately determine the location of the excluded strip running up to high masses in fig.~\ref{fig:mcp_constraint}. Using the results of Ref.~\cite{Fernandez:2021iti}, we have checked that internal chemical equilibrium for $T_{{\rm HS}} > m/3$ is indeed necessary if the relic MCPs are not to overclose the universe.

	\subsection{Dark radiation production in extended MCP models and implications for EDGES}\label{mcp:constraint}
	
	We have so far considered a minimal MCP model where only one fermion is charged under the dark $U(1)$ gauge symmetry. More generally,  the hidden sector may contain  multiple particles with dark charges. A full evaluation of the resulting dark radiation density in these models depends on the detailed spectrum of the hidden sector, including properties such as the number of particles and the values of their various dark charges. However, in this section, we show that a conservative lower bound on the dark photon density at recombination can be estimated  that is insensitive to such details. This conservative lower bound can then be used to place general constraints on the allowed parameter space of these models.
	
	The conservative lower bound on the dark photon density is obtained by considering only $s$-channel energy transfer processes (annihilations or decays) and considering only one MCP and one dark photon in the HS bath. On the one hand, if the HS equilibrates with the SM plasma, the final dark radiation density is largely insensitive to the specifics of the energy transfer processes but remains proportional to the degrees of freedom in the HS bath, $g_{\rm HS}$. Thus, minimizing the particles in the HS also minimizes the final value obtained for the dark radiation density. On the other hand, if the HS remains out-of-equilibrium with the SM plasma, the dark radiation density is determined by the energy transfer  from  the SM. While the energy transferred by $t$-channel scattering processes decreases as we increase $g_{\rm HS}$ (holding the total $\rho_{\rm HS}$ fixed), the energy transferred by $s$-channel processes is insensitive to $g_{\rm HS}$ as long as it is dominated by a single mediating species. 	Consequently, the dark radiation density cannot be smaller than that following from $s$-channel processes alone for an out-of-equilibrium hidden sector.
	
	This conservative lower bound on the dark radiation density can be translated directly into a lower bound on $\Delta N_{\rm eff}$. This lower bound has an immediate application to the MCP model proposed by Ref.~\cite{Liu:2019knx} to explain the anomalously small hydrogen spin temperature 	as measured by the EDGES experiment \cite{Bowman:2018yin}. Their model consists of two fermions that are charged under a dark $U(1)$ gauge symmetry. One fermion is the main component of dark matter, $\chi_1$, and the other fermion, $\chi_2$, constitutes a small fraction of dark matter. The particle $\chi_2$ is responsible for cooling hydrogen atoms via millicharge interactions and then transfers that heat to the dark matter bath via dark long range interactions. The dark photon mediating the long-range interaction is a light relativistic relic that contributes to $\Neff$.
	
	\begin{figure}
		\centering
		\includegraphics[width=\textwidth]{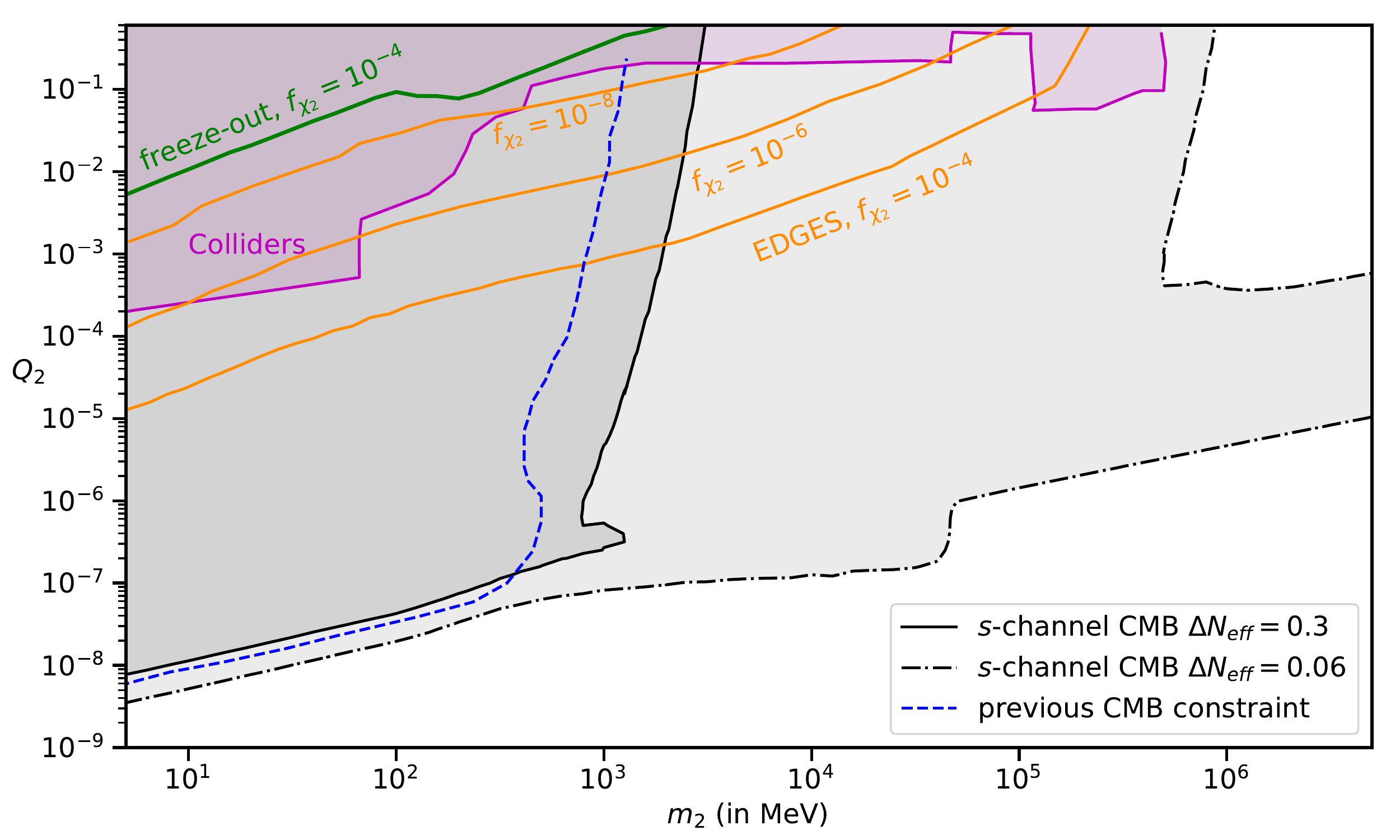}
		\caption{Constraints on the mass and millicharge of the millicharged particle in the context of extended models.  The orange lines mark the values of charge and mass of the MCP $\chi_2$  for which the model given in Ref.~\cite{Liu:2019knx} resolves the EDGES anomaly. The orange lines have been plotted after fixing the dark matter mass to 10 MeV and setting the fraction of $\chi_2$ density relative to dark matter to $f_{\chi_2}=10^{-4}$, $10^{-6}$, and $10^{-8}$ as indicated. The green line marks the values of $Q_2$  for which $\chi_2$ would obtain $f_{\chi_2}=10^{-4}$ in the absence of dark annihilation channels. The black solid and dot-dashed contours mark the parameter space that yields $\Delta N_{\rm eff}= 0.3$ and $\Delta N_{\rm eff}= 0.06$, respectively, after neglecting energy transfer from Coulomb scattering processes and assuming one millicharged particle in the hidden sector bath. The blue dashed contour is the CMB constraint derived in Ref.~\cite{Vogel:2013raa} for $\Delta N_{\rm eff}= 0.8$. The pink shaded regions marks the parameter space ruled out by SLAC \cite{Prinz:1998ua}, MiniBooNE \cite{Magill:2018tbb}, LEP \cite{Davidson:2000hf} and LHC \cite{Jaeckel:2012yz}.}
		\label{fig:mcp_edges}
	\end{figure}
	
	In figure~\ref{fig:mcp_edges} we show the parameter space in the model of Ref.~\cite{Liu:2019knx} that is consistent with various current and projected CMB measurements of $N_{\rm eff}$.  The orange lines in figure~\ref{fig:mcp_edges} show the values of the millicharge, $Q_2$, and mass, $m_2$, of the $\chi_2$ particle required to resolve the EDGES anomaly, as calculated in Ref.~\cite{Liu:2019knx}. The lines are plotted for fixed values of  dark charges and $\chi_1$ masses chosen such that the cooling of hydrogen atoms is maximized while remaining consistent with cosmological bounds from the CMB and BBN. 
	The black solid and dashed lines show the values of $\Neff$ computed using the conservative method described above that saturate the Planck and projected CMB-S4 $2\sigma$ bounds, respectively. The  contours below the thermalization threshold are well described by eq.~\eqref{eq:mcp_analytical_Neff}. Current measurements of $N_{\rm eff}$ already limit $m_2>$2 GeV, while future CMB experiments can completely rule out the MCP model proposed by Ref.~\cite{Liu:2019knx}. Since the dark radiation constraints we show here are largely insensitive to the details of the specific extended MCP model, they offer a powerful way to constrain model-building in this direction to explain the EDGES anomaly.

	Naively one might imagine that the dark radiation constraints on $Q_2$ can be circumvented if $\chi_2$ predominantly annihilates into SM particles rather than dark photons. However, for $\chi_2$ to resolve the EDGES anomaly, it must have significant couplings with a lighter dark particle in order to avoid being overproduced in the early universe. For instance, the solid green line in figure~\ref{fig:mcp_edges} marks the values of $Q_2$  that produce $\chi_2$ constituting a fraction $f_{\chi_2}=10^{-4}$ of dark matter density today if $\chi_2$ only has annihilation channels to SM fermions.  Assuming SM-only freezeout, the relic abundance of  $\chi_2$ increases below the green line by a factor of $1/Q_2^2$. Consequently, the values of $Q_2$  required to resolve the EDGES anomaly result in a $\chi_2$ relic abundance multiple orders of magnitude larger than what is required unless  $\chi_2$ has an additional annihilation channel.  The minimal possibility is that $\chi_2$ dominantly annihilates into the dark mediator that sources the requisite long-ranged interaction between $\chi_1$ and $\chi_2$.\footnote{An alternative non-minimal method to dilute the $\chi_2$ abundance is to have an unstable field preferentially reheat the SM plasma at some temperature $T_{\rm rh}<m_2$.} The produced dark mediator is then constrained by the $\Neff$ measurements, which consequently restricts $Q_2$ as shown in figure~\ref{fig:mcp_edges}. 
	
	Finally, applying the CMB $\Neff$ constraint to any MCP model assumes that the dark photon is free-streaming during recombination. 
	If the dark photon and the MCPs have sufficiently large self-interactions during recombination, they can instead form a fluid, and the dark photon would accordingly contribute to $N_{\rm fluid}$ instead of producing a neutrino-like signal. 
	The ability to form a fluid depends on the MCP relic abundance as well as the interaction between the MCP and the dark photon, both of which are determined by the dark coupling constant, $e'$. A more detailed analysis would be required to find the relevant values of $e'$ that can produce a self-interacting radiation bath without violating either unitarity or cosmological bounds. For such values of $e'$, one would instead have to look to $N_{\rm fluid}$ measurements, which are factors of $2-3$ less sensitive than measurements of $\Neff$ \cite{Baumann:2015rya}.  A future CMB-S4 constraint of $\Delta N_{\rm fluid}\lesssim 0.16$ would yield a thermalization threshold of $m_{2, {\rm th}}\sim$ GeV.

	\section{$B-L$ right-handed neutrinos}\label{sec:RH_nu}

	In this section we derive dark radiation constraints on the scenario where the global SM symmetry of baryon number minus lepton number ($B-L$) is promoted to a gauge symmetry.  This promotion requires the addition of three right-handed neutrinos to cancel gauge anomalies.  When these three additional neutrinos are light, they contribute to the energy budget of the Universe as dark radiation. Consequently, their energy density and the parameter space of the model are constrained by measurements of $N_{\rm eff}$.
	
	The gauged $B-L$ model is also constrained by fifth-force searches \cite{Wagner:2012ui}, stellar evolution \cite{Redondo:2013lna}, supernova 1987A \cite{Croon:2020lrf}, and collider experiments \cite{BaBar:2014zli, LHCb:2017trq, ATLAS:2017fih, Escudero:2018fwn, Bauer:2018onh, Bjorken:2009mm, Andreas:2012mt, Blumlein:2013cua}. Constraints on this model from  $\Neff$ measurements have been studied previously  in Ref.~\cite{Heeck:2014zfa} and updated in Ref.~\cite{Abazajian:2019oqj}.  
	Here we improve over previous studies by taking into account the out-of-equilibrium production of right-handed neutrinos. 
	
	This section is organized as follows. We begin in section \ref{sec:BMLmodel} by introducing the model and our conventions. In section~\ref{Bl:numerical}  we describe the relevant Boltzmann equations, detailing the approximations within which we work. We then solve the Boltzmann equations to find the model parameter space that saturates the $\Neff$ bounds from current and upcoming CMB experiments. Next, in section~\ref{BL:density} we analyze the evolution of the energy density in right handed neutrinos, $\rho_{\nu_R}$, and show that its final value is qualitatively changed depending on the lifetime of the $Z'$ boson.  Finally in section~\ref{BL:constraint}, we provide an analytical explanation of the features of the dark radiation constraint on the model parameter space.
	
	\subsection{The model}\label{sec:BMLmodel}
	
	The Lagrangian describing the interactions of the SM with the $B-L$ gauge boson $Z'$ and the right-handed neutrinos is given by 
	\begin{align}
		\label{eq:bmlmodel}
		\mathcal{L}=&-\frac{1}{4}F'_{\mu\nu}F^{\mu\nu}{}'+\frac{1}{2}M_{Z'}^2Z'_{\mu}Z'{}^{\mu}+g'Z'_{\mu}\sum_{i}\left[\frac{1}{3}(\bar{u}_i\gamma^{\mu}u_i+\bar{d}_i\gamma^{\mu}d_i)-\bar{e}_i\gamma^{\mu}e_i-\bar{\nu}_{L,i}\gamma^{\mu}\nu_{L,i}\right]\nonumber\\
		&-g'Z'_{\mu}\sum_{i}\bar{\nu}_{R,i}\gamma^{\mu}\nu_{R,i}.
	\end{align}
	Here, the index $i$ runs over the three generations of SM fermions, while $u$, $d$, $e$, $\nu_L$ and $\nu_R$ denote the up quark, down quark, electron, left-handed neutrino and right-handed neutrino counterparts of each generation. Above we have explicitly separated the interaction of the $Z'$ gauge boson with the $\nu_R$ from its interactions with the known SM fermions.  We consider the minimal version of the model where the three right-handed neutrinos form Dirac particles with the left-handed neutrinos after electroweak symmetry breaking.   Because the neutrinos are always relativistic during and prior to recombination, we ignore neutrino masses in the subsequent analysis and treat $\nu_L$ and $\nu_R$ as distinct Weyl fermions.
	The $Z'$ gauge boson has mass $M_{Z'}$, which can come from a Stueckelberg or a Higgs mechanism.  To remain as model-independent as possible, we ignore potential contributions to the dark radiation density arising from possible Higgs fields associated with $B-L$ breaking and focus on the irreducible contribution from the $Z'$ itself.\footnote{This is an excellent approximation when a $B-L$ Higgs is more massive than the $Z'$, and conservative in the case when it is not; this treatment is also applicable to the technically natural scenario where the $Z'$'s only interactions are the Stueckelberg mass and the coupling to the SM $B-L$ current as given in eq.~\ref{eq:bmlmodel}.}   
	
	Right-handed neutrinos are produced in this model as a result of the $B-L$ interactions with the Standard Model in the early Universe. Because they are approximately massless and sterile at late times, after the $Z'$ freezes out, these right-handed neutrinos are dark radiation and contribute to $\Neff$. Furthermore, for values of $g'$ allowed by current $\Neff$ constraints, the $B-L$ interactions with $\nu_L$ are significantly weaker than the weak interactions with $\nu_L$ prior to neutrino decoupling. We focus on the region of parameter space where dark radiation is produced prior to BBN, and thus before the weak interactions freeze out and the $\nu_L$ leave equilibrium. In this region of parameter space, the production of $\nu_R$ provides the major contribution to $\Delta \Neff$,
	\begin{align}\label{eq:DeltaN_nuR}
		\Delta\Neff=\frac{8}{7}\left(\frac{11}{4}\right)^{4/3}\frac{\rho_{\nu_R}}{\rho_{\gamma}}.
	\end{align}

	\subsection{Boltzmann equations and constraints for the $B-L$ model}\label{Bl:numerical}
	
	Right-handed neutrinos in this model are dominantly produced by $Z'$-mediated SM fermion annihilation. In part of the relevant parameter space, the $Z'$ bosons are long-lived, i.e., they do not decay within a Hubble time. Consequently, the energy transferred into $\nu_R$ can depend on the cosmic evolution of the on-shell $Z'$ density.  The relevant Boltzmann equations for this system need to track the evolution of both $Z'$ and $\nu_R$, and read
	\begin{align}\label{eq:Boltz_eps_full1}
		\frac{d\rho_{\rm SM}}{dt}+3H(1+w_{{\rm SM}})\rho_{\rm SM}=&-\mathcal{C}_{ff\rightarrow Z'}-\mathcal{C}^{\rm off}_{ff\rightarrow \nu_R\nu_R},\\
		\frac{d\rho_{Z'}}{dt}+3H(1+w_{Z'})\rho_{Z'}=&\mathcal{C}_{ff\rightarrow Z'}-\mathcal{C}_{Z'\rightarrow \nu_R\nu_R}, \label{eq:Boltz_eps_full2}\\
		\frac{d\rho_{\nu_R}}{dt}+4H\rho_{\nu_R}=&\mathcal{C}_{Z'\rightarrow \nu_R\nu_R}+\mathcal{C}^{\rm off}_{ff\rightarrow \nu_R\nu_R}.\label{eq:Boltz_eps_full3}
	\end{align}
	Here the Hubble rate is $H=\sqrt{\rho_{\nu_R}+\rho_{\rm SM}+\rho_{Z'}}/[\sqrt{3}M_{\rm Pl}]$, and the various $\mathcal{C}_i$ denote energy transfer collision terms from three processes: $\mathcal{C}_{ff\rightarrow Z'}$, for the inverse decay of SM fermions into $Z'$s;  $\mathcal{C}_{Z'\rightarrow \nu_R\nu_R}$, describing the decay of $Z'$s into right handed neutrinos; and $\mathcal{C}^{\rm off}_{ff\rightarrow \nu_R\nu_R}$, which 
	describes contact interactions between SM fermions and $\nu_R$, mediated by off-shell $Z'$s (see also Refs.~\cite{Pilaftsis:2003gt,Giudice:2003jh}). We include the quantum phase space distributions for initial state particles but ignore final state quantum effects in the evaluation of the collision terms.\footnote{Ignoring final-state quantum effects is an excellent approximation as long as the $Z'$ is out of equilibrium with both the $\nu_R$ and SM plasma. On the other hand, if the $B-L$ interactions are strong enough to thermalize the $Z'$ and the $\nu_R$ with the SM, the precise value of the collision term has only a marginal impact on the final densities of $Z'$ and $\nu_R$ and hence final-state quantum effects are not quantitatively important.} The detailed expressions for the collision terms are given in appendix~\ref{sec:BL_coll}.
	
		While evaluating the backward collision terms describing $\nu_R\nu_R \rightarrow ff$ and $\nu_R\nu_R\rightarrow Z'$, we assume that right-handed neutrinos are internally thermalized with a temperature $T_{\nu_R}\equiv [\rho_{\nu_R}/(g_{\nu_R} \pi^2/30)]^{1/4}$, where $g_{\nu_R}=2\times3\times 7/8$.   On the one hand, this assumption is unimportant if the two sectors do not thermalize as the backward collision term is negligibly small in comparison to the forward collision term for $T_{\nu_R} \ll T_{{\rm SM}}$. On the other hand, if they do thermalize then the assumption is automatically satisfied.  The transition regime, where the backward collision term can be important, is relevant for the current $\Neff$ constraints in the mass range $1\ {\rm GeV}<M_{Z'}<2$ GeV;  in this range a differential treatment of the phase space distribution of the right-handed neutrinos would be required to improve on our treatment.  For $\Delta\Neff<0.06$ we expect to be well into the out-of-equilibrium regime where the backward collision term is unimportant. 
		
	When right-handed neutrinos are in equilibrium with the SM plasma, the decoupling temperature is determined by either $Z'$ decays or contact interactions. When the right-handed neutrinos are out-of-equilibrium with the SM plasma, the energy transferred through $Z'$ decays and inverse decays is orders of magnitude larger than that via contact interactions.
	The forward energy transfer collision terms for $Z'\rightarrow ff$ and $Z'\rightarrow \nu_R\nu_R$ are given by
	\begin{align}
		\mathcal{C}^{\rm f}_{Z'\rightarrow ii}=M_{Z'}\Gamma_{Z'\rightarrow i}n_{Z'},
	\end{align}
	where $n_{Z'}$ is the number density of the $Z'$ and $\Gamma_{Z'\rightarrow i}$ is the $Z'$ decay width into particle species $i$. While computing $\Gamma_{Z'\rightarrow SM}$ we neglect the decays of $Z'$ bosons into hadrons for $M_{Z'}<2T_{\rm QCD}$, where we set the QCD transition scale at $T_{\rm QCD}=200$ MeV. For $M_{Z'}>2T_{\rm QCD}$, we include $Z'$ decays into free quarks in $\Gamma_{Z'\rightarrow SM}$. 
	
	We approximate $\rho_{Z'}=M_{Z'}n_{Z'}$ and $w_{Z'}=0$ in the Boltzmann equations, as appropriate for non-relativistic $Z'$ bosons. Most of the energy injection into $\nu_R$ occurs when the $Z'$ bosons are non-relativistic, as demonstrated explicitly below, and  therefore this approximation has a minimal effect on the final dark radiation density and the ensuing constraints.
	
	\begin{figure}
		\centering
		\includegraphics[width = \textwidth]{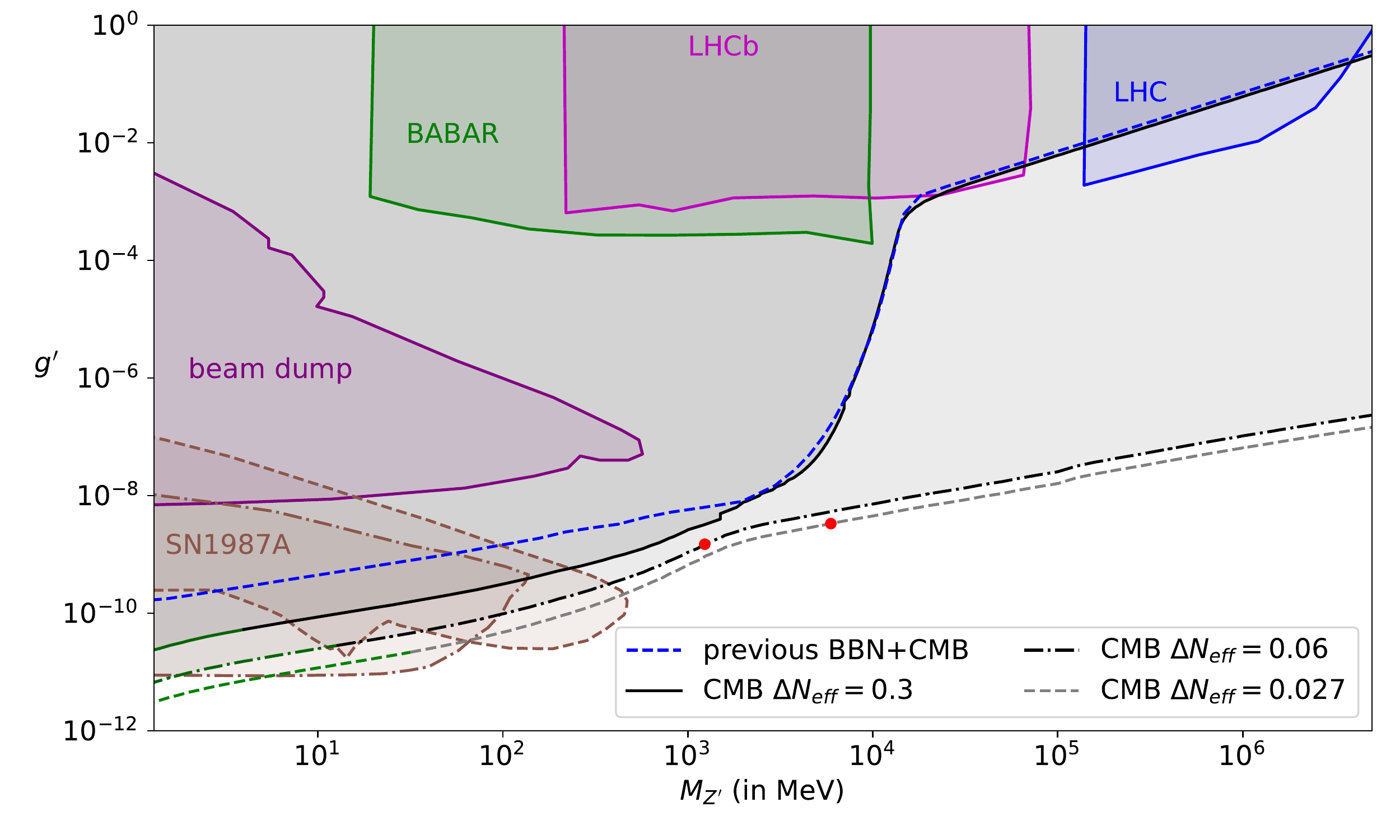}
		\caption{Constraints on $B-L$ gauge coupling and gauge boson mass. The black solid, black dot-dashed, and gray dashed contours mark the parameter space that yields $\Delta N_{\rm eff}= 0.3$, $\Delta N_{\rm eff}= 0.06$, and $\Delta\Neff=0.027$ respectively. These bounds correspond to the $2\sigma$ upper limit for Planck \cite{Aghanim:2018eyx}, $2\sigma$ upper limit for CMB-S4 \cite{Abazajian:2019eic}, and the sensitivity goal for future CMB experiments, and update the BBN+CMB constraints derived in Ref~\cite{Abazajian:2019oqj}, which are shown with the blue dashed contour. The red dots mark the points on the constant $\Neff$ curves below and to the left of which $\Gamma_{Z'}$ is smaller than the Hubble rate at $T_{\rm SM}=M_{Z'}/2$. The green color on our $\Delta\Neff$ contours marks the region where we expect $Z'$ decays into $\nu_L$ to contribute to $\Delta\Neff$ and alter our results by an $\mathcal{O}(1)$ factor. Brown lines show constraints from supernova 1987A from  Refs.~\cite{Croon:2020lrf} (dashed) and~\cite{Shin:2021bvz} (dot-dashed).  We also show constraints from BABAR \cite{BaBar:2014zli}, LHCb \cite{LHCb:2017trq}, LHC \cite{ATLAS:2017fih, Escudero:2018fwn}, and beam dump experiments \cite{Bauer:2018onh, Bjorken:2009mm, Andreas:2012mt, Blumlein:2013cua}.}
		\label{fig:BL_constraint}
	\end{figure}
	
	We are now ready to compute the final dark radiation density in $\nu_R$ by solving the Boltzmann equations given in eq.~\eqref{eq:Boltz_eps_full1}-\eqref{eq:Boltz_eps_full3}. We begin the evolution at an initial SM temperature $T_{\rm SM}\gg M_{Z'}$, setting $\rho_{Z'}=\rho_{\nu_R}=0$, and evolve forward until the end of energy injection.   In figure~\ref{fig:BL_constraint}, we show the contours of $g'$ as a function of $M_{Z'}$ that saturate the current one-tailed 2$\sigma$ upper limit from Planck \cite{Aghanim:2018eyx}, $\Delta\Neff = 0.3$ (black solid); the projected $2\sigma$ upper limit from CMB-S4 \cite{Abazajian:2019eic}, $\Delta\Neff = 0.06$ (black dot-dashed); and the threshold goal for future CMB experiments $\Delta\Neff =0.027$ (gray dashed). 
	
	The curves of constant $\Delta N_{\rm eff}$ in figure~\ref{fig:BL_constraint} have a number of key features. As in the MCP model, these curves have a thermalization threshold beyond which they are only logarithmically sensitive to $g'$. For the $\Delta \Neff = 0.3$ curve, the threshold is at $M_{Z'}\sim 1.7$ GeV, while for other contours displayed, there is no threshold. This is because $\Delta \Neff = 0.3$ allows three BSM Weyl fermions to decouple from the SM plasma before the QCD phase transition, but the smaller values $\Delta N_{\rm eff}= 0.06$ and $\Delta\Neff=0.027$ cannot accommodate so many new degrees of freedom ever thermalizing with the SM.   For $\Delta \Neff = 0.3$, the logarithmic sensitivity to $g'$  becomes a power law again above $g'\lesssim M_{Z'}/(16\ {\rm TeV})$ (see also \cite{FileviezPerez:2019cyn}), as the decoupling temperature goes from being determined by $Z'$ decays and inverse decays to being determined by contact interactions, described by $\mathcal{C}^{\rm off}_{ff\rightarrow \nu_R\nu_R}$. 
	
	The curves corresponding to  $\Delta\Neff = 0.06$ and lower (as well as the curve for $\Delta\Neff = 0.3$ below the thermalization threshold) are controlled by the out-of-equilibrium production of right-handed neutrinos. As we describe below, there are two qualitatively different out-of-equilibrium production mechanisms depending on the  ratio $\Gamma_{Z'}/H$ at $T_{\rm SM}\sim M_{Z'}/2$, where $\Gamma_{Z'}$ is the total decay width of $Z'$ bosons. The red dots on the curves indicate where $\Gamma_{Z'}$ is equal to the Hubble rate at $T_{\rm SM}=M_{Z'}/2$. Along the contours below and to the left of the red dots, the $Z'$ bosons become long-lived and we need to track their number density to evaluate dark radiation production. This key result, together with the usual out-of-equilibrium production of $\nu_R$, accounts for the difference between the results in this work and those previously obtained in Ref.\ \cite{Abazajian:2019oqj}, shown  in figure~\ref{fig:BL_constraint} as the blue dashed curve.
	
	Constraints on the $B-L$ gauge boson can also be derived by considering the production of $\nu_R$ in colliders or in supernova. In figure~\ref{fig:BL_constraint} we also show the regions of parameter space that are excluded by measurements from these other sources.  Current CMB constraints are already the leading probe of this hidden sector across much of parameter space, with LHC constraints taking over for masses above 100 GeV.  The $N_{\rm eff}$ measurements from future CMB experiments along with existing supernova measurements will provide the strongest constraint on $g'$  for all masses $M_{Z'}\gtrsim 1$ MeV.
	
	For $g'\lesssim10^{-10}\sqrt{{\rm MeV}/M_{Z'}}$, the $Z'$ bosons decay after neutrino decoupling.  In this part of parameter space, decays to both $\nu_L$ and $\nu_R$ contribute to $\Delta\Neff$ during recombination, while our analysis only considers the contribution from $\nu_R$. We estimate that the additional production of $\nu_L$ provides no more than an  $\mathcal{O}(1)$ correction to the $\Neff$ constraints calculated in this study. We indicate this region in figure~\ref{fig:BL_constraint} by coloring the $\Neff$ contours green. Furthermore, for $M_{Z'}<2m_e$, the dominant energy transfer occurs between $\nu_L$ and $\nu_R$, while our Boltzmann equations assume energy injection from a thermal SM plasma with all species at the photon temperature. Thus below the MeV scale, our analysis no longer applies, and hence we restrict our attention here to $M_{Z'}>2m_e$. Meanwhile, stellar cooling places powerful constraints on this theory for $M_{Z'}<0.1$ MeV  \cite{Knapen:2017xzo,Hong:2020bxo}.  A full treatment of early universe constraints on the $B-L$ model in the mass range between $0.1{\ \rm MeV}<M_{Z'}<(10^{-20}/g'^2){\ \rm MeV}$ requires a detailed treatment of neutrino decoupling as well as light element formation during BBN, and is beyond the scope of this work.

	\subsection{Dark radiation density in the out-of-equilibrium regime}\label{BL:density}
	In the out-of-equilibrium (OOE) regime,  the final energy deposited into $\nu_R$ depends on whether or not the total decay width of  the $Z'$,  $\Gamma_{Z'}$, is less than the Hubble rate at SM temperatures around $T_{\rm SM}\sim M_{Z'}$, where the production rate of $Z'$s is maximized. 	In the case where $\Gamma_{Z'}/H \gg 1$ at $T_{\rm SM}\sim M_{Z'}$, the large population of on-shell $Z'$ bosons produced at resonance decay almost immediately into $\nu_R$.  However when $\Gamma_{Z'}/H \ll 1$ at $T_{\rm SM}\sim M_{Z'}$, the on-shell $Z'$ bosons produced at resonance are cosmologically long-lived and, because they are non-relativistic at production, their energy density redshifts like matter. The right-handed neutrinos are then dominantly produced at some SM temperature $T_{\rm decay}\ll M_{Z'}$ when the population of massive $Z'$ bosons decays, $\Gamma_{Z'}= H(T_{\rm decay})$.  Numerically, we find that setting $(\Gamma_{Z'}/H)_{T_{\rm SM}= M_{Z'}/2}= 1$ is a convenient criterion to separate the long- and short-lived regimes.
	
\begin{figure}
		\begin{subfigure}{.5\textwidth}
			\includegraphics[width=1.00\textwidth]{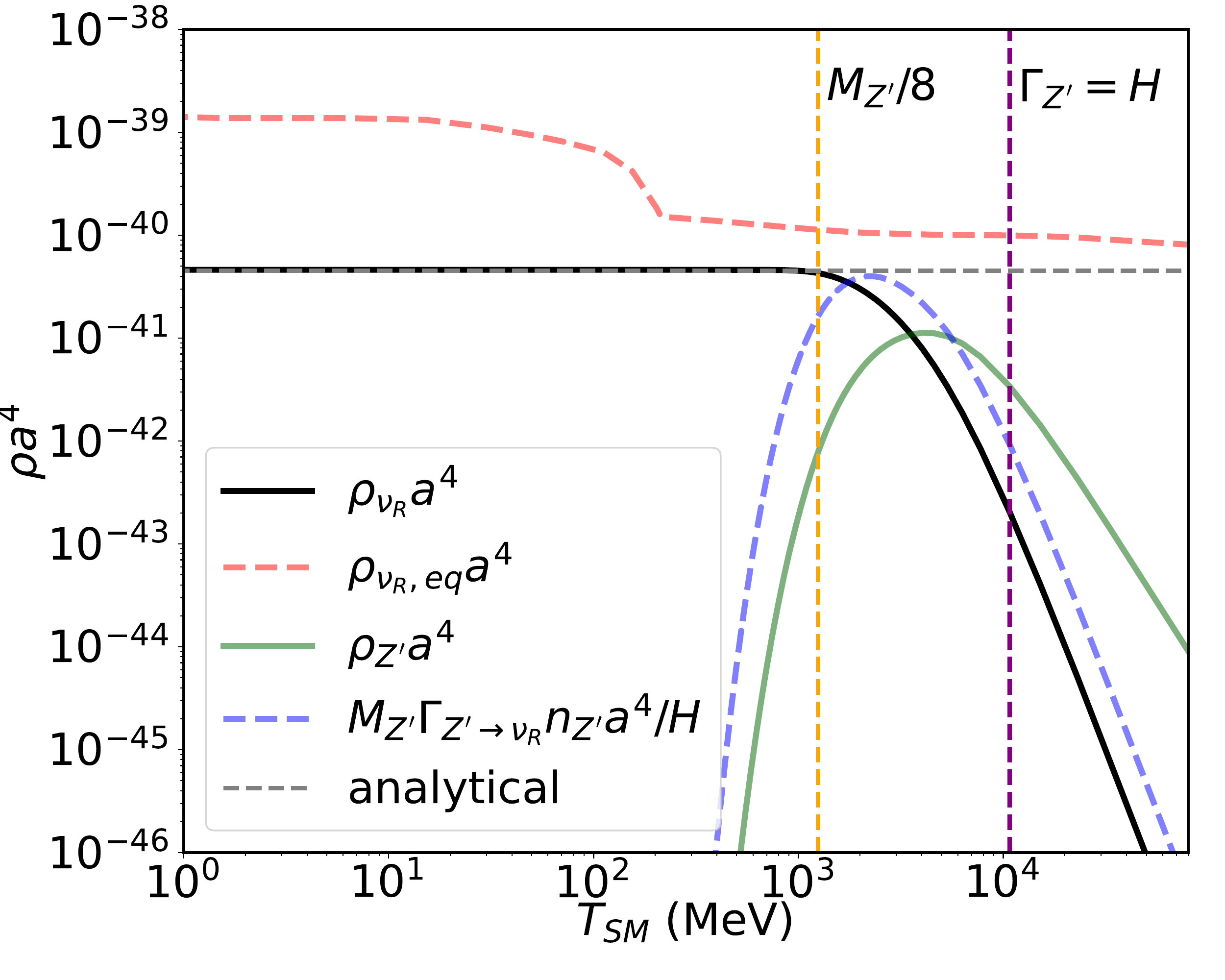}
		\end{subfigure}
		\begin{subfigure}{.5\textwidth}
			\includegraphics[width=1.00\textwidth]{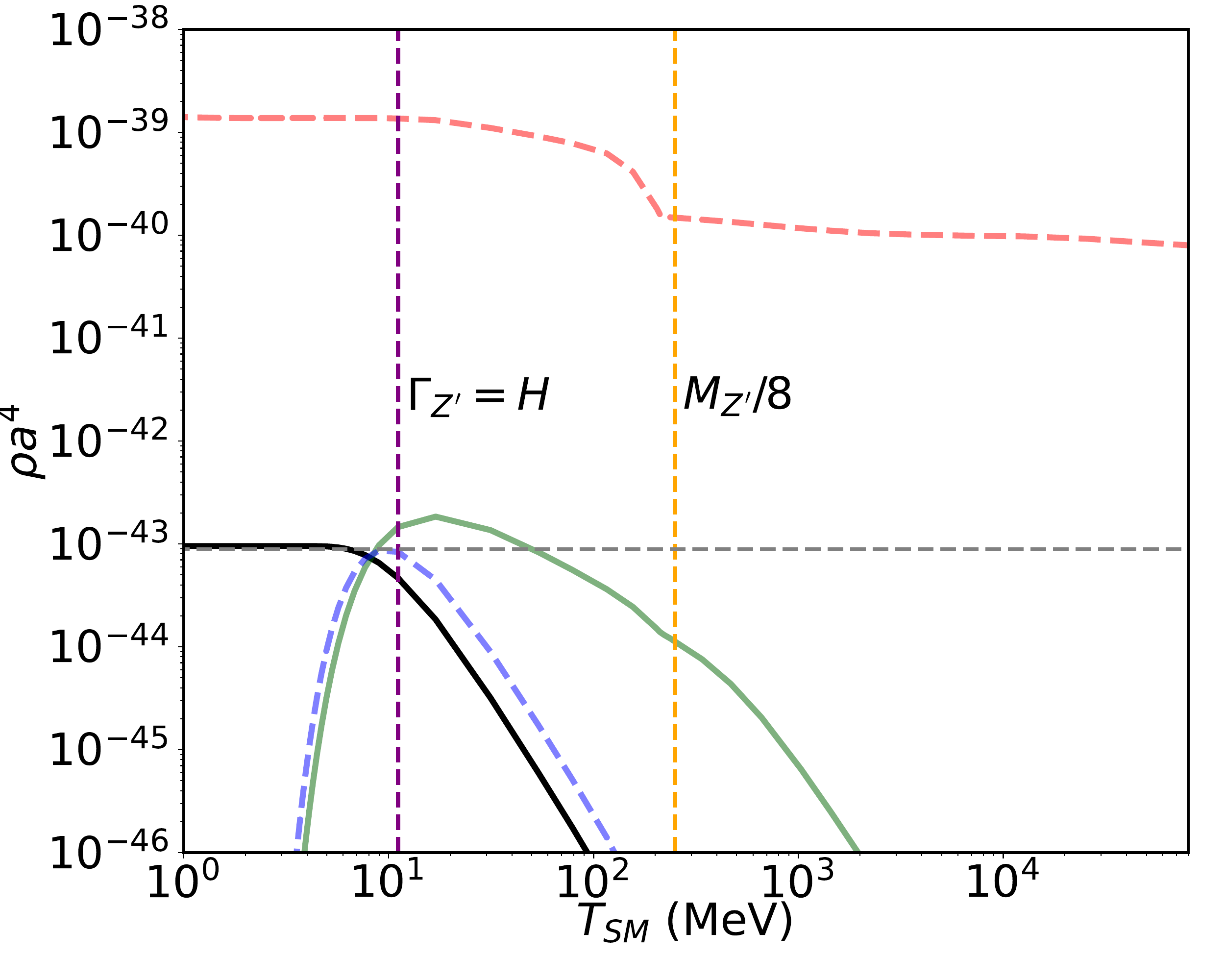}
		\end{subfigure} 
		\caption{Evolution of comoving $\nu_R$ energy density (black line) and comoving $Z'$ energy density (light green line) as a function of SM temperature for $\{M_{Z'},g'\}=\{10\ \textrm{GeV}, 10^{-8}\}$ and $\{M_{Z'},g'\}=\{2\times \ \textrm{GeV}, 2\times 10^{-11}\}$ in the left and right panels, respectively. The red line shows the evolution of $(\pi^2/30)g_{\nu_R}T_{\rm SM}^4a^4$, and the blue dot-dashed line is $M_{Z'}\Gamma_{Z'\rightarrow\nu_R}n_{Z'}a^4/H$. The vertical orange dashed line marks the point where $T_{\rm SM}=M_{Z'}/8$. The vertical purple dashed line marks the point where the $Z'$ decay rate equals the Hubble rate. The gray dashed line shows the analytical estimate of the asymptotic value of $\rho_{\nu_R}a^4$, which is calculated using eq.~\eqref{eq:density_ratio_an_BL} in the left panel and eq.~\eqref{eq:density_ratio_an_BL2} in the right panel.}
		\label{fig:BL_phasespace}
	\end{figure}

	We illustrate these two regimes with two representative parameter points in figure \ref{fig:BL_phasespace}. Here in both panels  the black line shows the comoving energy density of $\nu_R$, while the red dashed line indicates the energy density of $\nu_R$ after setting $T_{\nu_R}=T_{\rm SM}$ (similarly to the red line in figure~\ref{fig:mcp_densityevolve}).  The $\nu_R$ do not thermalize with the SM for either the parameter points shown, and correspondingly the black line remains below the red line in both panels. The blue dot-dashed line shows the evolution of $M_{Z'}\Gamma_{Z'\rightarrow \nu_R}n_{Z'}a^4/H$, which indicates the amount of comoving energy injected into $\nu_R$ in a Hubble time from the decay of on-shell $Z'$ bosons. The energy injected by SM fermions annihilating to $\nu_R$ through off-shell $Z'$ bosons, given by $\mathcal{C}^{\rm off}_{ff\rightarrow \nu_R\nu_R}a^4/H$, is below the range covered in figure~\ref{fig:BL_phasespace} and is not shown. The vertical orange dashed line marks when $T_{\rm SM}= M_{Z'}/8$, after which temperature we find empirically that the production of $Z'$ bosons from the SM plasma is negligible.

	The left panel in figure \ref{fig:BL_phasespace} corresponds to a parameter point  where $\Gamma_{Z'}$ exceeds the Hubble rate at some $T_{\rm SM}> M_{Z'}/2$. The $Z'$ bosons produced after $\Gamma_{Z'}=H$ are short-lived and decay within a Hubble time. The SM plasma keeps producing $Z'$ bosons until $T_{\rm SM}\sim M_{Z'}/8$, and thus the energy injection into $\nu_R$ ends once $T_{\rm SM}< M_{Z'}/8$. The right panel of figure~\ref{fig:BL_phasespace} corresponds to a parameter point where $(\Gamma_{Z'}/H)_{T_{\rm SM}= M_{Z'}/2}\ll 1$. In this scenario, the SM plasma first produces $Z'$ bosons via inverse decays. The production of $Z'$ bosons ends once $T_{\rm SM}<M_{Z'}/8$. Subsequently, $n_{Z'}$ evolves adiabatically until $\Gamma_{Z'}$ becomes of the order of $H$, after which $Z'$ decays into SM particles as well as $\nu_R$. 
	
	We now develop analytic approximations to the final value of $\rho_{\nu_R}$ for the short- and long-lived $Z'$ cases separately.
	
	\paragraph{Dark radiation production  for short-lived $Z'$ bosons:}
	
	In the regime where the $Z'$s are cosmologically short-lived, $(\Gamma_{Z'}/H)_{T_{\rm SM}= M_{Z'}/2}>1$, the Boltzmann equations can be simplified by noticing that after $\Gamma_{Z'}=H$ the abundance of $Z'$ bosons follows a quasi-static equilibrium where the production rate of $Z'$ bosons balances its decay rate. Setting the RHS of eq.~\eqref{eq:Boltz_eps_full2} to zero and replacing $\mathcal{C}_{ff\rightarrow Z'}$ and $\mathcal{C}_{Z'\rightarrow \nu_R\nu_R}$ using eq.~\eqref{eq:Z-arrow-aa} gives the quasi-static equilibrium abundance of $Z'$ bosons,
	\begin{align}\label{eqn:z'qs}
		n_{Z'}^{qs}=\frac{\Gamma_{Z'\rightarrow SM}}{\Gamma_{Z'}}\tilde{n}_{\zeta}(T_{\rm SM})+\frac{\Gamma_{Z'\rightarrow \nu_R}}{\Gamma_{Z'}}\tilde{n}_{\zeta}(T_{\nu_R}),
	\end{align}
	where $\tilde{n}_{\zeta}$ is defined in eq.~\eqref{def:n_tilde}.
	Substituting this quasi-static abundance $n_{Z'}^{qs}$ into eq.~\eqref{eq:Boltz_eps_full3}, we obtain an effective collision term describing energy injection into $\nu_R$ given by
	\begin{align}\label{eq:Bl_coll_eff}
		\mathcal{C}_{ff\rightarrow \nu_R\nu_R}\!\!=\!\!\frac{3M_{Z'}^3}{2\pi^2} \left[\frac{\Gamma(Z'\rightarrow \nu_R)\Gamma(Z'\rightarrow SM)}{\Gamma_{Z'}}\right]\!\!\!\left(T_{\rm SM}G_{1}\!\!\left(\frac{M_{Z'}}{T_{\rm SM}}\right)-T_{\nu_R}G_{1}\!\!\left(\frac{M_{Z'}}{T_{\nu_R}}\right)\right)+\mathcal{C}^{\rm off}_{ff\rightarrow \nu_R\nu_R},
	\end{align}
	where $G_{1}$ is a dimensionless function given by eq.~\eqref{eq:G_def}.
	In this regime, the collision term in eq.~\eqref{eq:Bl_coll_eff} reproduces the collision term  calculated using the complete $ff\rightarrow \nu_R\nu_R$ cross-section, including the on-shell $Z'$ bosons, as we show explicitly in appendix~\ref{sec:BL_coll}.  As this collision term no longer depends on $\rho_{Z'}$, we need only solve for $\rho_{\nu_R}$ and $\rho_{\rm SM}$ to find the contribution of $\nu_R$ to $\Delta\Neff$. Thus, in the short-lived $Z'$ limit, the resulting system of Boltzmann equations is similar to that for the MCP model, eq.~\eqref{eq:Boltz_eps_full_mcp}.
	
	We can determine the asymptotic value of $\rho_{\nu_R}$ by following  steps similar to those in section~\ref{mcp:physics} to obtain eq.~\eqref{eq:density_ratio_an_mcp}.   We can neglect the contribution from $\mathcal{C}^{\rm off}_{ff\rightarrow \nu_R\nu_R}$ because the net energy transferred to out-of-equilibrium $\nu_R$ through contact interactions is much smaller than the resonantly-enhanced contribution from the on-shell collision term. The fraction of SM energy transferred into $\nu_R$ is then given by
	\begin{align}\label{eq:density_ratio_an_BL}
		\left(\frac{\rho_{\nu_R}}{\rho_{\rm SM}}\right)_{\rm leak}=\frac{15\sqrt{3}}{64\pi^3[g_{*}(4\Lambda)\pi^2/30]^{3/2}}\frac{M_{\rm Pl}}{\Lambda}\times L,
	\end{align}
	where $\Lambda = M_{Z'}/8$, and 
	\begin{align}\label{eq:def_leak_BL}
		L=6\pi^2\kappa_{1}\left[\frac{\Gamma_{Z'\rightarrow \nu_R}\Gamma_{Z'\rightarrow SM}}{\Gamma_{Z'}}\right]=\frac{3\pi\kappa_{1}}{4} g'^2\left[\frac{\Gamma_{Z'\rightarrow SM}}{\Gamma_{Z'}}\right].
	\end{align}
	This limiting result for the comoving density of $\nu_R$,  $\rho_{\nu_R}a^4=\left(\rho_{\nu_R}/\rho_{\rm SM}\right)_{\rm leak}(a^4\rho_{\rm SM})_{T_{\rm SM}=M_{Z'}/8}$, is shown by the gray dashed line in the left panel of figure~\ref{fig:BL_phasespace}, which demonstrates its agreement with the numerically evaluated asymptote of $\rho_{\nu_R}a^4$.
	Eq.~\eqref{eq:density_ratio_an_BL} is only valid as long as the $\nu_R$ do not thermalize with the SM plasma. Numerically we find that for $(\Gamma_{Z'}/H)_{T_{\rm SM}= M_{Z'}/2}\gtrsim 30$, the $\nu_R$ thermalize with the SM plasma and the final density ratio is simply given by  $(\rho_{\nu_R}/\rho_{\rm SM})_f=g_{\nu_R}/g_{*}$.

	\paragraph{Out-of-equilibrium dark radiation production from long-lived $Z'$ bosons:}
	
	To solve for the dark radiation density in $\nu_R$ in the case where the $Z'$ bosons are long-lived, we first need to calculate the freeze-in abundance of $Z'$. To proceed, we make two simplifications. First, since the $Z'$ bosons are long-lived and, until they decay, the $\nu_R$ abundance is negligible, we can neglect the decays of $Z'$ as well as the inverse decays of $\nu_R$ into $Z'$ when calculating the freeze-in $Z'$ abundance. Second, because  $\rho_{\rm SM}\gg\rho_{Z'}, \,\rho_{\nu_R}$ we  neglect the contributions of the $Z'$ and $\nu_R$ in determining the Hubble rate. 	Assuming the SM degrees of freedom remain constant until the production of $Z'$ ends at  $T_{\rm SM}\lesssim M_{Z'}/8$, we can then simply integrate  eq.~\eqref{eq:Boltz_eps_full2} for $\rho_{Z'}$. With $\rho_{Z'} = M_{Z'} n_{Z'}$ and $\mathcal{C}_{ff\rightarrow Z'}$ given by eq.~\eqref{eq:Z-arrow-aa}, the frozen-in abundance of $Z'$ bosons is then
	\begin{align}
		a^3n_{Z'}^{\rm fz-in}=(aT_{\rm SM})_{T_{\rm SM}=M_{Z'}/8}^3\times \frac{3\lambda}{8\pi^2}\left[\frac{\Gamma_{Z'\rightarrow {\rm SM}}}{H(T_{\rm SM}=M_{Z'}/2)}\right],
	\end{align}
	where $\lambda=\int G_1(1/x)x^{-5} dx\approx 5.93$. 
	
	The frozen-in population of $Z'$ boson eventually decays, and accordingly  the comoving number density evolves as 
	\begin{align}\label{eqn:z'evolve} 
		n_{Z'}= n_{Z'}^{\rm fz-in}\, e^{-\Gamma_{Z'}t}. 
	\end{align}
	Note that the final number density of $\nu_R$ is not affected by whether the $Z'$ bosons decay before or after achieving their freeze-in abundance. This is because the number density of $\nu_R$ is set by the branching ratio of $Z'$ decays into $\nu_R$ and the number of $Z'$ bosons produced by the SM plasma, neither of which depend on when the $Z'$ bosons decay. In contrast, the energy density of the $\nu_R$ does depend on the timing of the $Z'$ boson decay because the $\nu_R$ are produced with a fixed energy of $M_{Z'}/2$, which subsequently redshifts as $1/a$. Consequently,  $Z'$ bosons that decay later result in more energetic $\nu_R$ at recombination, and thus a larger contribution to $N_{\rm eff}$.
	
	The asymptotic value of $\rho_{\nu_R}$ is found by substituting the evolution of the massive $Z'$s, eq.\ \eqref{eqn:z'evolve}, into the Boltzmann equation for $\nu_R$, eq.~\eqref{eq:Boltz_eps_full3}. Once again, both inverse decays of $\nu_R$ into $Z'$ and off-shell contributions to SM fermion  annihilation  can be ignored 
	in comparison to the contribution from $Z'$ decays. The resulting $\rho_{\nu_R}$  is given by
	\begin{align}
		\left(\frac{\rho_{\nu_R}}{\rho_{\rm SM}}\right)_{\rm decay}=&\frac{M_{Z'}\Gamma_{Z'\rightarrow \nu_R}}{a^4\rho_{\rm SM}}\int_0^a\frac{\tilde{a}^3n_{Z'}^{\rm fz-in} e^{-\Gamma_{Z'}t}}{H}d\tilde{a}\\
		\approx&\frac{2\sqrt{2}\lambda}{15\sqrt{\pi}\kappa_1} \left(\frac{\rho_{\nu_R}}{\rho_{\rm SM}}\right)_{\rm leak} \left[\frac{H(T_{\rm SM}=M_{Z'}/2)}{\Gamma_{Z'}}\right]^{1/2}\left[\frac{g_*^{3}(M_{Z'}/2)g_{*}(T_{\rm decay})}{g_{*}^4(M_{Z'}/8)}\right]^{1/4},\label{eq:density_ratio_an_BL2}
	\end{align}
	where $T_{\rm decay}$ is the SM temperature at which $H(T_{\rm decay})=\Gamma_{Z'}$ and $\left(\rho_{\nu_R}/\rho_{\rm SM}\right)_{\rm leak}$ is  defined in eq.~\eqref{eq:density_ratio_an_BL}. In the second line we approximated $g_*$ to be constant around $T_{\rm decay}$ and set $g_{*s}= g_*$.
	The numerical coefficient in eq.~\eqref{eq:density_ratio_an_BL2} and the ratio of $g_*$ factors in the square brackets are both $\mathcal{O}(1)$. Consequently, $\rho_{\nu_R}$ is enhanced by a factor of $(\sqrt{H/\Gamma_{Z'}})_{T_{\rm SM}= M_{Z'}/2}$ if the $Z'$ bosons are long-lived compared to the cases where the $Z'$ bosons decay instantaneously. The right panel of figure~\ref{fig:BL_phasespace}, shows the analytical estimate of the comoving density of $\nu_R$, given by $\rho_{\nu_R}a^4=\left(\rho_{\nu_R}/\rho_{\rm SM}\right)_{\rm decay}(a^4\rho_{\rm SM})_{T_{\rm SM}=T_{\rm decay}}$, as the gray dashed line. At late times, this analytical estimate is in close agreement with the numerically evaluated $\rho_{\nu_R}a^4$, as shown by the black solid line.

	\subsection{Dark radiation production and analytical approximations to the $\Neff$  constraint}\label{BL:constraint}

	In this section we provide analytical expressions for the curves of constant $\Delta \Neff$ in the $B-L$ parameter space. 	We consider the out-of-equilibrium and equilibrated regions of parameter space separately.

	\paragraph{Out-of-equilibrium dark radiation production}

	In the case when $\nu_R$ remains out-of-equilibrium with the SM plasma, the dependence of the final dark radiation density, $\rho_{\nu_R}$,  on the $B-L$ coupling, $g'$, depends  on whether $\Gamma_{Z'}$ is larger or smaller than the Hubble rate at $T_{\rm SM}=M_{Z'}/2$. We find that the $g'$ and $M_{Z'}$ values on the $\Delta\Neff = 0.3$ constraint contour typically satisfy $(\Gamma_{Z'}/H)_{T_{\rm SM}= M_{Z'}/2}\gtrsim 1$. Consequently, we use eq.~\eqref{eq:density_ratio_an_BL} to evaluate the constraint on $g'$ and $M_{Z'}$ for $\Delta\Neff < 0.3$. In particular, we adiabatically evolve $\rho_{\nu_R}$ given in eq.~\eqref{eq:density_ratio_an_BL} from the end of energy injection at $T_{\rm SM}=M_{Z'}/8$ to recombination and restrict the $\Delta\Neff$ shift given in eq.~\eqref{eq:DeltaN_nuR} to remain below the $(\Delta\Neff)_{\rm max}$ upper bound set by CMB measurements. Doing so yields
	\begin{multline}\label{eq:BL_analytical_Neff}
		g'^2<5.8\times 10^{-19}\left(\frac{g_{*}(M_{Z'}/2)}{10}\right)^{3/2}\left(\frac{g_{*}(M_{Z'}/8)}{10}\right)^{1/3}\left( \frac{\Gamma_{Z'}}{\Gamma_{Z'\rightarrow {\rm  SM}}}\right) \left(\frac{(\Delta\Neff)_{\rm max}}{0.3}\right)\left(\frac{M_{Z'}}{\rm GeV}\right).
	\end{multline}
	The ratio of decay widths here is typically an $\mathcal{O}(1)$ number depending on the value of $M_{Z'}$.
	
	For the $\Delta\Neff=0.06$ and $\Delta\Neff=0.027$ constraint contours,  the condition \newline $(\Gamma_{Z'}/H)_{T_{\rm SM}= M_{Z'}/2}> 1$ is satisfied above and to the right of the red dot in figure~\ref{fig:BL_constraint}. Consequently, the analytical result for short-lived $Z'$s in eq.~\eqref{eq:BL_analytical_Neff} also applies to the $\Delta\Neff=0.06$ and $\Delta\Neff=0.027$ contours in this region. To find an analytical result applicable below and to the left of the red dot, we start from the expression for $\rho_{\nu_R}$ given in eq.~\eqref{eq:density_ratio_an_BL2}. We then  evolve  $\rho_{\nu_R}$ adiabatically from the end of $Z'$ decays at $T_{\rm decay}$ to recombination. The corresponding constraint on $g'$ is then given by 
	\begin{multline}\label{eq:BL_analytical_Neff2}
		g'^2 <8.1\times 10^{-21}\left(\frac{g_{*}(M_{Z'}/2)}{10}\right)\left(\frac{g_{*}(M_{Z'}/8)}{10}\right)^2\left(\frac{g_{*}(T_{\rm decay})}{10}\right)^{1/6}\left(\frac{\Gamma_{Z'}}{\Gamma_{Z'\rightarrow \nu_R}}\right)\left( \frac{\Gamma_{Z'}}{\Gamma_{Z'\rightarrow {\rm SM}}}\right)^2\\ \times \left(\frac{(\Delta\Neff)_{\rm max}}{0.06}\right)^2\left(\frac{M_{Z'}}{\rm GeV}\right).
	\end{multline}
	Note that the constraint on $g'$ for short-lived $Z'$s, given in eq.~\eqref{eq:BL_analytical_Neff}, is proportional to $(\Delta\Neff)_{\rm max}$ while the long-lived $Z'$ result in eq.~\eqref{eq:BL_analytical_Neff2} is proportional to $(\Delta\Neff)_{\rm max}^2$. The delayed $Z'$ decays parametrically enhance the ultimate dark radiation density and hence the sensitivity of $\Neff$ measurements to the model parameters.
	
	\paragraph{Dark radiation production in the equilibrium regime:}

	If the right-handed neutrinos thermalize with the SM, then the final comoving energy density in $\nu_R$ depends on the decoupling temperature, $T_d$, which is only logarithmically sensitive to $g'$. The thermalization threshold for the $B-L$ model can be calculated in a similar manner to the MCP model in section \ref{mcp:physics} above, see eq.~\eqref{eq:thresh_mcp}. Since the ratio of the energy injection rate to the Hubble rate, $\Gamma_E/H= \mathcal{C}_{\rm f}/(\rho_{\nu_R,eq}H)$, is negligible for temperatures below  $T_{\rm SM}\sim M_{Z'}/8$ in the $B-L$ model (as compared to to  $T_{\rm SM}\sim m/4$ in the MCP model), the thermalization threshold in eq.~\eqref{eq:thresh_mcp} is for the $B-L$ model 
	\begin{align}
		M_{Z', {\rm th} }=8T_{d}[(\Delta\Neff)_{\rm max} ,g_{\nu_R}],
	\end{align}
	where $T_{d}$ is defined through eq.~\eqref{eq:def_Td}. In figure~\ref{fig:BL_constraint}, this thermalization threshold for $\Delta \Neff =0.3$  occurs around $M_{Z', {\rm th}}\sim 1.7\times10^3$. For $\Delta  \Neff < 0.14$, the thermalization threshold is pushed to arbitrarily large values of $M_{Z',  {\rm th}}$ because restricting $\Delta N_{\rm eff}<0.14$ rules out ever having three right-handed neutrinos in thermal equilibrium with the SM plasma, assuming no new degrees of freedom in the SM.
	
	The exponential weakening of the $\Delta \Neff = 0.3$ curve in figure~\ref{fig:BL_constraint} stops around $M_{Z'}\sim 15$ GeV, after which the constraint follows $g'\lesssim M_{Z'}/(16\ {\rm TeV})$ (see also Ref.\ \cite{Abazajian:2019oqj,FileviezPerez:2019cyn}).  At these large masses the contact-operator-mediated annihilations, described by $\mathcal{C}^{\rm off}_{ff\rightarrow \nu_R\nu_R}$, dominate over on-shell $Z'$ production, $M_{Z'}\Gamma_{Z'\rightarrow\nu_R}n_{Z}$, in determining decoupling temperatures that are much smaller than $M_{Z'}$. 
	
	Note that, unlike the MCP model, the $B-L$ model does not have an excluded strip in parameter space extending up to high masses. 
	For such a strip to exist, the mediator between the SM particles and the BSM relativistic particles must have stronger couplings to the BSM relativistic particles than it does to the SM.  In the $B-L$ model, the decay width of $Z'$ into SM particles is larger than its decay width into $\nu_R$, in contrast to the MCP model, where the mediating MCP has much stronger interactions with the dark photons than with the SM photon. 

	\section{Dark radiation constraints on classes of hidden sectors}\label{sec:wide_application}

	We have so far considered dark radiation constraints on specific, minimal BSM models where a particle  $\phi$  with mass $m_{\phi} \gtrsim $ MeV has renormalizable couplings to both the SM and new relativistic particles. While in general the heavy particle could be the SM Higgs boson (or indeed the $Z$ boson),  in this work we focus on the case where the heavy particle is a new SM gauge singlet particle. Additionally we focus on mass scales $m_{\phi} \gtrsim $ MeV because for lighter masses the constraints from stellar cooling observations generically become important. We have seen in two specific examples that such models will be stringently tested by upcoming CMB experiments that promise to measure $\Neff$ to an accuracy of $(\Delta\Neff)_{\rm max} = 0.06$ at 95\% confidence. In particular, we have demonstrated that, while detailed constraints on the parameter space require numerical evaluation of a coupled system of Boltzmann equations, a conservative, semi-analytic estimate of the allowed parameter space can be made by making a number of simplifying assumptions. 
	
	In this section we highlight the general methodology and assumptions required to estimate this conservative constraint and argue that the constraint holds even when the new relativistic particles are part of a much larger hidden sector (HS). We then explore the restrictions on HS model building that will be placed by upcoming CMB measurements of $\Neff$.
	
	\subsection{Hidden sector models}
	
	We consider classes of HS models that contain light degrees of freedom that are relativistic during recombination. These degrees of freedom may be required by symmetries (as the $\nu_R$ were above), or they may be required to sequester entropy to facilitate the freezeout of HS dark matter. Stellar cooling observations strongly constrain direct renormalizable couplings of these light degrees of freedom to stable SM particles, and we assume they couple to the SM through new, heavy, SM  gauge-singlet particles, $\phi$, with masses which we take to be $m_\phi >$ MeV. The existence of this portal coupling enables the production of the mediator $\phi$ particles in the early universe via annihilation or decay of SM particles. These mediator particles then lead to the production of the light degrees of freedom in the HS, whose energy density is constrained by measurements of $\Neff$.  	These $\Neff$ constraints are applicable as long as these light degrees of freedom remain relativistic during recombination. For masses larger than $0.1$ eV, the combined constraints from the large scale structure and the CMB measurements are more stringent \cite{DePorzio:2020wcz}.

	We seek to derive conservative constraints on the couplings of such sectors to the SM by estimating the production of dark radiation.  The precise computation of the dark radiation density depends on the details of the HS, such as the number of degrees of freedom, masses of the particles, and interactions between them. However, we argue that  a lower bound on the dark radiation density can be estimated from the production of mediator particles  by the SM plasma, provided the energy  in the mediator is preferentially transferred to the HS degrees of freedom. This amounts to assuming that the mediator interacts more strongly with the HS than with the SM. Any HS energy density that subsequently becomes freestreaming dark radiation at the decoupling of the CMB is  minimized if all the energy dilutes as radiation as soon as it is produced. Therefore, assuming that all the energy that is transferred to the mediator particles by the SM plasma is rapidly deposited into light degrees of freedom in the HS provides a lower bound on the resulting dark radiation density. This lower bound provides  a conservative estimate on the shift in $\Delta \Neff$. Below we quantify this conservative estimate for different types of couplings between the mediator and the Standard Model. As above, we separate the estimates into the regions where the HS thermalizes, and those where it remains out of equilibrium.

	\subsubsection{Out-of-equilibrium dark radiation production}
	
	Practically, in the out-of-equilibrium regime, our conservative estimate of $\Delta \Neff$ is obtained by assuming a hidden sector consisting of a  single massless particle together with a massive mediator $\phi$ that couples to the SM. The HS equation of state is taken to be that of radiation, $w_{{\rm HS}}= 1/3$. The dark radiation density is determined  by solving the Boltzmann equation given by eq.~\eqref{eq:Boltz_eps_full_mcp}, where the collision term is determined by all the energy transfer processes from the SM plasma into the HS bath enabled by the portal coupling. While taking into account all the energy transfer process depends on the specifics of a particular HS model under consideration, energy transfer through the production of $\phi$ is common in the vast majority of HS models. Consequently, to obtain a conservative estimate of the asymptotic energy density in dark radiation, we evaluate the collision term only for processes involving on-shell production of $\phi$. For renormalizable interactions with the SM, these processes are annihilations of SM particles, $aa\rightarrow\phi\phi$; decays of SM particles $a\rightarrow\phi\phi$; or inverse decays of SM particles $aa\rightarrow\phi$. In all cases $a$ denotes a SM particle coupled to $\phi$ via a renormalizable portal coupling.  We further restrict our attention to $s$-channel processes, which are independent of the properties of the HS radiation bath as long as the interaction proceeds well out of equilibrium.
	
	In the out-of-equilibrium regime,  we can analytically find the energy transferred into the HS by taking the SM temperature to evolve as $ T\propto 1/a$ 
	(a good approximation away from mass thresholds). The calculation is analogous to that in section \ref{mcp:physics} leading to eq.~\eqref{eq:C_integrate}, and we obtain
	\begin{align}\label{eq:Cf_integrate2}
		\left(\frac{\rho_{\rm HS}}{\rho_{\rm SM}}\right)_{T_{\rm SM}= \Lambda}\approx\frac{\sqrt{3}M_{\rm Pl}}{[g_{*}(4\Lambda)\pi^2/30]^{3/2}}\int_{0}^{\infty}\frac{dT_{\rm SM}}{T_{\rm SM}^7}\mathcal{C}_{\rm f},
	\end{align}
	where $\Lambda$ is the energy injection decoupling temperature, below which the production of $\phi$ ends and $\mathcal{C}_{\rm f}$ is the forward energy transfer collision term for production of $\phi$. 
	
	The integral over the forward collision term can be carried out given a specific model for the cross-section, allowing us to express the energy density injected during out-of-equilibrium scattering  in terms of a leak factor $L$, 
	\begin{align}\label{eq:density_ratio_an2}
		L = \frac{64\pi^3}{15} \Lambda \int_0^{\infty} \frac{dT_{\rm SM}}{T_{\rm SM}^7}\mathcal{C}_{\rm f}.
	\end{align}

	\paragraph{Annihilation of SM particles into the HS.}
	
	For annihilations of SM particles into the HS, $aa\rightarrow\phi\phi$, the forward collision term is
	\begin{align}
		\mathcal{C}_{\rm f}=&\sum_f\frac{1}{32\pi^4}\int_{4\max(m_a,m_{\phi})^2}^{\infty} ds (s-4m_{a}^2)s\sigma_{aa\rightarrow \phi\phi}(s)T_{\rm SM}G_{\zeta_f}(\sqrt{s}/T_{\rm SM}).
	\end{align}
	This production process occurs in the millicharged particle model when the SM fermions annihilate into millicharged particles.
	The corresponding energy injection decoupling temperature and the leak factor are given by
	\begin{align}
		\Lambda=\frac{1}{4}\max(m_a,m_{\phi}), && L= \kappa_{\zeta_a} \Lambda\int_{64\Lambda^2}^{\infty} ds \frac{(s-4m_a^2)}{s\sqrt{s}}\sigma_{aa\rightarrow \phi\phi}(s).
	\end{align}
	Here $\sigma_{aa\rightarrow \phi\phi}(s)$ is the spin-summed CM frame cross-section and $\kappa_{\zeta_a}$ is determined by the quantum statistical distribution of $a$, as described below eq.~\eqref{eq:mcp_temp}.
	
	\paragraph{Inverse decay of SM particles into the HS.}
	
	For the inverse decay process, $aa\rightarrow\phi$, the collision term is of the form 
	\begin{align}
		\mathcal{C}_{\rm f}=m_{\phi}\Gamma_{\phi\rightarrow a}\tilde{n}_{\zeta_a}(T_{\rm SM}),
	\end{align}
	where $\Gamma_{\phi\rightarrow a}$ is the decay width of $\phi$ to $a$, and $\tilde{n}_{\zeta_a}$ is given by eq.~\eqref{def:n_tilde}. We encountered this production process for the gauged $B-L$ model in section~\ref{Bl:numerical}. After integrating the RHS of eq.~\eqref{eq:Cf_integrate2} for the process $aa\rightarrow\phi$, the final result can can be written in the form of eq.~\eqref{eq:density_ratio_an2} with
	\begin{align}\label{eq:inv_dec_leak}
		\Lambda=\frac{m_{\phi}}{8}, && L= 2\pi^2g_{\phi}\kappa_{\zeta_a}\frac{\Gamma_{\phi\rightarrow a}}{m_{\phi}},
	\end{align}
	where $g_{\phi}$ is the number of spin degrees of $\phi$. 
	
	The final HS energy density calculated using eq.~\eqref{eq:inv_dec_leak} is different from the one we obtained in the case of the gauged $B-L$ model for two reasons. First, the decay width of $Z'$ into $\nu_R$ is smaller than its total decay width into SM particles for $M_{Z'}>2m_e$. Thus most of the energy transferred into $Z'$ bosons does not end up in $\nu_R$ but is rather returned to the SM plasma. If we consider the $Z'$ bosons to couple much more strongly with additional HS particles, then the above calculation would accurately reflect the minimum energy transferred into the HS. Second, the final energy density in $\nu_R$ is enhanced when the $Z'$ bosons are long lived.

	\paragraph{Decays of SM particles into the HS.} 
	
	Finally for decays of SM particles into the HS, $a\rightarrow\phi\phi$, the collision term is of the form 
	\begin{align}
		\mathcal{C}_{\rm f}=m_{a}\Gamma_{a\rightarrow \phi}n_{{\rm eq},a}(T_{\rm SM}), 
	\end{align}
	where 
	\begin{align}
		n_{{\rm eq},a}=g_a\int \frac{d^3p}{(2\pi)^{3}}\frac{1}{ [e^{E/T_{\rm SM}}+\zeta_a]}
	\end{align} 
	is the equilibrium number density of particle $a$. This production process occurs in the millicharged particle model when the $Z$ bosons decay into millicharged particles.\footnote{This production process can also be realized in the case of a SM singlet scalar coupling through the Higgs portal, in which case the Higgs boson can decay  into pairs of scalar fields; for a specific recent application producing dark radiation through this coupling, see \cite{Biswas:2022vkq}.} Again the energy transferred into the HS can be expressed by eq.~\eqref{eq:density_ratio_an2} with
	\begin{align}
		\Lambda=\frac{m_{a}}{8}, && L= \pi^2g_{a}{\tilde{\kappa}_{\zeta_a}}\frac{\Gamma_{a\rightarrow \phi}}{m_{a}},
	\end{align}
	where $\tilde{\kappa}_{1}=31\pi^6/30240$, $\tilde{\kappa}_{0}=1$, and $\tilde{\kappa}_{-1}=\pi^6/945$.

	When the HS remains out of equilibrium with the SM plasma, we can find the contribution to $\Delta\Neff$  by starting from eq.~\eqref{eq:density_ratio_an2} and then adiabatically evolving $\rho_{\rm HS}$ as radiation from the end of energy injection until recombination. Requiring $\Delta\Neff$ to be less than the CMB sensitivity, $(\Delta\Neff)_{\rm max}$, yields
	\begin{align}\label{eq:gen_constraint}
		L< g^{3/2}_*(4\Lambda)g_{*}^{1/3}(\Lambda)(\Delta\Neff)_{\rm max}\frac{\Lambda}{M_{\rm Pl}}.
	\end{align}
	The above calculations assume that  all  produced $\phi$ particles decay rapidly into relativistic HS particles. This assumption holds if $\phi$ has sufficiently strong couplings with HS particles. This is a conservative assumption because a long-lived $\phi$ would result in a larger density in the HS, and a larger shift in $\Delta \Neff$.
	
	\subsubsection{Equilibrium dark radiation production}
	
	If the HS thermalizes with the SM plasma, then the final energy density in the HS depends on the decoupling temperature, $T_d$, which is only logarithmically sensitive to the strength of the portal coupling. Consequently, the $\Neff$ constraint on the portal coupling become exponentially weak once the HS thermalizes. Similar to the case of the $B-L$  and millicharged particle models, the weakening of constraints occur for values of the energy injection decoupling temperature, $\Lambda$, larger than 
	\begin{align}\label{eq:gen_lambda}
		\Lambda_{\rm th}\equiv T_{d}[(\Delta \Neff)_{\rm max},g_{\rm HS}],
	\end{align}
	where $T_{d}$ is given by eq.~\eqref{eq:def_Td}. Thus, the $\Neff$ constraint given by eq.~\eqref{eq:gen_constraint} is only valid for $\Lambda<\Lambda_{\rm th}$. Note that a larger $g_{\rm HS}$ would push the thermalization threshold given in eq.~\eqref{eq:gen_lambda} to larger $\Lambda_{\rm th}$. Thus, eq.~\eqref{eq:gen_constraint} together with a thermalization threshold scale $\Lambda_{\rm th}$ calculated assuming $g_{\rm HS}=1$ provides a conservative constraint on the portal coupling that is independent of details within the hidden sector.

	\subsection{Implications for HS model building}
	
	In this section, we have argued that under fairly generic conditions, a conservative lower bound on the energy density in dark radiation in a generic HS may be estimated. This lower bound can in turn be used to place bounds on the couplings between a mediator that couples the dark sector to the SM.  As future measurements of $\Neff$ become more and more precise, increasing pressure will be placed on models of BSM physics that contain  light states contributing to dark radiation. 
	
	From a different perspective, this analysis also points at ways such models may be brought into agreement with future data. There are a number of possibilities. In particular, one may simply be able to arrange the couplings so that the mediator interacts more strongly with the SM than the HS, and thereby energy is transferred back into the SM from the mediator. Another possibility is to have new degrees of freedom in equilibrium with the SM plasma that become non-relativistic after $T_{{\rm SM}}<\Lambda$. Consequently, the annihilation of the new degrees of freedom heat the SM plasma relative to the HS,  diluting the dark radiation today. Similarly, if a massive field comes to dominate the universe and subsequently decays predominantly into the SM at some temperature $T_{\rm rh}<\Lambda$, the resulting reheating of the SM relaxes constraints from $\Neff$; this mechanism was invoked, for instance, to  ameliorate dark radiation constraints on Twin Higgs models \cite{Chacko:2016hvu,Craig:2016lyx}.  The entropy of the SM plasma can also increase if the SM comes into equilibrium with a new light species at a temperature $T_{\rm eq} < \Lambda$ that later becomes nonrelativistic and deposits its entropy into the SM.  This mechanism generally requires a light BSM field with couplings to the SM that become cosmologically important at late times.  While stellar cooling constraints are typically prohibitive for models that realize equilibration {\em after} SM neutrino decoupling \cite{Berlin:2019pbq}, the BSM dark radiation considered here has a thermal decoupling scale $\Lambda>$ MeV and thus suppressing $\Neff$ using this mechanism can be much simpler.  Finally, $\Neff$ also decreases if one or more of the states contributing to the dark radiation at $T \sim \Lambda$ can decay back into the SM prior to recombination; in this case the decay can produce visible signatures in light element abundances and/or CMB spectral distortions, depending on the details of the decay.   
	
	If one considers a minimal extension of the SM, where the SM has renormalizable interactions with a single massive particle in the HS and the cosmological evolution of the SM plasma is not otherwise altered, then one cannot completely evade the bounds set by $\Neff$ measurements. 
	However, the constraints can be somewhat ameliorated if the relic energy density in the HS does not always evolve as free-streaming dark radiation. For instance, if the relativistic HS particles have strong self-interactions, such that they behave as an ideal fluid during recombination, then they would instead contribute to $N_{\rm fluid}$, the constraints on which are are weaker by factors of 2-3 compared to $\Neff$ \cite{Baumann:2015rya}. Examples of this scenario include interacting neutrino models, recently surveyed in \cite{Berryman:2022hds}.  Alternatively, while a single hot HS relic that subsequently becomes nonrelativistic is more stringently constrained than if it remains relativistic \cite{DePorzio:2020wcz}, the combination of $\Neff$ and large-scale structure constraints may be mitigated in a system with more than one hot relic if one HS species becomes non-relativistic before recombination while at least one other species remains relativistic.  
	In principle, one can obtain a conservative constraint on portal interactions between the SM and a HS containing light degrees of freedom that can accommodate such variations in the spectrum of the HS by combining both CMB and large-scale structure measurements. We leave this to future work.

	\section{Summary and discussion}\label{sec:summary}

	In this work, we have studied the production of dark radiation in scenarios where the SM has renormalizable  interactions with a heavy ($m_{\phi}>$ MeV) gauge singlet mediator that annihilates or decays into dark radiation prior to BBN.  We  have focused on two  specific minimal models:  (i) a MCP model with a massless dark photon, and (ii) a gauged $B-L$ model with light right-handed neutrinos. By numerically solving the relevant Boltzmann equations, we have computed the resulting dark radiation abundance and determined the corresponding shifts in $\Neff$ in the regions of parameter space relevant for upcoming CMB experiments.  We present updated CMB constraints for the MCP model, and have shown that future CMB measurements will be sensitive enough to either rule out or discover the extended MCP model invoked to explain the EDGES anomaly \cite{Liu:2019knx}. In the case of the gauged $B-L$ model, our computations extend and improve previous analyses by taking into account all relevant out-of-equilibrium processes, including the potentially out-of-equilibrium decays of the $B-L$ gauge boson. As a result, our  projected constraints on the allowed parameter space of the $B-L$ model are stronger than previous studies. 	In both models we take into account the quantum statistical phase space distribution for Standard Model particles, which was not done in previous studies. We find that quantum statistics provide a correction of about 10\% to the predicted shift in  $\Neff$.
	
	The relation between dark radiation production and the model parameters depends crucially on whether or not the HS comes into thermal equilibrium with the SM.  We have provided simple semi-analytical recipes to obtain the predicted shift in $\Neff$ in both cases.
	When the HS remains out of equilibrium with the SM, we have demonstrated that the resulting dark radiation density is determined by the energy transfer rate from the SM into the HS at temperatures of order the mediator mass, $(\Gamma_E/H)_{T_{\rm SM}\sim m_{\phi}}$. The energy transfer rate typically goes like $\Gamma_E(T_{\rm SM}\sim m_{\phi})\propto g_{\phi}^2m_{\phi}$, where $g_{\phi}$ is the Standard Model coupling with the heavy mediator particle with mass $m_\phi$. Consequently, the contour of constant $\Delta \Neff$ relates  $g_{\phi} \propto \sqrt{m_{\phi}/M_{\rm Pl}}$, which accounts approximately for the shape of the contours in the regions where the sectors are out-of equilibrium in figures \ref{fig:mcp_constraint} and \ref{fig:BL_constraint}.  We provide a simple formula for evaluating the resulting $\Neff$ constraint, given an input cross-section. CMB $\Delta\Neff$ constraints are already the leading limit on both models in most of the out-of-equilibrium parameter space, along with constraints from SN1987A; these astrophysical and cosmological constraints far exceed terrestrial accelerator constraints in the sub-GeV regime.	
	
	As one increases $m_{\phi}$ at a fixed value of the dark radiation density, the coupling $g_\phi$ can increase to a point where the HS comes into thermal equilibrium with the SM.	
	When the HS thermalizes with the SM, the resulting dark radiation density is determined by the temperature at which the HS and SM decouple. This decoupling temperature is primarily determined by Boltzmann suppression of the collision term. Consequently, the decoupling temperature is mainly set by the mass of the mediator, $m_{\phi}$, and only depends logarithmically on the coupling $g_{\phi}$; once the sectors are in thermal equilibrium, increasing the coupling only marginally decreases the resulting decoupling temperature, and thus marginally increases the resulting dark radiation density. Because of the weak sensitivity to $g_{\phi}$, the constraint imposed by $\Neff$ measurements on $g_{\phi}$ is exponentially weakened if the HS thermalizes with the SM. This effect gives rise to a thermalization mass threshold, $m_{\rm th}$, beyond which the constraint curves in figures \ref{fig:mcp_constraint} and \ref{fig:BL_constraint} are exponentially weakened.

	The example models discussed above consider a minimal hidden sector that is coupled to the SM via a heavy mediator. More generally, one can consider the mediator to communicate with a hidden sector that may have a nonminimal internal spectrum. 
	While the exact evaluation of dark radiation production in extended models would require a numerical computation of the Boltzmann equations that take into account all internal hidden sector interactions, we have shown  how to obtain a simple analytical lower bound on the relic dark radiation that depends only on the mass and coupling of the mediator, and is independent of the number of particles in the hidden sector or their internal interactions. This minimum  dark radiation abundance is obtained by considering that energy transfer into the HS occurs through the production of heavy mediators by the SM plasma, and assuming that any energy transferred to the mediator is promptly deposited in the relativistic HS degrees of freedom. This amounts to assuming that the mediator is more strongly coupled to the HS than to the SM.  In the regime where CMB constrains the HS to remain out of equilibrium with the SM in the early universe, this is a very mild requirement on the mediator coupling.  This model-insensitive lower bound on $\Delta\Neff$ assumes there are no BSM contributions to the entropy of the SM plasma, and that the relic dark radiation remains a free-streaming relativistic relic throughout the formation of the CMB.  Relaxing these assumptions can evade our lower bound.
	
	We have shown that future CMB measurements of $\Neff$ have the potential to constrain portal couplings to values which typically are orders of magnitude weaker than those probed by collider experiments, and provided simple semi-analytic recipes to evaluate their reach.  If future CMB observations do not find any deviation from the Standard Model prediction for $\Neff$,  hidden sector models with light species will also be out of reach for accelerator experiments, unless there are departures from the standard cosmology. This work highlights the potential of future CMB missions to significantly narrow down the space of observationally relevant BSM theories.

	\section*{Acknowledgments}
	
	We thank Katelin Schutz for clarifying some aspects of her work and Nicolas Fernandez for useful conversations. The work of PA and PR was supported in part by NASA Astrophysics Theory Grant NNX17AG48G. The work of JS and PR was supported in part by DOE Early Career grant DE-SC0017840.
	
	\appendix
	\section{Cross-sections for the millicharged particle model}\label{sec:mcp_other_channels}

	In this section, we give the cross-sections for various processes contributing to energy transfer into dark photons for the MCP model. We discuss some of the approximations we use and in particular highlight the simplifications we employed in modeling the electroweak and QCD phase transitions.
	In what follows we first discuss energy injection via SM fermion annihilations into MCPs in section~\ref{mcp_app_fermion}. 
	In sections~\ref{mcp_app_Z} and \ref{mcp_app_plasmon}, we describe energy injection into HS due to $Z$-boson and plasmon decays into MCPs, respectively. 
	Finally, in section~\ref{mcp_app_scatt} we describe energy transfer via Coulomb scattering between SM fermions and MCPs. 
	
	Other processes  also contribute to the energy transfer, such as electroweak boson annihilations into MCPs, Compton scattering of MCPs with photons and dark photons, and photon-dark photon fusion into MCPs. We have verified that energy transfer through SM boson annihilations and fusion processes are around two orders of magnitude weaker than that through SM fermion annihilations and Coulomb scattering.  Compton scattering depends on both the dark coupling constant, $e'$, and the millicharge, $Q$, while all other processes only depend on $Q$. We have checked that provided $e'<0.9$, the collision term for  Compton scattering is subdominant to Coulomb scattering. Consequently in our study we neglect the contribution from Compton scattering, SM boson annihilations and photon dark-photon fusion.	
	
	\subsection{Fermion annihilations}\label{mcp_app_fermion}
	
	The center-of-momentum (CM) frame spin-summed cross-section for SM fermion annihilation into MCPs is given by
	\begin{align}\label{eq:cross_ff_ann_full}
		\sigma_{ff\rightarrow \psi\psi}=&\frac{4\pi Q^2N_c(f)\alpha^2}{s^3}\frac{\sqrt{s-4m^2}}{\sqrt{s-4m_f^2}}\nonumber\\
		&\times\bigg\{\frac{4}{3}(2m^2+s)(2m_f^2+s)\left[Q_f^2-\frac{Q_fC_V}{\cos^2\theta_W}\frac{s(s-m_Z^2)}{(s-m_Z^2)^2+m_Z^2\Gamma_Z^2}\right]\nonumber\\
		&+\frac{1}{4\cos^4\theta_W}\frac{s^2(s+2m^2)}{(s-m_Z^2)^2+m_Z^2\Gamma_Z^2}\bigg[\frac{4}{3}(C_V^2+C_A^2)(s-m_f^2)+4(C_V^2-C_A^2)m_f^2\bigg]\bigg\},
	\end{align}
	where $C_V$ and $C_A$ are the vector and axial couplings of the SM fermion $f$ to the $Z$ boson, respectively, given by $C_V=T^3_f-2Q_f\sin^2\theta_W$ and $C_A=T^3_f$. Here the term proportional to $Q_f^2$ comes from the photon-mediated interaction. The terms proportional to $C_V^2$ and $C_A^2$ comes from the $Z$-mediated interaction while the term proportional to $C_V Q_f$ comes from the interference between photon and $Z$-mediated terms. 
	
	The  cross-section in eq.\ \eqref{eq:cross_ff_ann_full} has a pole at $s=M_{Z'}^2$, which can be seen explicitly in the narrow width limit,
	\begin{align}
		\frac{1}{(s-M_{Z}^2)^2+\Gamma_{Z}^2M^2_{Z}}&\approx \frac{1}{M_{Z}^4}\Theta(M_{Z}^2-s)+\frac{\pi \delta(s-M_{Z}^2)}{M_{Z}\Gamma_{Z}}+\frac{1}{s^2}\Theta(s-M_{Z}^2),
	\end{align}
	where $\Theta$ is the Heaviside function. The contribution to the collision integral from the Dirac delta term  gives an identical contribution to the collision term due to $Z$-boson decays (see appendix~\ref{sec:BL_coll} or Refs.~\cite{Pilaftsis:2003gt,Giudice:2003jh}),  discussed in the next subsection. To avoid double-counting we subtract the Dirac delta piece. Additionally, we also neglect the terms proportional to $\Theta(M_{Z}^2-s)$ as the  contribution from those terms is heavily suppressed compared to others. This yields the effective off-shell cross-section
	\begin{align}\label{eq:cross_ff_ann_Z}
		\sigma_{ff\rightarrow \psi\psi}^{\rm off}=&\frac{4\pi Q^2N_c(f)\alpha^2}{s^3}\frac{\sqrt{s-4m^2}}{\sqrt{s-4m_f^2}}\nonumber\\
		&\times\bigg\{\frac{4}{3}(2m^2+s)(2m_f^2+s)\left[Q_f^2+\Theta(s-M_Z^2)\left(\frac{(C_V^2+C_A^2)}{4\cos^4\theta_W}-\frac{C_VQ_f}{\cos^2\theta_W}\right)\right]\nonumber\\
		&-\Theta(s-M_Z^2)\frac{(C_V^2+3C_A^2)m_f^2}{2\cos^4\theta_W}(s+2m^2)\bigg\}.
	\end{align}
	We find that the cross-section from photon contributions alone (i.e., retaining only terms proportional to $Q_f^2$) to be at least an order of magnitude larger than the contribution from the remaining terms that involve at least one coupling to the $Z$. Thus, in the analytical calculation of the leak factor in eq.~\eqref{eq:leak_mcp} we neglect the $Z$-mediated contributions  for simplicity.  
	
	The forward energy transfer collision term, $\mathcal{C}_{\rm f}$, corresponding to fermion annihilations into MCPs is calculated by using the cross-section in eq.~\eqref{eq:cross_ff_ann_Z} inside the generic collision term derived in eq.~\eqref{eq:s_quant_sim} and summing over all SM fermions. The total energy transfer collision term is then evaluated through $\mathcal{C}=\mathcal{C}_{\rm f}(T_{{\rm SM}})-\mathcal{C}_{\rm f}(T_{{\rm HS}})$. We include quarks, treated as free fermions, for  $T_{\rm SM}>T_{\rm QCD}$, where we take $T_{\rm QCD}=200$ MeV;  for $T_{\rm SM}<T_{\rm QCD}$ we neglect hadronic contributions, as they are generally Boltzmann-suppressed. 
	
\begin{figure}
		\begin{subfigure}{.32\textwidth}
			\includegraphics[width=1\textwidth]{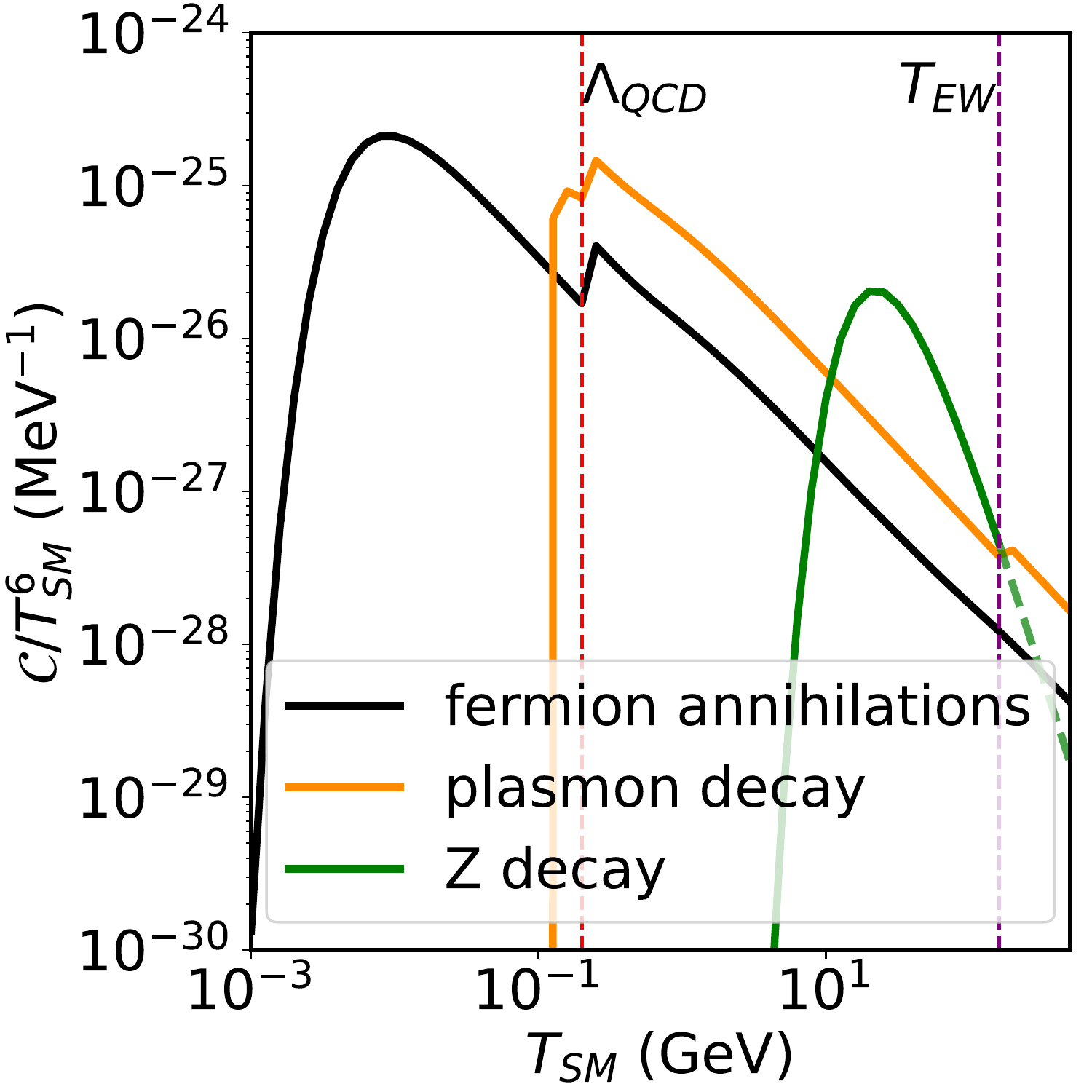}
		\end{subfigure}
		\begin{subfigure}{.32\textwidth}
			\includegraphics[width=1\textwidth]{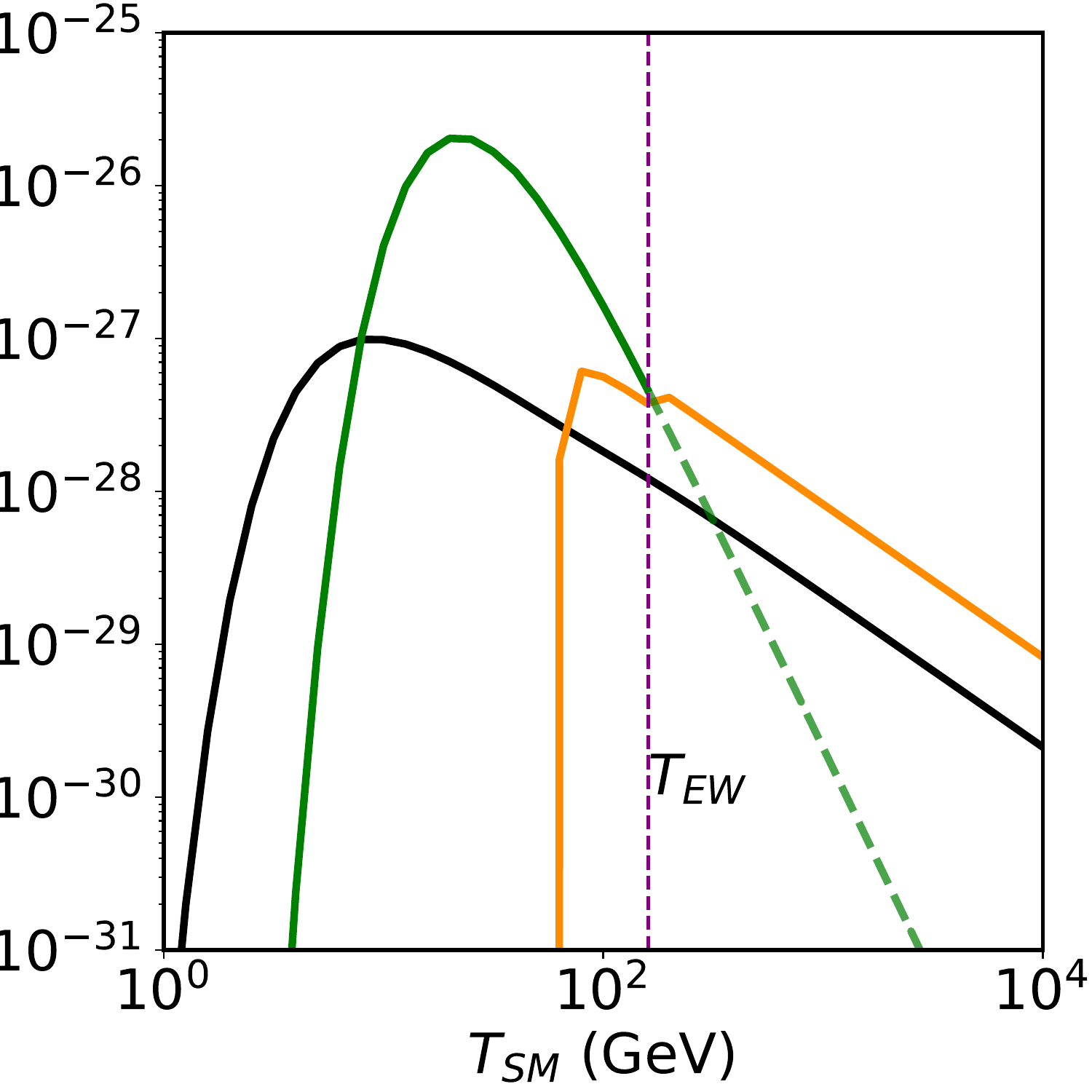}
		\end{subfigure}
		\begin{subfigure}{.32\textwidth}
			\includegraphics[width=1\textwidth]{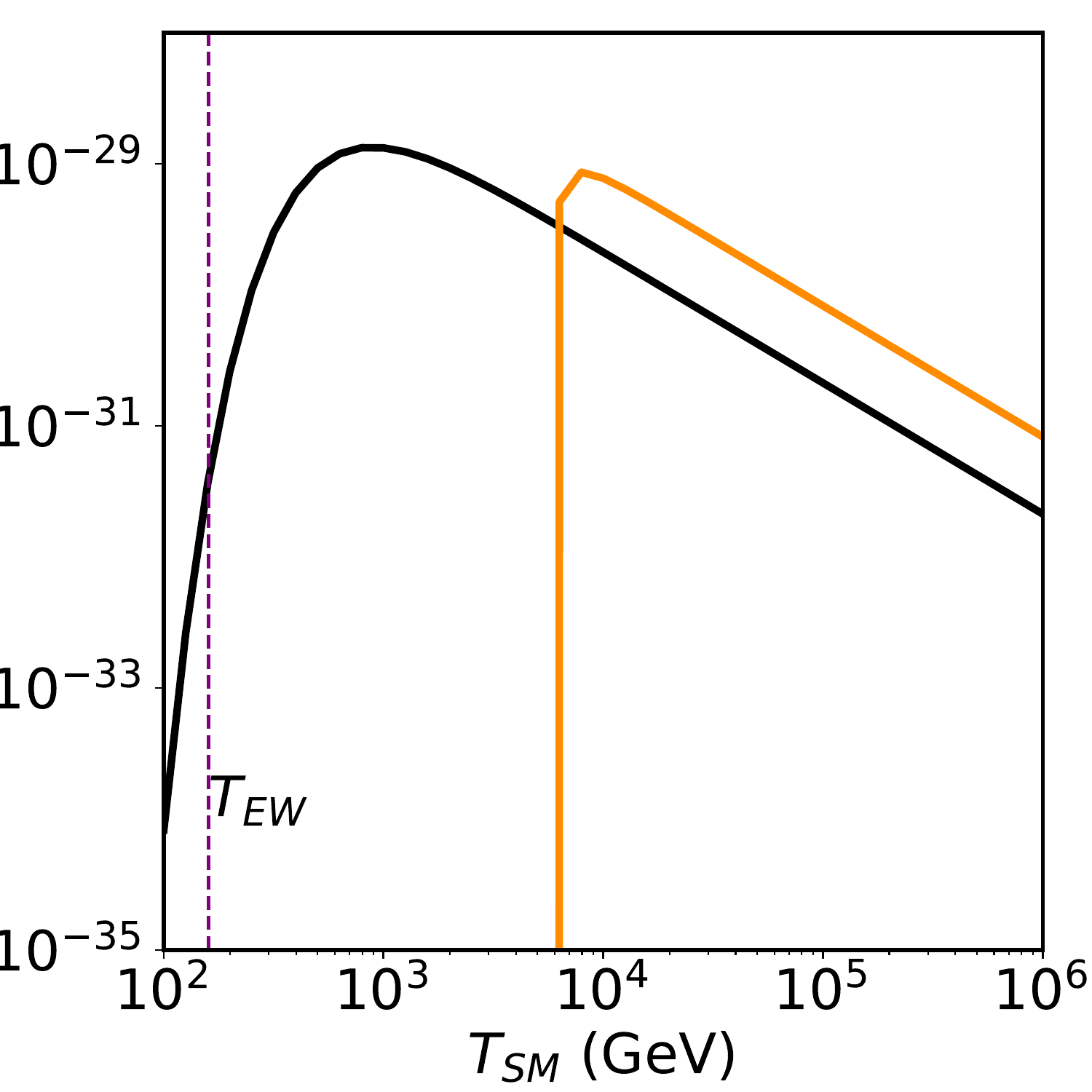}
		\end{subfigure}
		\caption{Collision terms from the three $s$-channel processes (fermion annihilations, $Z$ decays, and plasmon decays) for $Q=10^{-9}$ and $m=10$ MeV (left), $m=10$ GeV (center), and $m=1$ TeV (right).  We plot $\mathcal{C}/T_{\rm SM}^6$, the maximum value of which provides the dominant contribution to the energy injected into the hidden sector. The black line corresponds to the collision term from SM fermion annihilations,  the green line to $Z$ decays (eq.~\eqref{eq:Zdecay}), and the orange line to plasmon decays (eq.~\eqref{eq:plasmon2}). Vertical red dashed and purple dashed lines mark the temperatures used for the QCD and electroweak phase transitions, respectively. Fermion annihilations into MCPs provide the dominant $s$-channel energy transfer process except in the mass range $0.3{\rm\ GeV}\lesssim m\lesssim 40$ GeV where $Z$-boson decays dominate.}
		\label{fig:Zboson_rates}
	\end{figure}

	In section~\ref{mcp:boltz}, we found that the maximum of $\mathcal{C}_{\rm f}a^4/H$ roughly determines the final comoving energy density of dark radiation, where $\mathcal{C}_{\rm f}$ is the forward collision term and $H$ is the Hubble rate. As $\mathcal{C}_{\rm f}a^4/H\propto \mathcal{C}_{\rm f}/T_{\rm SM}^6$, in figure~\ref{fig:Zboson_rates} we plot $\mathcal{C}_{\rm f}/T_{\rm SM}^6$ for fermion annihilation into MCPs (black line) for different values of $m$. The collision term $\mathcal{C}_{\rm f}/T_{\rm SM}^6$ reaches its maximum around $T_{\rm SM}\sim m/2$ below which it becomes Boltzmann suppressed.  (We focus on the regime with $m>m_e$.)
	
	Above the electroweak phase transition, the dark photon mixes with the hypercharge gauge boson.  In the $s\gg M_Z^2$ limit, the cross-section in eq.~\eqref{eq:cross_ff_ann_full} reduces to the  cross-section describing annihilation through a hypercharge boson.  For simplicity, we neglect the temperature dependence of the Higgs vev, and thus the (tree-level) $Z$ mass, through the electroweak phase transition, as $Z$ contributions are subleading below the transition and negligible above it.

	\subsection{$Z$-boson decay}\label{mcp_app_Z}
	
	Next, we discuss energy transfer from $Z$ decays into MCPs. The collision term due to $Z$ decays is given by
	\begin{align}\label{eq:Zdecay}
		\mathcal{C}_{Z\rightarrow\psi\psi}=\Gamma_{Z\rightarrow \psi\psi}m_Z(n_{Z}-n_{eq,Z}(T_{\rm HS})),
	\end{align}
	where
	\begin{align}
		\Gamma_{Z\rightarrow \psi\psi}=\frac{Q^2\alpha\tan^2\theta_W}{3}m_Z\sqrt{1-\frac{4m^2}{m_Z^2}}\left(1+\frac{2m^2}{m_Z^2}\right)\Theta(m_Z-m/2),
	\end{align}
	$n_Z$ is the $Z$ number density, $n_{Z, {\rm eq}}$ is the equilibrium number density of bosons with three degrees of freedom, and $\Theta$ is the Heaviside function. As $Z$ bosons are always in equilibrium with the SM plasma, we have $n_Z=n_{Z,{\rm eq}}(T_{\rm SM})$.
	
	At temperatures above  $T_{\rm EW}=160$ GeV \cite{DOnofrio:2015gop}, electroweak symmetry is unbroken and thus the contribution from $Z$ decays is absent. We use a simple model of the electroweak phase transition, where we set $\mathcal{C}_{Z\rightarrow\psi\psi}=0$ for $T>T_{\rm EW}$ but neglect any temperature variation in the $Z$ mass for $T<T_{\rm EW}$.  As the main contribution from $Z$ decays arises at temperatures significantly smaller than $T_{\rm EW}$, this is a sufficient approximation for our purposes.  In figure~\ref{fig:Zboson_rates}, we plot the resulting collision term, normalized by $T_{\rm{SM}}^6$, as the green line.  The net energy injected into the HS is dominated by the peak values of $\mathcal{C}/T_{\rm SM}^6$, which occurs around $T_{\rm SM}\sim M_{Z}/4\ll T_{\rm EW}$.

	The contribution from $Z$ decays to the total energy injected into the HS can dominate over the contribution from  photon-mediated fermion annihilation when $0.3{\rm\ GeV}\lesssim m\lesssim 40$ GeV.  Elsewhere, $Z$ decays provide a sub-leading contribution.

	\subsection{Plasmon decay}\label{mcp_app_plasmon}
	
	Below the electroweak phase transition, $T_{\rm SM}<T_{\rm EW}$, the thermal effects in the plasma cause photons to acquire an in-medium plasma mass. The corresponding plasmons can decay into MCPs with the collision term given by \cite{Raffelt:1996wa,Vogel:2013raa,Dvorkin:2019zdi}
	\begin{align}\label{eq:plasmon_decay_int}
		\mathcal{C}_{\gamma\rightarrow \psi\psi}=\sum_{\rm pol}\int\frac{d^3k}{(2\pi)^3}\left(\frac{1}{e^{\omega/T_{\rm SM}}-1}-\frac{1}{e^{\omega/T_{\rm HS}}-1}\right)\omega \Gamma_{\gamma\rightarrow \psi\psi},
	\end{align}
	where 
	\begin{align}
		\Gamma_{\gamma\rightarrow \psi\psi}=\frac{\alpha Q^2}{3\omega}Z(m_{\gamma}^2+2m^2)\sqrt{1-\frac{4m^2}{m_{\gamma}^2}}.
	\end{align}
	Here $Z$ is a wave-function renormalization factor  and $m_{\gamma}$ is the plasmon mass, both of which differ for transverse and longitudinal polarizations. For a relativistic photon, where $\omega (k) \equiv \sqrt{m_\gamma^2+k^2} \gg m_{\gamma}$, the decays from the longitudinal polarization are negligible compared to the decays from the transverse polarizations \cite{Raffelt:1996wa,Dvorkin:2019zdi}. Moreover, for the transverse polarization at relativistic energies we have $Z\approx 1$ and 
	\begin{align}\label{eq:plasmon_mass}
		m_{\gamma}^2=&\sum_fg_fQ_{f}^2\frac{4\alpha}{\pi}\int_0^{\infty}dp\ p \, f_f(p),
	\end{align}
	where $f_f$ is the phase space distribution of the SM fermion $f$ and the summation runs over all fermions;  $g_f$ counts the spin degrees of freedom of each fermion. Since $m_{\gamma}\ll T_{\rm SM}$, approximating $\omega\gg m_{\gamma}$  in eq.~\eqref{eq:plasmon_decay_int} is valid as the integrand  is dominated by momenta with $\omega\sim T_{\rm SM}$.
	Thus the collision term simplifies to
	\begin{align}\label{eq:plasmon}
		\mathcal{C}_{\gamma\rightarrow \psi\psi}&=\frac{2\alpha Q^2}{3}(m_{\gamma}^2+2m^2)\sqrt{1-\frac{4m^2}{m_{\gamma}^2}}\times \left(n_{\gamma}(T_{\rm SM})-n_{\gamma}(T_{\rm HS})\right),
	\end{align}
	where $n_{\gamma}$ is the equilibrium number density of photons.
	Energy transfer from plasmon decay is prohibited when $m_{\gamma}<2m$. Since $m_{\gamma}\sim 0.1T_{\rm SM}$, energy injection via plasmon is only efficient at high temperatures where $T_{\rm SM}> 10m$. 
	
	Above the electroweak phase transition, $T_{\rm SM}>T_{\rm EW}$, we need to evaluate the decay of hypercharge bosons into MCPs. The collision term for this process is similar to that for photon decay, up to the replacement of the electric charge $e$ by the hypercharge coupling $e/\cos\theta_W$ and the fermion electric charges $Q_f$ by their hypercharges $Q_Y$. Consequently, we obtain
	\begin{align}\label{eq:plasmonB}
		\mathcal{C}_{B\rightarrow \psi\psi}&=\frac{2\alpha Q^2}{3\cos^4\theta_W}(m_{B}^2+2m^2)\sqrt{1-\frac{4m^2}{m_{B}^2}}\times \left(n_{B}(T_{\rm SM})-n_{B}(T_{\rm HS})\right),
	\end{align}
	where $n_B$ is the equilibrium number density of hypercharge gauge bosons and $m_B$ is the thermal mass, given by 
	\begin{align}\label{eq:plasmon_massB}
		m_{B}^2=&\frac{11\alpha\pi}{3\cos^2\theta_W}T_{\rm SM}^2 ,
	\end{align}
	for large temperatures.
	We take the plasmon decay contribution to be given by	
	\begin{align}\label{eq:plasmon2}
		\mathcal{C}_{\rm plasmon}=\begin{cases}\mathcal{C}_{\gamma\rightarrow \psi\psi} & T_{\rm SM}<T_{\rm EW}\\
			\mathcal{C}_{B\rightarrow \psi\psi} & T_{\rm SM}>T_{\rm EW}.\end{cases}
	\end{align}
	In figure~\ref{fig:Zboson_rates} we compare the resulting collision term $\mathcal{C}_{\rm f}/T_{\rm SM}^6$ (orange line) to the collision term describing photon-mediated SM fermion annihilations. The collision term $\mathcal{C}_{\rm f}/T_{\rm SM}^6$ from fermion annihilation is maximized around $T_{\rm SM}=m/4$, while that from plasmon decay is maximized around $T_{\rm SM}=m/10$. Since the maximum value of $\mathcal{C}/T_{\rm SM}^6$ controls the final energy injected into the HS, the energy injected into the HS via plasmon decay is subdominant to the energy injected via fermion annihilations, even though at high temperatures the collision term for plasmon decay is larger than the collision term for fermion annihilation. Thus, the approximations used in $\mathcal{C}_{\rm plasmon}$ near the electroweak and QCD phase transitions are of marginal consequence in evaluating the resulting dark radiation density. 	
	
	\subsection{Coulomb scattering}\label{mcp_app_scatt}
	
	SM particles can also inject energy into the HS through the Coulomb scattering of MCPs with SM particles, $\psi+f\rightarrow \psi+f$. The cross-section for Coulomb scattering has a forward singularity, which we regulate by adding a plasmon mass in the propagator \cite{Vogel:2013raa}. 
	
	Below the electroweak scale, the Coulomb scattering is mediated by photons, with the plasmon mass given by eq.~\eqref{eq:plasmon_mass}. The relevant spin-summed matrix element for SM fermion scattering with MCPs is given by
	\begin{align}
		|\mathcal{M}|^2_{f\psi\rightarrow f\psi}=&\frac{8Q^2N_c(f)Q_f^2e^4}{(t-m_{\gamma}^2)^2}\left(2(s-m_f^2-m^2)^2+2st+t^2\right),
	\end{align}
	where $Q$, $Q_f$, $m$ and $m_f$ are the charge and mass of the MCP and the SM fermion, respectively, $N_c(f)$ is the color factor of the SM fermion, $m_{\gamma}$ is the plasmon mass given by eq.~\eqref{eq:plasmon_mass}, and $s$ and $t$ are the Mandelstam variables.  The collision term for the above process, including quantum statistics, is given by eq.~\eqref{eq:t_channel_full} with $m_{\phi} \to m_{\gamma}$ and the coefficients of non-zero $c_{nm\lambda}$, defined in eq.~\eqref{eq:cnml_def}, given by
	\begin{align}\label{eq:cnml_qed}
		\frac{c_{222}}{16\pi Q^2Q_f^2e^4}=\frac{3}{4}, && \frac{c_{202}}{16\pi Q^2Q_f^2e^4}=-\frac{1}{4}, && \frac{c_{022}}{16\pi Q^2Q_f^2e^4}=-\frac{1}{4}, && \frac{c_{002}}{16\pi Q^2Q_f^2e^4}=\frac{3}{4},\nonumber\\
		\frac{c_{001}}{16\pi Q^2Q_f^2e^4}=(m_f^2+m^2), && \frac{c_{201}}{16\pi Q^2Q_f^2e^4}=m^2, && \frac{c_{021}}{16\pi Q^2Q_f^2e^4}=m_f^2,\nonumber\\ \frac{c_{000}}{16\pi Q^2Q_f^2e^4}=4m_f^2m^2.
	\end{align}
	Additionally, we multiply the resulting collision term by a factor of four to account for all combinations of particles and antiparticles.
	While solving the Boltzmann equations in section~\ref{sec:MCP}, we sum over the contribution from all SM fermions.  Again, we include quarks for $T_{\rm SM}>T_{\rm QCD}$, and neglect hadron contributions for $T_{\rm SM}<T_{\rm QCD}$.
	
	Above the electroweak scale the Coulomb scattering is mediated by the hypercharge boson. For a (Weyl) SM fermion scattering with MCPs, the spin-summed matrix element is
	\begin{align}
		|\mathcal{M}|^2_{f\psi\rightarrow f\psi}=&\frac{4Q^2N_c(f)Q_Y(f)^2e^4}{\cos^4\theta_W(t-m_{B}^2)^2}\left(2(s-m_f^2-m^2)^2+2st+t^2\right),
	\end{align}
	where $\theta_W$ is the weak mixing angle, $Q_Y(f)$ is the hypercharge of the fermion, and $m_B$ is the thermal mass of hypercharge gauge boson (eq.~\eqref{eq:plasmon_massB}). The coefficients $c_{nm\lambda}$ for the above matrix element are the same as those given in eq.~\eqref{eq:cnml_qed} up to an overall rescaling by the factor $Q_Y^2/(2Q_f^2\cos^4\theta_W)$. Additionally, the Higgs doublet can also scatter with MCPs, with the corresponding matrix element being
	\begin{align}
		|\mathcal{M}|^2_{H\psi\rightarrow H\psi} =&2\frac{Q^2e^4}{4\cos^4\theta_W}\frac{1}{(t-m_{B}^2)^2}\times8[s^2+st-m^2(t+2s)+m^4].
	\end{align}
	The corresponding coefficients $c_{nm\lambda}$ are
	\begin{align}
		\frac{c_{222}}{4\pi Q^2e^4/\cos^4\theta_W}=\frac{3}{4}, && \frac{c_{202}}{4\pi Q^2e^4/\cos^4\theta_W}=-\frac{1}{4}, &&
		\frac{c_{022}}{4\pi Q^2e^4/\cos^4\theta_W}=-\frac{1}{4}, \nonumber\\
		\frac{c_{002}}{4\pi Q^2e^4/\cos^4\theta_W}=-\frac{1}{4} ,&&
		\frac{c_{201}}{4\pi Q^2e^4/\cos^4\theta_W}=m^2, &&
		\frac{c_{001}}{4\pi Q^2e^4/\cos^4\theta_W}=-m^2.
	\end{align}
	
	Unlike the $s$-channel processes, for Coulomb scattering the forward collision term describing energy transfer into the HS  is sensitive to the distributions of both HS and SM particles. Moreover, the backward collision term for Coulomb scattering is of the same order of magnitude as the forward collision term for $T_{\rm HS}>0.1T_{\rm SM}$, while the backward term for $s$-channel processes is almost negligible compared to the forward term for $T_{\rm HS}<0.9T_{\rm SM}$.
	
	\begin{figure}
		\begin{subfigure}{.5\textwidth}
			\includegraphics[width=1\textwidth]{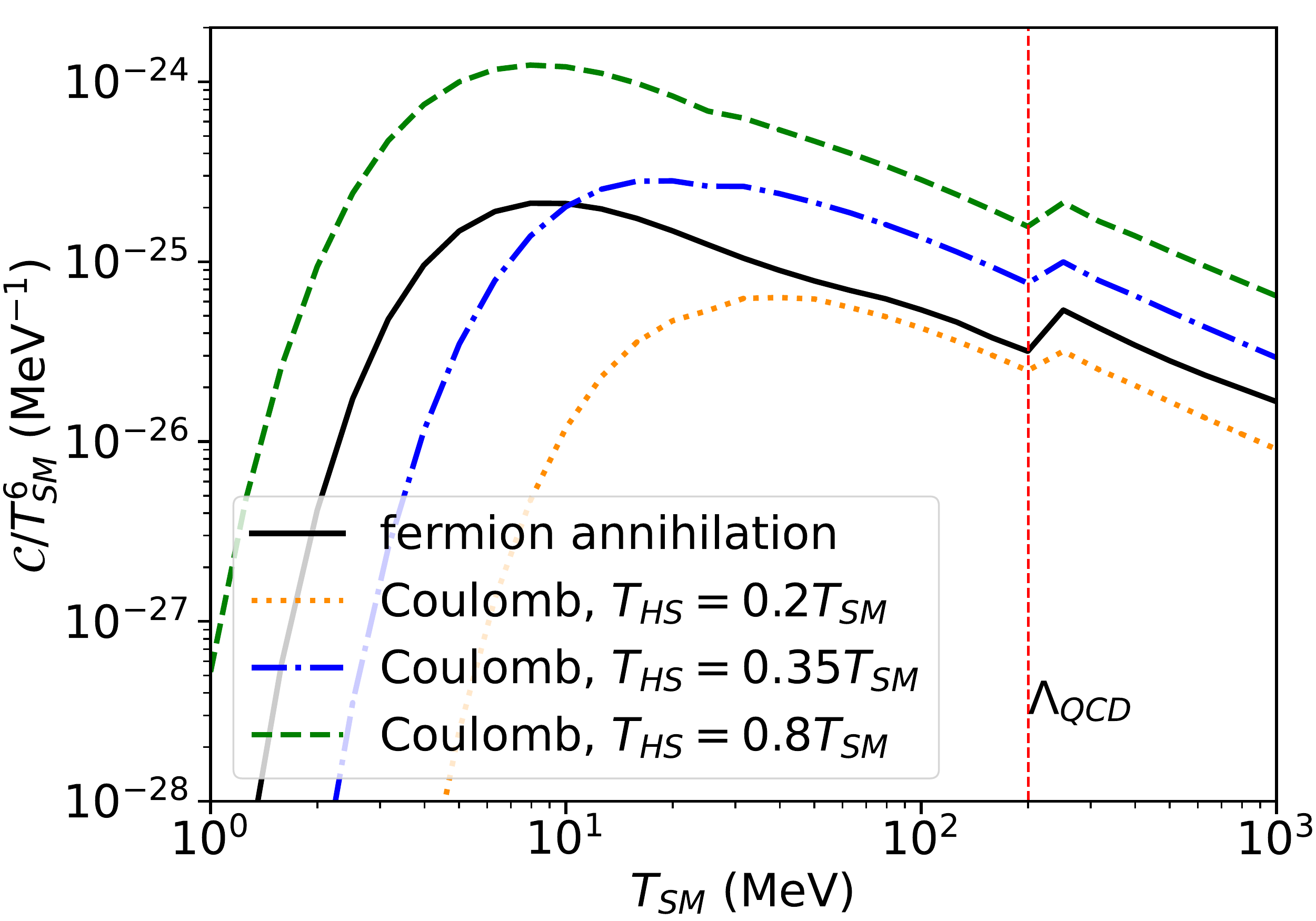}
		\end{subfigure}
		\begin{subfigure}{.5\textwidth}
			\includegraphics[width=1\textwidth]{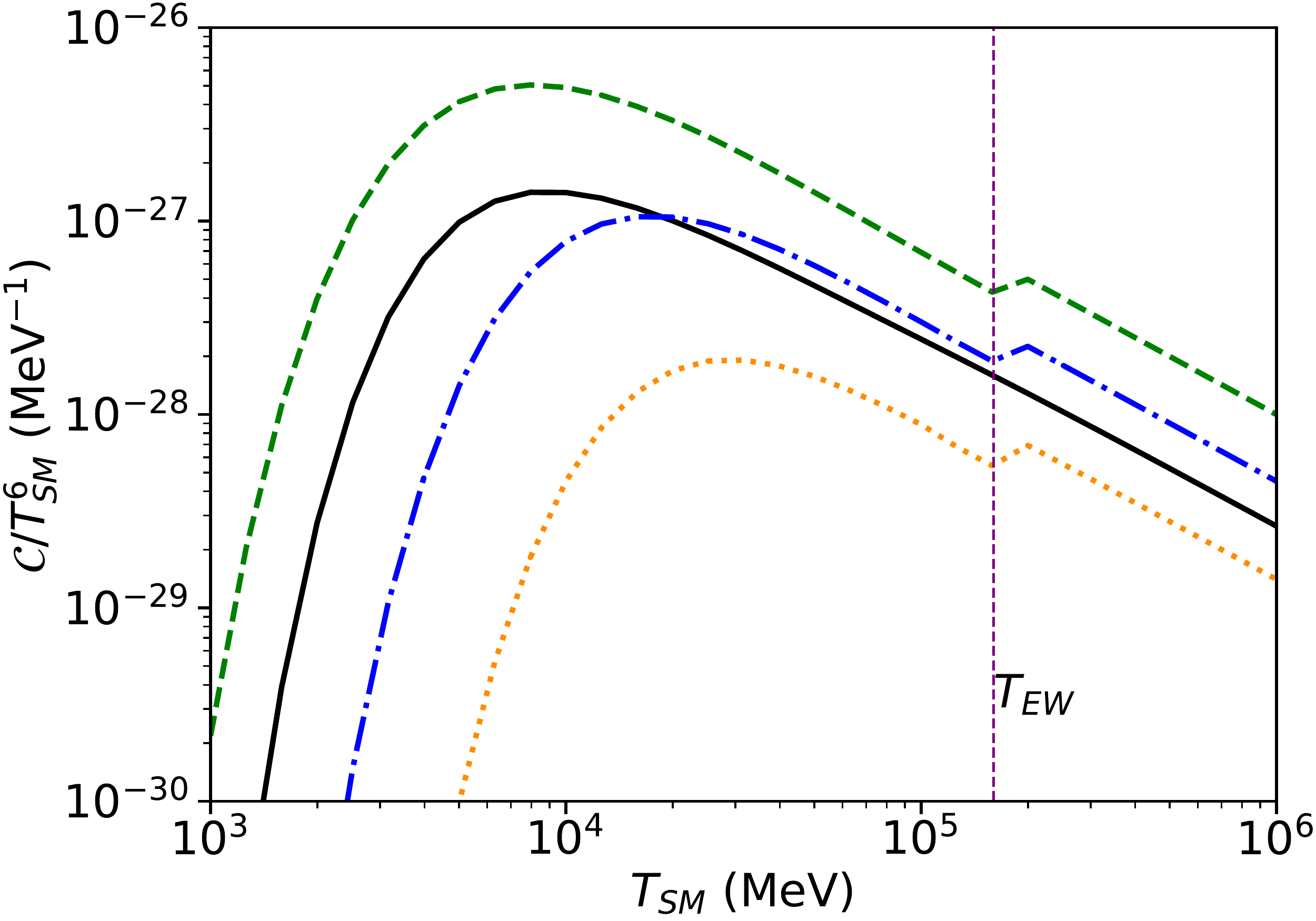}
		\end{subfigure}
		\caption{Comparison between Coulomb scattering and fermion annihilation for $Q=10^{-9}$ and MCP masses $m=10$ MeV (left) and $m=10$ GeV (right). Green, blue and orange  lines show the collision term from Coulomb scattering, normalized by $T_{\rm SM}^6$, at different values of $T_{\rm HS}/T_{\rm SM}$ as indicated in the legend, while the black line shows the forward collision term due to SM fermion annihilation.  The energy transferred via Coulomb scatterings dominates over that via annihilation for $T_{\rm HS}>0.35T_{\rm SM}$.}
		\label{fig:collision_mcp}
	\end{figure}
	
	In figure~\ref{fig:collision_mcp} we compare the total collision term for Coulomb scattering between MCPs and SM particles with the forward collision term for SM fermion annihilation into MCPs. The collision term for Coulomb scattering decreases for smaller $T_{\rm HS}$ as the number density of HS particles in the initial state drops. The Coulomb scattering collision becomes the dominant process for $T_{\rm HS}/T_{\rm SM} > 0.35$.

	%
	\section{Cross-sections for the $B-L$ model}\label{sec:BL_coll}
	%
	The dominant energy injection from the Standard Model plasma into right-handed neutrinos occurs through the annihilations of the Standard Model fermions into right-handed neutrinos: $f + \bar{f}\rightarrow Z'\rightarrow \bar{\nu}_R +\nu_R$.	
	The cross-section for this process is
	\begin{align}\label{eq:BmLfullxsec}
		{\sigma}_{f\bar{f}\rightarrow \nu_R\bar{\nu}_R}&=\frac{1}{2\pi}Q_f^2g'^4N_c(f)\frac{s}{(s-M_{Z'}^2)^2+\Gamma_{Z'}^2M^2_{Z'}}\sqrt{\frac{s}{s-4m_f^2}}\left(1+\frac{2m_f^2}{s}\right),
	\end{align}
	where $Q_f$ is the $B-L$ charge of SM species $f$, $N_c(f)$ is the number of colors, $m_f$ is the mass of the fermion, and $\Gamma_{Z'}$ is the total decay width of the $Z'$ boson,
	\begin{align}\label{eq:Z_decay}
		\Gamma_{Z'}=\frac{g'^2}{12\pi} M_{Z'}\left[3+\sum_{f}^{2m_f<M_{Z'}} Q_f^2N_c(f)\left(1+\frac{2m_f^2}{M_{Z'}^2}\right)\sqrt{1-\frac{4m_f^2}{M_{Z'}^2}}\right].
	\end{align}
	Above the first factor of $3$ comes from decays into both left- and right-handed neutrinos, while the summation runs over all charged SM fermions with mass $m_f<M_{Z'}/2$.
	
	In the narrow width limit we can approximate
	\begin{align}\label{eq:narro_width}
		\frac{s}{(s-M_{Z'}^2)^2+\Gamma_{Z'}^2M^2_{Z'}}&\approx \frac{s}{M_{Z'}^4}\Theta(M_{Z'}^2-s)+\frac{\pi M_{Z'}\delta(s-M_{Z'}^2)}{\Gamma_{Z'}}+\frac{1}{s}\Theta(s-M_{Z'}^2),
	\end{align}
	where $\Theta$ is the Heaviside function. As for the MCP model above, we separate this cross-section into resonant and nonresonant pieces to avoid double-counting.  
	The resonant part of the cross-section, which corresponds to the term with the Dirac delta function, is proportional to $g'^2$ due to the total decay width in the denominator, while the cross-section due to contact interactions, which correspond to terms with Heaviside functions, is proportional to $g'^4$. As we typically have $g'\ll 1$, the energy injection is dominated by the resonant cross-section. The resonant contribution is accounted for in the on-shell production and decay of $Z'$s while the cross-section corresponding to contact interactions is 
	\begin{align}
		\label{eq:BmLcontact}
		{\sigma}_{ff\rightarrow \nu_R\nu_R}^{\rm off}=\frac{1}{2\pi}Q_f^2g'^4N_c(f)\sqrt{\frac{s}{s-4m_f^2}}\left(1+\frac{2m_f^2}{s}\right)\left[\frac{s}{M_{Z'}^4}\Theta(M_{Z'}^2-s)+\frac{1}{s}\Theta(s-M_{Z'}^2)\right].
	\end{align}
	
	The collision term describing describing decay and inverse decays of the $Z'$ boson is given by (see appendix~\ref{sec:decay_quant} for derivation)
	\begin{align}\label{eq:Z-arrow-aa}
			\mathcal{C}_{ii\rightarrow Z'}=-\mathcal{C}_{Z'\rightarrow ii}=M_{Z'}\Gamma_{Z'\rightarrow i}(\tilde{n}_{\zeta_i}(T_{i})-n_{Z'}),
	\end{align}
	where $\Gamma_{Z'\rightarrow i}$ is the $Z'$ decay width into particle species $i$, $n_{Z'}$ is the number density of $Z'$ bosons, and \begin{align}\label{def:n_tilde}
			\tilde{n}_{\zeta_i}(T)=\frac{3}{2\pi^2}M_{Z'}^2TG_{\zeta_i}(M_{Z'}/T).
	\end{align}
	Here $G_{\zeta_i}$ is a dimensionless function given by eq.~\eqref{eq:G_def} with $\zeta=\pm 1$ depending on whether the particles producing the $Z'$ are fermions or bosons. In the limit $\zeta_i\rightarrow 0$ or $T\ll M_{Z'}$, $\tilde{n}_{\zeta_i}$ asymptotes to the equilibrium number density of $Z'$ in the Maxwell-Boltzmann limit.
		
	The collision term describing forward energy transfer via fermion annihilations into $\nu_R$ through contact operators, including Fermi-Dirac distributions for SM fermions, is obtained by using the cross-section in eq.~\eqref{eq:BmLcontact} inside the generic collision term derived in eq.~\eqref{eq:s_quant_sim} and summing over all SM fermions. We include quarks, treated as free fermions, for  $T_{\rm SM}>T_{\rm QCD}$, where we take $T_{\rm QCD}=200$ MeV;  for $T_{\rm SM}<T_{\rm QCD}$ we neglect hadronic contributions, as they are generally Boltzmann-suppressed. 
	The total collision term is then given by
	\begin{align}\label{eq:Cf_gondolo_BL}
		\mathcal{C}^{\rm off}_{ff\rightarrow \nu_R\nu_R}=&\frac{1}{32\pi^4}\sum_f\int_{4m_f^2}^{\infty} ds (s-4m_{f}^2)s\sigma_{f\bar{f}\rightarrow \nu_R\bar{\nu}_R}^{\rm off}(s)\left[T_{\rm SM}G_{\zeta_f}(\sqrt{s}/T_{\rm SM})-T_{\nu_R}G_{\zeta_f}(\sqrt{s}/T_{\nu_R})\right],
	\end{align}
	where $G_{\zeta}$ is a dimensionless function given by eq.~\eqref{eq:G_def} and is determined by the quantum statistical distribution $f(p)=[e^{-E/T}+\zeta]^{-1}$, where $\zeta=1$ for fermions and $\zeta=-1$ for bosons. Since here the initial state particles are fermions, we have $\zeta_f=1$. In the limit  where the fermion $f$ can be approximated to follow a Maxwell-Boltzmann distribution ($\zeta_f\rightarrow0$), $G$ asymptotes to the second order modified Bessel function of the second kind, $K_2$, and eq.~\eqref{eq:Cf_gondolo_BL} then matches with the well-known result in Ref.~\cite{Gondolo:1990dk}.
	
	In the limit when $\Gamma_{Z'}$ is much larger than the Hubble rate, the on-shell $Z'$ bosons  are in equilibrium. Consequently, we can calculate the forward collision term contributed by on-shell $Z'$s, $\mathcal{C}^{\rm on}_{ff\rightarrow \nu_R\nu_R}$, using eq.~\eqref{eq:Cf_gondolo_BL} with the resonant piece of the total cross-section in place the off-shell piece.  The resonant cross-section follows from  eq.~\eqref{eq:BmLfullxsec} by retaining only the term containing the Dirac delta function in eq.~\eqref{eq:narro_width}. After summing over all contributing SM fermions, we obtain
	\begin{align}\label{eq:BL_Cfploe}
		\mathcal{C}^{\rm on}_{ff\rightarrow \nu_R\nu_R}=&\frac{3M_{Z'}^3}{2\pi^2} \left[\frac{\Gamma(Z'\rightarrow \nu_R)\Gamma(Z'\rightarrow {\rm SM})}{\Gamma_{Z'}}\right]\left[T_{\rm SM}G_{\zeta_f}(M_{Z'}/T_{\rm SM})-T_{\nu_R}G_{\zeta_f}(M_{Z'}/T_{\nu_R})\right],
	\end{align}
	where $\Gamma(Z'\rightarrow \nu_R)$ and $\Gamma(Z'\rightarrow {\rm SM})$ are the partial decay widths into right-handed neutrinos and SM fermions, respectively. Note that this expression matches the effective collision term we derived using Boltzmann equations in eq.~\eqref{eq:Bl_coll_eff}.
	
	In the limit when $\Gamma_{Z'}$ is much smaller than the Hubble rate, the on-shell $Z'$ bosons experience significant evolution in an expanding universe and hence we can no longer use eq.~\eqref{eq:BL_Cfploe}. Instead, we need to independently solve for the evolution of the number density of $Z'$s using the Boltzmann equations given in eqs.~\eqref{eq:Boltz_eps_full1}-\eqref{eq:Boltz_eps_full3}.

	%
	\section{Energy transfer collision term with quantum statistics}\label{sec:schannel_quant}
	%
	
	In this section we calculate collision terms describing energy transfer between two baths at different temperatures via both $s$- and $t$-channel processes, incorporating quantum statistical thermal distributions (both Bose-Einstein and Fermi-Dirac).
	
	\subsection{Annihilation}\label{subsec:ann}
	In this section we simplify the forward energy transfer collision term for particle $a$ annihilating into particle $b$,
	\begin{align}
		1(a)+2(\bar{a})\rightarrow 3(b)+4(\bar{b}),
	\end{align}
	where in general we consider particle $b$ to have a temperature $T_b$ different from the temperature $T_a$ of particle $a$.
	We start with the forward collision term given by
	\begin{align}
		\mathcal{C}_{\rm f}&=\int \left[\prod_{i=1}^4\frac{d^4p_i}{(2\pi)^3}\delta(p_i^2-m_i^2)\Theta(p_i^0)\right](2\pi)^4\delta^4(p_1+p_2-p_3-p_4)  S|\mathcal{M}_f|^2(p_1^0+p_2^0)\nonumber\\
		&\times \left[f_{a}(p_1)f_{a}(p_2)(1\pm f_{b}(p_3))(1\pm f_{b}(p_4))\right],
	\end{align}
	where $\Theta$ is the Heaviside function and $|\mathcal{M}_f|^2$ is the spin-summed matrix element for the process. Here $S$ is a potential symmetry factor, accounting for the potential presence of identical particles in the initial or final state.
	
	In the limit where the density of $b$ is much smaller than the density of $a$, terms depending on the  phase space distribution of final state particles can be neglected regardless of the statistics obeyed by $b$.  If $b$ is part of a thermal bath with $T_b\ll T_a$, then $f_b$ is peaked at momenta $p\sim T_b$, where $T$ is the temperature. As the $b$ particles produced in annihilations have momenta $p\sim T_a$, the values of $f_b$ probed by the collision integral are much smaller than one. (This approximation is also good in the case when particle $b$ does not thermalize; in this case $f_b$ is peaked at $p\sim T_a$ but its value is still much less than one as the overall number density of $b$ is small.)  
	
	By neglecting the final state effects, the final state phase space integration 
	can simply be absorbed into the definition of the Lorentz-invariant cross-section $\sigma$ \cite{Gondolo:1990dk},
	\begin{align}
		4F\sigma_{aa\rightarrow bb}\equiv\int \frac{d^3p_3}{(2\pi)^32E_3}\frac{d^3p_4}{(2\pi)^32E_4}(2\pi)^4\delta^4(p_1+p_2-p_3-p_4)  S|\mathcal{M}_f|^2,
	\end{align}
	where $F=\sqrt{(p_1\cdot p_2)^2-m_1^2m_2^2}$. If the masses of the initial state particles are equal, which is true for the processes we consider in this paper, then in the CM frame $F=[(2E_1)(2E_2)|\vec{v}_3-\vec{v}_4|]$. Thus for $m_1=m_2$, $\sigma$ reduces to the spin-summed center-of-mass (CM) frame cross-section. With this simplification, the collision term becomes
	\begin{align}
		\mathcal{C}_{\rm f}=\int \left[\prod_{i=1}^2\frac{d^4p_i}{(2\pi)^3}\delta(p_i^2-m_a^2)\Theta(p_i^0)\right]4F(p_1^0+p_2^0) \sigma_{aa\rightarrow bb, CM}f_{a}(p_1)f_{a}(p_2).
	\end{align}
	This integral can be further simplified if we make the change of variables (see also \cite{Birrell:2014uka})
	\begin{align}
		p= p_1+p_2, \quad  q= p_1-p_2,
	\end{align}
	to obtain
	\begin{align}
		\mathcal{C}_{\rm f}(T_f)=&\int\frac{1}{2^4} \frac{d^4p}{(2\pi)^2}\left[\frac{d^4q}{(2\pi)^4}\delta((p+q)^2/4-m_a^2)\delta((p-q)^2/4-m_a^2)\Theta(p^0-|q^0|)\right]\nonumber\\
		&\times4Fp^0 \sigma_{aa\rightarrow bb, CM}f_{a}((p^0+q^0)/2)f_{a}((p^0-q^0)/2)\\
		\equiv &\int\frac{1}{2^4} \frac{d^4p}{(2\pi)^2} dI_q\times4Fp^0 \sigma_{aa\rightarrow bb, CM}f_{a}((p^0+q^0)/2)f_{a}((p^0-q^0)/2),\label{eq:dIq_schannel}
	\end{align}
	where the phase space element $dI_q$ is given by the quantity in square brackets in the first line.

	Next we simplify $dI_q$. The delta functions in $dI_q$ together impose the constraints
	\begin{align}\label{eq:qxyz_replace}
		q_z=\frac{q^0p^0}{|\vec{p}|}, &&	|\vec{q}_{xy}|^2=p^2\left(1-\frac{(q^0)^2}{|\vec{p}|^2}\right)-4m_a^2,
	\end{align}
	where $q_z$ is the component of $\vec{q}$ along $\vec{p}$, while $\vec{q}_{xy}$ is the component of $\vec{q}$ perpendicular to $\vec{p}$. Consequently, we can perform the integral over $q_z$ and $|\vec{q}_{xy}|$ in $dI_q$ to integrate over the delta functions, yielding
	\begin{align}
		dI_q&=\bigg[\delta((p+q)^2/4-m_a^2)\delta((q-p)^2/4-m_a^2)\Theta(p^0-|q^0|)dq^3\ |\vec{q}_{xy}|d|\vec{q}_{xy}|\bigg]d\theta_{xy} \frac{dq^0}{(2\pi)^4}\\
		&=\frac{2}{|\vec{p}|}\Theta(p^0-|q^0|)\Theta\bigg(p^2\left[1-\frac{(q^0)^2}{|\vec{p}|^2}\right]-4m_a^2\bigg)d\theta_{xy} \frac{dq^0}{(2\pi)^4},\label{eq:dIqs_2}
	\end{align}
	where $\theta_{xy}$ is the azimuthal angle made by $\vec{q}_{xy}$ in the plane perpendicular to $\vec{p}$.
	The second theta function in the last line imposes the requirement that $|\vec{q}_{xy}|>0$. The two theta functions together rule out the region with $p^2 <4m_a^2$, as expected. Thus the arguments of the theta functions can be rewritten as
	\begin{align}
		dI_q&=\frac{2}{|\vec{p}|}\Theta(p^2-4m_a^2)\Theta(|\vec{p}|\beta_a-|q^0|)d\theta_{xy} \frac{dq^0}{(2\pi)^4},
	\end{align}
	where
	\begin{align}
		\beta_a=\sqrt{1-\frac{4m_a^2}{s}}.
	\end{align}
	
	Substituting the simplified $dI_q$ back in the collision term, we obtain
	\begin{align}
		\mathcal{C}_{\rm f}=&\int\frac{2\pi}{2^4} \frac{d^4p}{(2\pi)^6} \frac{2}{|\vec{p}|}\Theta(p^2-4m_a^2)\times4Fp^0 \sigma_{aa\rightarrow bb, CM}\int_{-|\vec{p}|\beta_a}^{|\vec{p}|\beta_a} dq^0f_{a}((p^0+q^0)/2)f_{a}((p^0-q^0)/2).
	\end{align}
	To integrate over the phase-space distribution, we assume the particles $a$ are in thermal equilibrium such that
	\begin{align}\label{eq:distribution}
		f_a(p)=\frac{1}{e^{p/T_a}+\zeta_a},
	\end{align}
	where $T_a$ is the temperature of particles $a$ and $\zeta_a=1$ (-1) if $a$ is a fermion (boson).
	
	For a thermal phase-space distribution, the integral over $q^0$ can be analytically performed to yield
	\begin{align}
		\mathcal{C}_{\rm f}=&\int\frac{2\pi}{2^4} \frac{d^4p}{(2\pi)^6} \frac{2}{|\vec{p}|}\Theta(p^2-4m_a^2)\times4Fp^0 \sigma_{aa\rightarrow bb, CM}\times \frac{4T_a}{e^{p^0/T_a}-\zeta_a^2}\ln\left(\frac{\exp{\frac{p^0+|\vec{p}|\beta_a}{2T_a}}+\zeta_a}{\exp{\frac{p^0}{2T_a}}+\zeta_a\exp{\frac{|\vec{p}|\beta_a}{2T_a}}}\right).
	\end{align}
	Rewriting the integration variable $p^{\mu}=(p^0,\vec{p})$ in terms of the Mandelstam $s=p^2$ and $y=\vp/\sqrt{s}$, and using $F=\sqrt{s(s-4m_a^2)}/2$, we obtain
	\begin{multline}
		\mathcal{C}_{\rm f}=\frac{T_a}{32\pi^4}\int_{4\max(m_a^2,m_b^2)}^{\infty} ds  s \sqrt{s(s-4m_a^2)}\sigma_{aa\rightarrow bb, CM} \\
		\times \left[2\int_0^{\infty} dy y \frac{1}{\bigg[\exp\bigg(\frac{\sqrt{y^2+1}}{T_a/\sqrt{s}}\bigg)-\zeta_a^2\bigg]}\ln\Bigg(\frac{\exp\Big(\frac{\sqrt{y^2+1}+\beta_ay}{2T_a/\sqrt{s}}\Big)+\zeta_a}{\exp\Big(\frac{\sqrt{y^2+1}}{2T_a/\sqrt{s}}\Big)+\zeta_a\exp\Big(\frac{\beta_a y}{2T_a/\sqrt{s}}\Big)}\Bigg)\right].\label{eq:s_quant_full}
	\end{multline}
	
	In the limit that the thermal distribution of particle $a$ can be approximated as Maxwell-Boltzmann, i.e.\ $\zeta_a\rightarrow 0$, the integral in the square brackets simplifies to $\beta_a K_2(\sqrt{s}/T_a)$, where $K_n$ is the modified Bessel function of second kind. Correspondingly, the collision term becomes
	\begin{align}
		\mathcal{C}_{\rm f}(T_a)=&\frac{T_a}{32\pi^4}\int_{4\max(m_a^2,m_b^2)}^{\infty} ds  s (s-4m_a^2)\sigma_{aa\rightarrow bb} K_2(\sqrt{s}/T_a),
	\end{align}
	recovering the result of Ref.~\cite{Gondolo:1990dk}.
	
	In the case where $m_a<m_b$, the energy injection into $b$ particles is mostly dominated by annihilations of $a$ particles when $a$ is relativistic. In the relativistic limit, we can approximate $\beta_a=1$, thus making the integral in square brackets in eq.~\eqref{eq:s_quant_full} only a function of $\sqrt{s}/T_a$. Defining
	\begin{align}\label{eq:G_def}
		G_{\zeta}(x)=2\int_0^{\infty} dt\ t \frac{1}{e^{x\sqrt{t^2+1}}-\zeta^2}\ln\Bigg(\frac{e^{x(\sqrt{t^2+1}+t)/2}+\zeta}{e^{x\sqrt{t^2+1}/2}+\zeta e^{xt/2}}\Bigg),
	\end{align}
	and approximating the term in square brackets in eq.~\eqref{eq:s_quant_full} as $\beta_a G_{\zeta_a}(\sqrt{s}/T_a)$ we obtain
	\begin{align}\label{eq:s_quant_sim}
		\mathcal{C}_{\rm f}(T_a)=&\frac{T_a}{32\pi^4}\int_{4\max(m_a^2,m_b^2)}^{\infty} ds  s (s-4m_a^2)\sigma_{aa\rightarrow bb} G_{\zeta_a}(\sqrt{s}/T_a).
	\end{align}
	The above collision term matches with eq.~\eqref{eq:s_quant_full} in the limit $T_a\gg m_a$. We use this simplified form of collision term while calculating the energy injection from particles $a$ into particles $b$ in the main body of our paper. The error in the final energy density introduced by using the simplified collision term  is maximized when $m_a\gg m_b$. This maximum error is about 2\% if particle $a$ is a fermion (as is typical for the models we consider) and 8\% if it is a boson.  However, this error is typically inconsequential because the energy injection into $b$ is dominated by annihilations of particles lighter than $b$.
	
	The function $G_{\zeta}$ can be computed analytically in the limit $x\gg1$ and $x\ll1$. In the large $x$ limit, $G_{\zeta}$ asymptotes to $K_2$ as expected. In the small $x$ limit, we find that
	\begin{align}
		G_{\zeta}(x)\xrightarrow{x\ll 1}\begin{cases}
			\frac{\pi^2}{6 x^2}\ln(2) & \zeta=1,\\
			\frac{2}{x^2} & \zeta=0,\\
			\frac{\pi^2}{3x^2}\ln(\frac{8\pi eA^{-12}}{x^2}) &\zeta=-1,
		\end{cases}
	\end{align}
	where   $A\approx 1.28$ is the Glaisher-Kinkelin constant. At high temperatures, $T_a\gg m_a, m_b$, the integrand in eq.~\eqref{eq:s_quant_sim} is dominated by $\sqrt{s}\ll T_a$. Thus the collision term computed with $\zeta_a=1$ (Fermi-Dirac statistics) is suppressed by a factor of $\sim2$ compared to the collision term calculated using $\zeta_a=0$ (Maxwell-Boltzmann statistics), while the collision term for $\zeta=-1$ (Bose-Einstein statistics) sees a non-trivial logarithmic enhancement compared to the $\zeta=0$ case.
	
	\begin{figure}
		\begin{subfigure}{.32\textwidth}
			\includegraphics[width=1\textwidth]{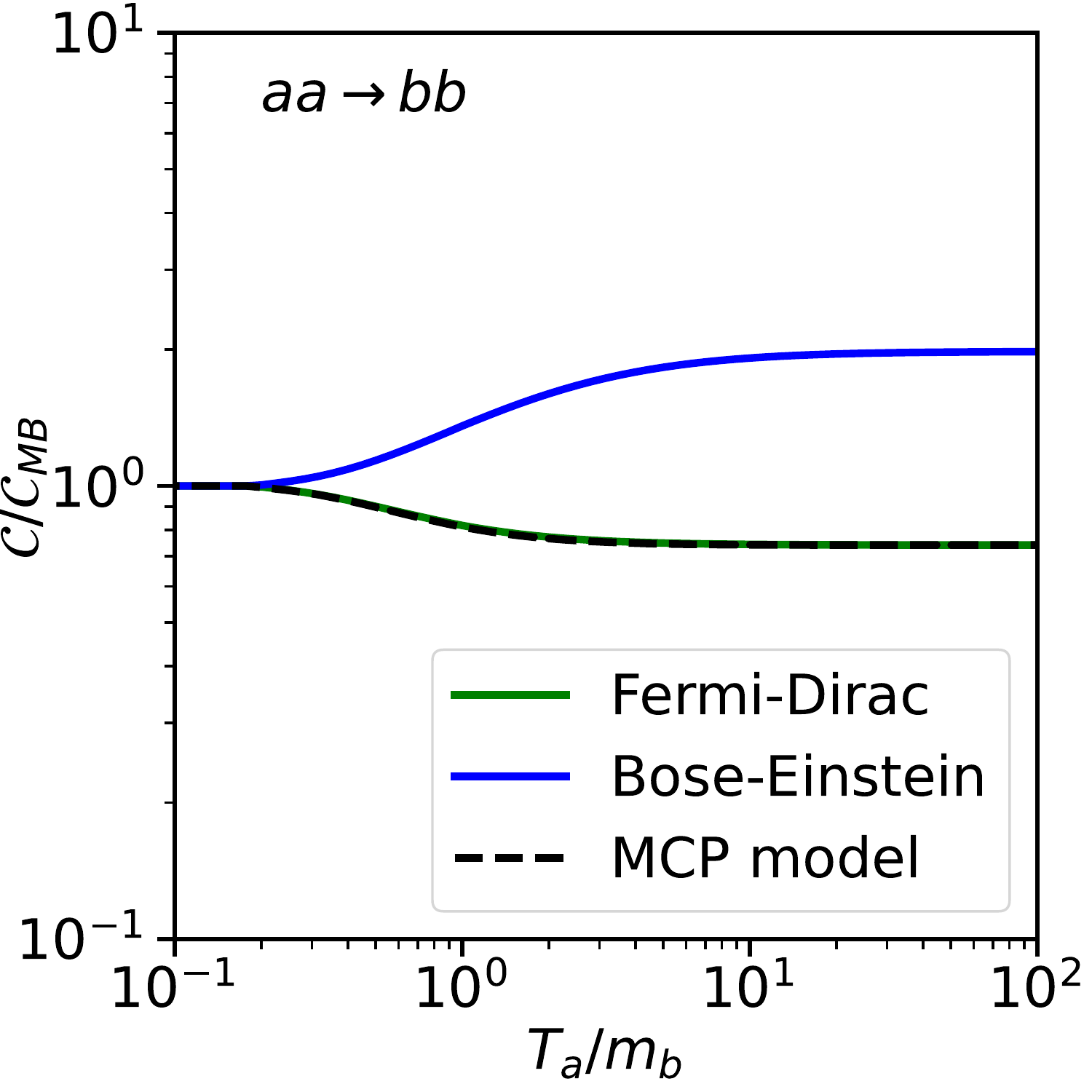}
		\end{subfigure}
		\begin{subfigure}{.32\textwidth}
			\includegraphics[width=1\textwidth]{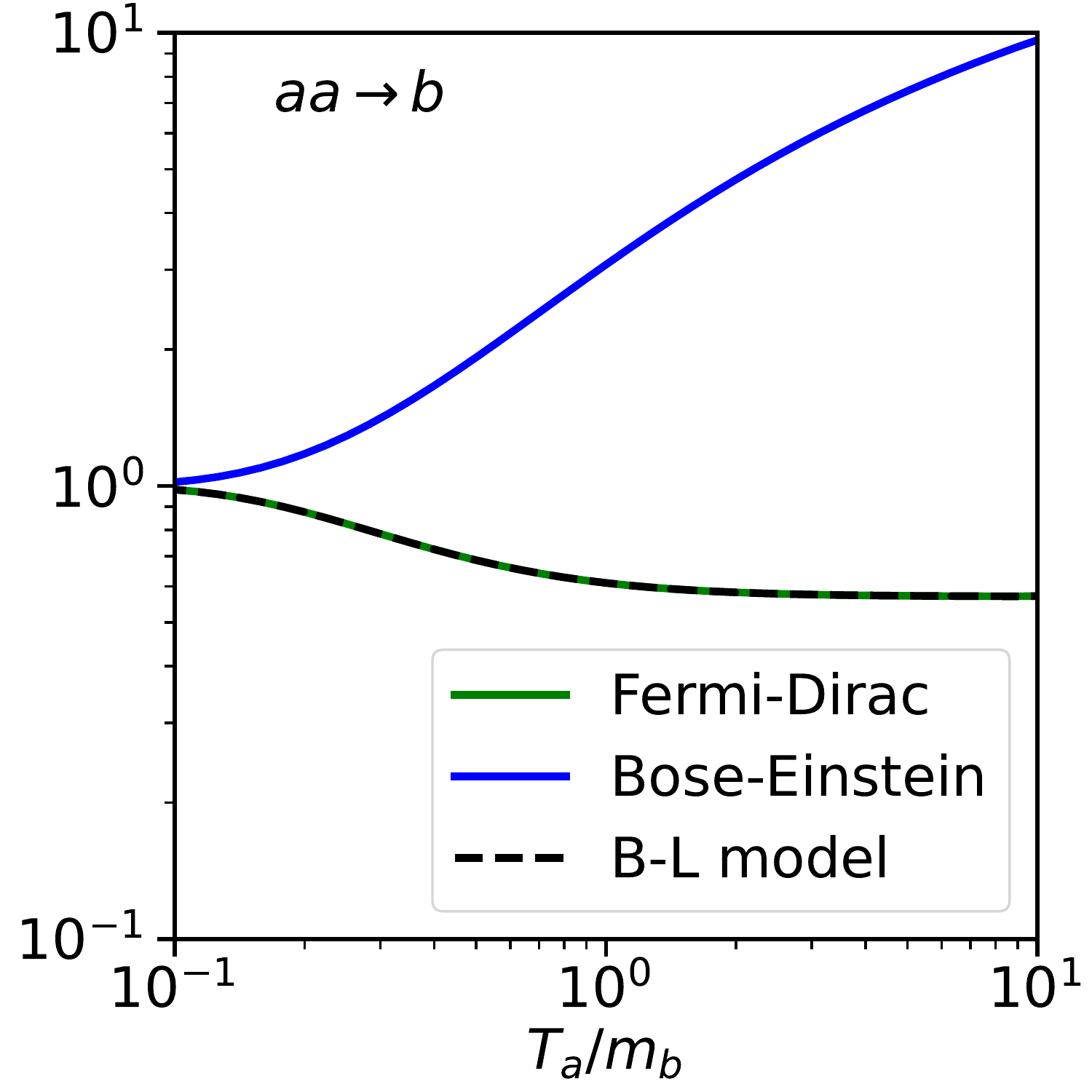}
		\end{subfigure}
		\begin{subfigure}{.32\textwidth}
			\includegraphics[width=1\textwidth]{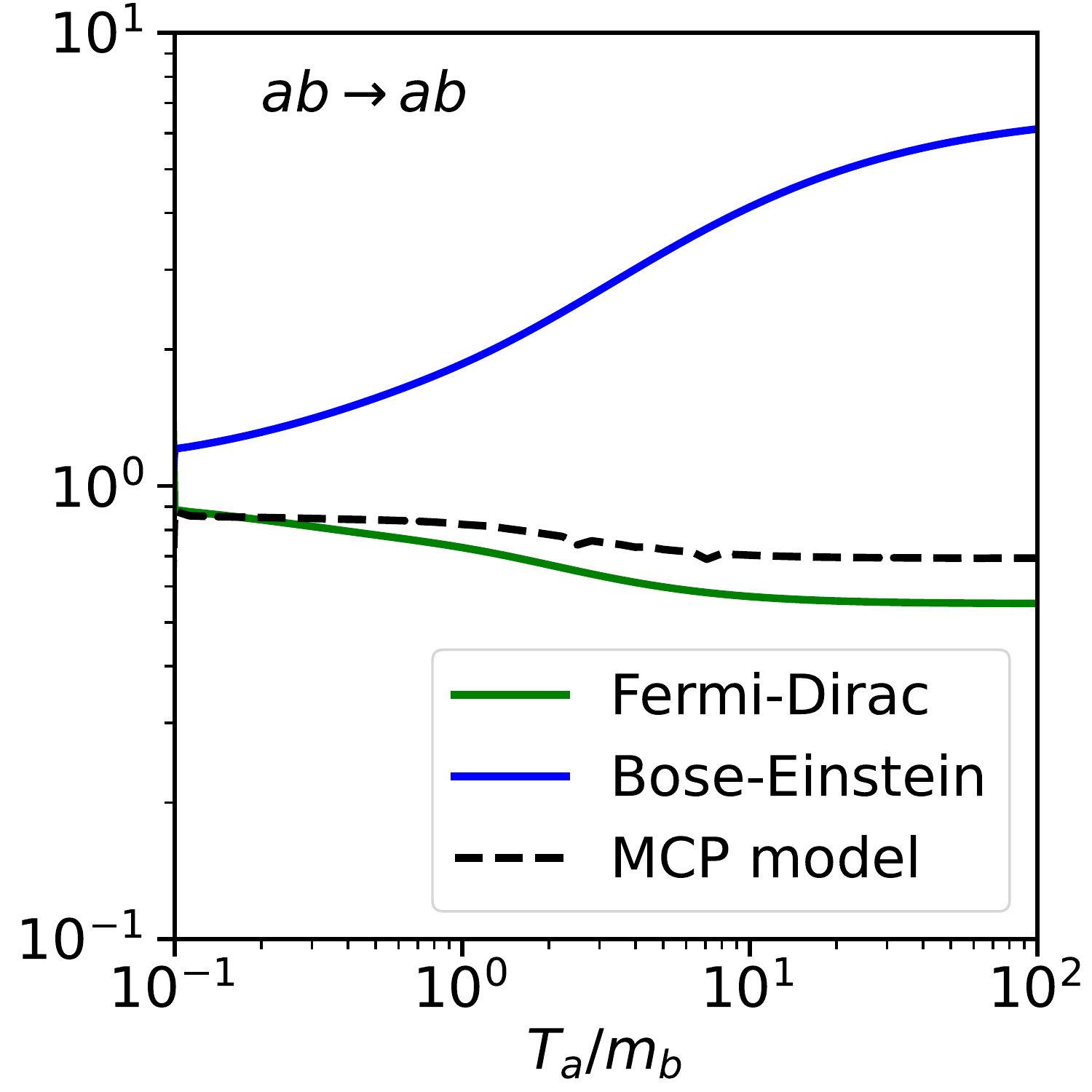}
		\end{subfigure}
		\caption{Ratio of collision terms with quantum statistics relative to the collision term calculated assuming Maxwell-Boltzmann statistics. We consider processes that transfer energy from relativistic particle $a$ with temperature $T_a$ to particle $b$ at temperature $T_b$ via annihilations (left), inverse decays (center) and elastic scattering (right). For the elastic scattering process we set $T_b=0.3T_a$ while for other processes the result is independent of $T_b$ for $T_b\ll T_a$. The colored solid lines are plotted for a constant matrix element and consider all particles to have Fermi-Dirac distribution (green) or Bose-Einstein distribution (blue). The black dashed lines indicate results for electron-positron annihilation into millicharged particles in the left panel, electron-positron fusion into a $B-L$ gauge boson in the central panel, and the Coulomb scattering of electrons with millicharged particles in the right panel. We set both the millicharged particle mass and $B-L$ gauge boson mass to 10 MeV. For $T_a\gg m_b$ incorporating quantum statistics results in a suppression of around 20-50\% for processes with initial state fermions, while for bosons it can provide significant enhancement.}
		\label{fig:quant_compare}
	\end{figure}
	
	In the left panel of figure~\ref{fig:quant_compare}, we show the ratio of collision terms, calculated using eq.~\eqref{eq:s_quant_sim} after setting $\xi_a=\pm1$ and $m_a=0$, relative to the Maxwell-Boltzmann collision term ($\xi_a=0$). One can see that the deviation from Maxwell-Boltzmann can be large for $T_a\gg m_a$, when all the particles are relativistic. The solid lines are plotted assuming a constant matrix element, or equivalently $\sigma_{aa\rightarrow bb}\propto 1/s$, while the black dashed line is plotted for electron-positron annihilation into millicharged particles using the cross-section in eq.~\eqref{eq:cross_ff_ann_Z}. As the cross-section in eq.~\eqref{eq:cross_ff_ann_Z} asymptotes to $\propto 1/s^2$ for $s\gg m^2$, the black-dashed line is almost indistinguishable from the solid green line.
	
	The total collision term is well approximated by $\mathcal{C}_{\rm f}$ as long as species $b$ remains more dilute than species $a$. However, close to thermalization where $T_b\approx T_a$, the phase-space distribution of particle $b$ can no longer be neglected in the computation of the total collision term. For this study, we approximate the total collision term using
	\begin{align}\label{eq:tot_coll}
		\mathcal{C}=\mathcal{C}_{\rm f}(T_a)-\mathcal{C}_{\rm f}(T_b),
	\end{align}
	where $T_b$ is the temperature of particle $b$ if it is in internal thermal equilibrium. The above collision term is not accurate close to thermalization because we have ignored the contribution of final state effects while calculating $\mathcal{C}_{\rm f}$. However, in the scenario when species $b$ thermalizes with the SM above some decoupling temperature, the value of the forward collision term for temperatures above the decoupling temperature is not important for the evaluation of the asymptotic dark radiation density.  Hence our approximated total collision is adequate for the analysis in this paper.  We have further verified that neglecting final state effects remains an excellent approximation by checking against results based on Ref.~\cite{Adshead:2019uwj}.
	
	\subsection{Decays}\label{sec:decay_quant}
	
	In this section we simplify the collision term describing energy transfer via decays and inverse decays, retaining quantum statistics for initial state particles. We start with decays
	$$a\rightarrow b+b.$$
	The collision term describing the forward energy transferred from $a$ to $b$ is given by
	\begin{equation}
		\label{eq:decay1}
		\mathcal{C}_{a\rightarrow b+b}^f=  \int d\Pi  d\Pi_1 d\Pi_2 \,(2\pi)^4\delta^4
		(p-p_1-p_2) \, S|{\cal M}_{\Gamma}|^2 f_{a}(p)(1\pm f_b(p_1))(1\pm f_b(p_2)) E,
	\end{equation} 
	where $f_a$ is the distribution function for particle $a$, $d\Pi_k = d^3p_k/[(2\pi)^3 2E_k]$, $|{\cal M}_{\Gamma}|^2$ is the spin-summed matrix element corresponding to the decay process, $S$ is the symmetry factor for potential identical particles in the final state, and variables with subscripts 1 and 2 correspond to the daughter particles while those with no subscripts correspond to $a$.
	
	Subsequently, we neglect the final-state phase space distribution of particle $b$. As  in section \ref{subsec:ann}, this approximation is valid as long as the density of particle $b$ is much more dilute than the density of particle $a$.
	By neglecting $f_b$, we can perform the phase space integration of the daughter particles in the rest frame of particle $a$ by using the definition of the rest frame decay width,
	\begin{align}\label{eq:gamma_def}
		\Gamma\equiv \frac{1}{2mg_{a}}\int d\Pi_1 d\Pi_2 \,(2\pi)^4\delta^4
		(p-p_1-p_2) \,S |{\cal M}_{\Gamma}|^2=\frac{S |{\cal M}_{\Gamma}|^2}{4\pi}\tilde{\beta}_b\frac{1}{2mg_{a}},
	\end{align}
	where $m$ is the mass of particle $a$, $g_a$ are the spin degrees of freedom of $a$, $\tilde{\beta}_b=\sqrt{1-4m^2/m_b^2}$ and $m_b$ is the mass of particle $b$.
	Doing so simplifies the collision term to
	\begin{equation}
		\mathcal{C}_{a\rightarrow b+b}^f=  m\Gamma g_a\int \frac{d^3p}{(2\pi)^3} f_{a}(p)=mn_{a}\Gamma.
	\end{equation} 
	
	Next, we simplify the collision term describing energy transferred by inverse decays of particle $b$ into particle $a$.  We start with
	\begin{align}
		\mathcal{C}_{b+b\rightarrow a}^f= \int d\Pi  d\Pi_1 d\Pi_2
		(2\pi)^4\delta^4(p_1+p_2-p)E\ |{\cal M}_{\Gamma}|^2S f_b(E_1)f_b(E_2);
	\end{align}
	this is of course the same expression as eq.~\eqref{eq:decay1}, up to the different phase space distribution factors.
	Again we  neglect the final state effect from particle $a$ under the assumption that particle $a$ is much more dilute than particle $b$.
	
	We perform the calculation in the rest frame of particle $a$.  Considering $U=(1,0,0,0)$ to denote the original isotropic frame, after changing frames such that $(\sqrt{\vec{p}^2+m^2},\vec{p})\rightarrow (m,\vec{0})$, we obtain $U=\frac{1}{m}(\sqrt{m^2+\vec{p}^2},-\vec{p})$. Consequently, the above collision term becomes
	\begin{align}
		\mathcal{C}_{b+b\rightarrow a}^f=& \int\frac{d^3p}{2(2\pi)^3}\bigg[ \int d\Pi_1 d\Pi_2
		(2\pi)^4\delta^3(p_1+p_2)\delta(2|\vec{p}_1|-m\tilde{\beta}_b)\ |{\cal M}_{\Gamma}|^2Sf_b(p_1\cdot U)f_b(p_2\cdot U)\bigg].
	\end{align}
	In the rest frame of particle $a$, $\vec{p}$ is the label of the boosted frame. Integrating over $\vec{p}_2$ and $|\vec{p}_1|$ yields
	\begin{align}
		\mathcal{C}_{b+b\rightarrow a}^f&=\frac{\tilde{\beta}_b^2}{8(2\pi)^5}\int d^3p\int d\Omega_{p_1}
		|{\cal M}_{\Gamma}|^2S  f_b(p_1\cdot U) f_b(p_2\cdot U).
	\end{align}
	
	Now, the spin-summed matrix element for a decay process is necessarily isotropic as well as independent of the momentum of $a$. Consequently, we can pull $|{\cal M}_{\Gamma}|^2$ outside of the integral. Taking  the $b$ particles to have a thermal distribution, 
	we can perform the angular integral over the distribution functions to yield
	\begin{align}
		\mathcal{C}_{b+b\rightarrow a}^f=&T_bm^2\tilde{\beta}_b\frac{|{\cal M}_{\Gamma}|^2\hat{S}}{8\pi^3}\int_0^{\infty} dt
		\frac{t}{\exp(x\sqrt{t^2+1})-\zeta_b^2}\log\left(\frac{\exp(\frac{x}{2}(\sqrt{1+t^2}+t\tilde{\beta}_b))+\zeta_b}{\exp(\frac{x\sqrt{1+t^2}}{2})+\zeta_b\exp(\frac{tx\tilde{\beta}_b}{2})}\right),
	\end{align}
	where $x=m/T_b$ and $t=|\vec{p}|/m$. In the limit $m_b\ll m$, we can approximate $\tilde{\beta}_b=1$ inside the integral, yielding
	\begin{align}\label{eq:dec_quant_sim}
		\mathcal{C}_{b+b\rightarrow a}^f\approx&m\Gamma\times\left[ m^2\frac{g_{Z'}}{2\pi^2}T_bG_{\zeta_b}(m/T_b)\right]\equiv m\Gamma\times\tilde{n}_{\zeta_b}(T_b),
	\end{align}
	where $G$ is as defined in eq.~\eqref{eq:G_def}. In the limit $\zeta_b=0$, we have $G_{\zeta_b}=K_2$, where $K_n$ is the modified Bessel function of the second kind. Consequently, $\tilde{n}_{0}(T_b)$ is the Maxwell-Boltzmann  equilibrium number density of particles as expected.
	
	In the center panel of figure~\ref{fig:quant_compare},  we show the ratio of collision terms calculated using eq.~\eqref{eq:dec_quant_sim} after setting $\xi_a=\pm1$ and $m_a=0$ relative to the collision term with $\xi_a=0$. The deviation from the Maxwell-Boltzmann result can be large at $T\gg m$, when all the particles are relativistic. The solid lines are plotted for a constant matrix element while the black dashed line is plotted for electron-positron inverse decay into $B-L$ gauge boson using the decay width in eq.~\eqref{eq:Z_decay}. As the ratio of collision terms is independent of decay width, the black-dashed line is the same as the solid green line.
	
	In the case of $B-L$ gauge bosons, the energy transfer from both decays and inverse decays are important in the computation of dark radiation production in the regions relevant for constraints from  $\Neff$. So we approximate the total collision term with
	\begin{align}
		\mathcal{C}=\mathcal{C}_{a\rightarrow b+b}^f-\mathcal{C}_{b+b\rightarrow a}^f=m\Gamma(n-\tilde{n}_{\zeta_b}(T_b)),
	\end{align}
	where $n$ is the (possibly out-of-equilibrium) number density of $B-L$ bosons.
	This collision term is accurate if either $n\gg \tilde{n}_{\zeta_b}(T_b)$ or $\tilde{n}_{\zeta_b}(T_b)\gg n$. However, once particles $a$ and $b$ thermalize, one would have to take into account final state effects for computation of the exact collision term. Consequently, $\tilde{n}_{\zeta_b}(T_b)$ is not the equilibrium number density of $b$ when it thermalizes. As the exact value of the collision term above the thermalization threshhold is unimportant for the resulting asymptotic dark radiation density in the scenarios where the HS thermalizes with the SM,  we use the above approximate collision term for the gauged $B-L$ model.
	
	In the case of $Z$ boson decays into MCPs, only the forward energy transfer from decays is important for the computation of $\Neff$ constraints. Thus we approximate the total collision term with
	\begin{align}
		\mathcal{C}=m\Gamma(n_{\rm eq}(T_a)-n_{\rm eq}(T_b)),
	\end{align}
	where $n_{\rm eq}$ is the equilibrium number density of $Z$ bosons. Again the above collision term is accurate for $T_a\gg T_b$. We use $n_{\rm eq}(T_b)$ and not $\tilde{n}_{\zeta_b}(T_b)$ to model the backward collision term so that we obtain $T_a=T_b$ after thermalization. 
	
	\subsection{Scattering}\label{sec:t_quant}
	
	In this section we simplify the phase space integral for the collision term describing the energy transfer from species $a$ to $b$ via a $t$-channel elastic scattering process:
	\begin{align}\label{eq:process}
		1(a)\ +\ 2(b)\ \rightarrow\ 3(a)\ +\ 4(b).
	\end{align}
	We consider particles $a$ and $b$ to be at different temperatures and additionally do not approximate their distribution as Maxwell-Boltzmann.
	
	The relevant energy transfer collision term for the process given in eq.~\eqref{eq:process} is 
	\begin{align}\label{eq:collC}
		\mathcal{C}=& 2\int \left[\prod_i\bigg(\frac{d^4p_i}{(2\pi)^3}\delta(p_i^2-m_i^2)\Theta(p_i^0)\bigg)
		(2\pi)^4\delta^4(p_1+p_2-p_3+p_4)\right]S|\mathcal{M}|^2(p_1^0-p_3^0)\nonumber\\
		&\times [f_a(p_1)f_b(p_2)(1\pm f_a(p_3))(1\pm f_b(p_4))]\\
		\equiv & 2\int dP\  S|\mathcal{M}|^2(p_1^0-p_3^0)[f_a(p_1)f_b(p_2)(1\pm f_a(p_3))(1\pm f_b(p_4))],
	\end{align}
	where $f_i$ are the distributions for the $i^{\mathrm{th}}$ particle, $\mathcal{M}$ is the (spin-summed) amplitude for the process, and $S$ includes potential identical particle factors for the process. 
	The overall factor of two due to the contribution from the backward process, which is the is the same as the forward process and we have summed over both processes in writing eq.\ \eqref{eq:collC}. We define the phase space element $dP$ as the factor in square brackets in the first line.
	
	Because of the energy transfer term $(p_1^0-p_3^0)$ in the integrand, we find that the computation of the collision term is simplified if we adopt the variables \cite{Adshead:2019uwj}
	\begin{align}
		p=p_1-p_3, && p'=p_2-p_4,\\
		q=p_1+p_3, && q'=p_2+p_4.
	\end{align}
	Correspondingly the Mandelstam variables are given by $s=(q+q')^2/4$, $t=p^2$ and $u=(q-q')^2/4$.
	
	After performing the above change of variables and integrating out $p'$ using the momentum-conserving delta function, we obtain
	\begin{align}
		dP=&\frac{(2\pi)^4}{2^8}\frac{d^4p}{(2\pi)^4}\bigg[\frac{d^4q}{(2\pi)^4}\delta((p+q)^2/4-m_a^2)\delta((q-p)^2/4-m_a^2)\Theta(q^0-|p^0|)\bigg]\nonumber\\
		& \times\bigg[\frac{d^4q'}{(2\pi)^4}\delta((p+q')^2/4-m_b^2)\delta((q'-p)^2/4-m_b^2)\Theta(q'^0-|p^0|)\bigg]\\
		\equiv&\frac{d^4p}{2^8}\ dI_q\ dI_{q'},
	\end{align}
	where $dI_q$ and $dI_{q'}$ are given by the first and second square brackets, respectively. Note that $dI_q$ and $dI_{q'}$ have identical functional forms up to the masses. Moreover, except for the argument of the theta function, $dI_q$ defined above has the same form as $dI_q$ defined in eq.~\eqref{eq:dIq_schannel}. Thus by performing the same steps as we did before to obtain eq.~\eqref{eq:dIqs_2}, we find
	\begin{align}
		dI_q&=\frac{2}{|\vec{p}|}\Theta(q^0-|p^0|)\Theta\bigg(p^2\left[1-\frac{(q^0)^2}{|\vec{p}|^2}\right]-4m_a^2\bigg)d\phi_{xy} \frac{dq^0}{(2\pi)^4},
	\end{align}
	where $\phi_{xy}$ is the azimuthal angle made by $\vec{q}$ in the plane perpendicular to $\vec{p}$.
	The second theta  function in the last line imposes the requirement that $|\vec{q}_{xy}|>0$. The two theta functions together rule out the region with $p^2>0$, which is expected as $t=p^2<0$. Thus the arguments of the theta  functions can be rewritten as
	\begin{align}
		dI_q&=\frac{2}{|\vec{p}|}\Theta(-p^2)\Theta(q^0-|\vec{p}|\beta_a)d\phi_{xy} \frac{dq^0}{(2\pi)^4},
	\end{align}
	where
	\begin{align}
		\beta_a=\sqrt{1-\frac{4m_a^2}{t}}.
	\end{align}
	A similar calculation gives the analogous result for  $dI_{q'}$ with the replacements  $m_a\to m_b$ and $q\to q'$. Thus the phase space element $dP$ simplifies to
	\begin{align}
		dP&=\frac{16\pi}{2^8(2\pi)^{8}}\Theta(-p^2)\bigg[\Theta(q^0-|\vec{p}|\beta_a)d\phi_{xy} dq^0\bigg]\bigg[\Theta(q'^0-|\vec{p}|\beta_b)d\phi_{xy}' dq'^0\bigg]d|\vec{p}|dp^0\frac{d\Omega_p}{4\pi}.
	\end{align}
	Replacing the above phase space element back in the collision term, and using the fact that the integrand is independent of the orientation of $\vec{p}$ as well as the overall phase $\phi_{xy}+\phi_{xy}'$, we obtain
	\begin{align}\label{eq:coll_t_temp}
		\mathcal{C}=&\frac{32\pi^2}{2^7(2\pi)^{8}}\int\bigg[\Theta(q^0-|\vec{p}|\beta_a) f_a(p_1)(1\pm f_a(p_3)) dq^0\bigg]\bigg[\Theta(q'^0-|\vec{p}|\beta_b)f_b(p_2)(1\pm f_b(p_4)) dq'^0\bigg]\nonumber\\
		&\times \left(\int |\mathcal{M}|^2d\theta\right) p^0 d|\vec{p}|dp^0\Theta(-p^2),
	\end{align}
	where $\theta=\phi_{xy}-\phi_{xy}'$.  Note that neither of the factors in square brackets depend on $\theta$ because the Boltzmann distributions are only functions of $p^0$ and $q^0$. Thus only the matrix element can have possible $\theta$ dependence.
	
	 The matrix element is a function of both $t=p^2=(p^0)^2-|\vec{p}|^2$ and $s=(q+q')^2/4$. Therefore the integrand does not generically factorize into functions of only single integration variables, and the integral in eq.~\eqref{eq:coll_t_temp} cannot be simplified by integrating over either $q$ or $q'$ independently.  
	A $t$-channel matrix element can generically be written as 
	\begin{align}\label{eq:temp_matrix}
		|\mathcal{M}|^2=\frac{\sum_{vw} c_{vw} s^vt^w}{(t-m_{\phi}^2)^2},
	\end{align}
	where $m_{\phi}$ is the mediator mass. The Mandelstam $t$ is simply equal to $p^2$ while $s$ has a complicated dependence on $q$, $q'$, and $\theta$ given by
	\begin{align}
		s=&\frac{1}{4}(q^2+q'^2+2q^0q'^0-2q_zq_z'-2q_{xy}q_{xy}'\cos\theta)\\
		=&\frac{1}{4}\bigg[4m_a^2+4m_b^2-2p^2-2q^0q'^0\frac{p^2}{|\vec{p}|^2}\nonumber\\
		&-2\left(p^2\left(1-\frac{(q^0)^2}{|\vec{p}|^2}\right)-4m_a^2\right)^{1/2}\left(p^2\left(1-\frac{(q'^0)^2}{|\vec{p}|^2}\right)-4m_b^2\right)^{1/2}\cos\theta\bigg],\label{eq:s_p_relation}
	\end{align}
	where in the second line we replaced $q_{z}, q_{xy}, q_z',$ and $q_{xy}'$ using
	eq.~\eqref{eq:qxyz_replace}. After integrating the matrix element over $\theta$, all terms with odd powers of $\cos\theta$ vanish. Hence the integrated matrix element is simply given by a polynomial of form
	\begin{align}\label{eq:cnml_def}
		\int |\mathcal{M}|^2 d\theta= \frac{1}{(p^2-m_{\phi}^2)^2}\sum_{nm\lambda}c_{nm\lambda}(q^0)^n(q'^0)^m\frac{p^{2\lambda}}{|\vec{p}|^{n+m}}.
	\end{align}
	The  exponents appearing here are restricted to $n,m,\lambda \in \{0,1,2\}$ because we require $v+w\leq 2$ in eq.~\eqref{eq:temp_matrix} for the matrix element to be unitary. Furthermore, since $\int d\theta s^v$ only depends on even combinations of $q^0$ and $q'^0$, $n+m$ is always even.
	
	Substituting the above expression for the matrix element into the collision integral eq.~\eqref{eq:coll_t_temp}, we obtain
	\begin{align}
		\mathcal{C}=&\sum_{nm\lambda} \frac{32\pi^2c_{nm\lambda}}{2^7(2\pi)^{8}}\int\frac{p^0}{(p^2-m_{\phi}^2)^2}p^{2\lambda}\bigg[\int \Theta(q^0-|\vec{p}|\beta_a) f_a(p_1)(1\pm f_a(p_3)) \frac{(q^0)^n}{\vp^n}dq^0\bigg]\nonumber\\
		&\times\bigg[\int \Theta(q'^0-|\vec{p}|\beta_b)f_b(p_2)(1\pm f_b(p_4)) \frac{(q'^0)^m}{\vp^m} dq'^0\bigg] d|\vec{p}|dp^0\Theta(-p^2)\\
		\equiv&\sum_{nm\lambda} \frac{32\pi^2c_{nm\lambda}}{2^7(2\pi)^{8}}\int \frac{p^0}{(p^2-m_{\phi}^2)^2}p^{2\lambda}\times I_{n,\zeta_a}(p) I'_{m,\zeta_b} (p) d|\vec{p}|dp^0\Theta(-p^2),
	\end{align}
	where the factors $I_{n,\zeta_a} (p)$ and $I_{n,\zeta_b}' (p)$ are the result of performing the integrals in the first and second square brackets in the first equation above.
	
	Considering a thermal distribution for particle $a$ as given in eq.~\eqref{eq:distribution}, the integral $I_{n,\zeta}(p)$ can be analytically evaluated to yield
	\begin{align}
		I_{n,\zeta_a} (p)=\frac{2T_a}{e^{p^0/T_a}-1}\left(\frac{2T_a}{|\vec{p}|}\right)^{n}L_{n,\zeta_a}\left(\frac{|\vec{p}|\beta_a}{2T_a},\frac{p^0}{2T_a}\right),
	\end{align}
	where
	\begin{align}
		L_{n,\zeta}(a,b)=\sum_{r=0}^n\frac{n!}{(n-r)!}a^{n-r}[-\zeta\textrm{Li}_{r+1}(-\zeta e^{-a+b})+\zeta\textrm{Li}_{r+1}(-\zeta e^{-a-b})],
	\end{align}
	and Li is the Polylogarithmic function. 
	(Recall that $\zeta=1$ if particle $a$ is a fermion and $\zeta=-1$ if particle $a$ is a boson.) In the Maxwell-Boltzmann limit, the terms in the square brackets in $L$ simplify to $e^{-a+b}$ for all $r$.
	
	The integral $I'_m$ is the same as for $I_n$ up to the replacements $p^0\to -p^0$, and  $a\to b$. Putting these results for $I_n$ and $I_m'$ back in the collision term yields
	\begin{multline}
		\mathcal{C}=\frac{32\pi^2}{2^7(2\pi)^{8}}4T_aT_b\sum_{nm\lambda}c_{nm\lambda} \int_{-\infty}^{\infty} dp^0 p^0\int_{|p^0|}^{\infty} d|\vec{p}|\frac{p^{2\lambda}}{(p^2-m_{\phi}^2)^2}\left(\frac{2T_a}{|\vec{p}|}\right)^{n}
		\left(\frac{2T_b}{|\vec{p}|}\right)^{m}\\\times\frac{L_{n,\zeta_a}\left(\frac{|\vec{p}|\beta_a}{2T_a},\frac{p^0}{2T_a}\right)}{e^{p^0/T_a}-1}\frac{L_{m,\zeta_b}\left(\frac{|\vec{p}|\beta_b}{2T_b},-\frac{p^0}{2T_b}\right)}{e^{-p^0/T_b}-1}.
	\end{multline}
	Note that $p^0>0$ indicates forward energy transfer from $a$ to $b$, while $p^0<0$ indicates backward energy transfer. Consequently, the forward energy transfer collision term can also be obtained from this expression by restricting the $p^0$ integral to the range $0 < p^0 <\infty$.
	
	Using the fact that $L_{m,\zeta}(a,-b)=-L_{m,\zeta}(a,b)$ we convert the integral over negative values of $p^0$ to positive values, yielding
	\begin{align}\label{eq:t_channel_full}
		\mathcal{C}=&\frac{32\pi^2}{2^7(2\pi)^{8}}4T_aT_b\int_{0}^{\infty} dp^0 p^0\left[\frac{1}{(e^{p^0/T_a}-1)(1-e^{-p^0/T_b})}-\frac{1}{(e^{p^0/T_b}-1)(1-e^{-p^0/T_a})}\right]\nonumber\\
		&\times\sum_{nm\lambda}c_{nm\lambda} \int_{|p^0|}^{\infty} d|\vec{p}|\frac{p^{2\lambda}}{(p^2-m_{\phi}^2)^2}\left(\frac{2T_a}{|\vec{p}|}\right)^{n}\left(\frac{2T_b}{|\vec{p}|}\right)^{m}L_{n,\zeta_a}\left(\frac{|\vec{p}|\beta_a}{2T_a},\frac{p^0}{2T_a}\right)L_{m,\zeta_b}\left(\frac{|\vec{p}|\beta_b}{2T_b},\frac{p^0}{2T_b}\right).
	\end{align}
	Notice that in the limit $T_a=T_b$, the term in  square brackets vanishes, as expected.
	
	In the limit where both particles can be described by Maxwell-Boltzmann distributions, the collision term simplifies to 
	\begin{align}\label{eq:t_channel_mb}
		\mathcal{C}=&\frac{32\pi^2}{2^7(2\pi)^{8}}4T_aT_b \int_{0}^{\infty} dp^0 p^0\left[e^{-p^0/2T_a}e^{p^0/2T_b}-e^{p^0/2T_a}e^{-p^0/2T_b}\right]\nonumber\\
		&\times\sum_{nm\lambda}c_{nm\lambda}\int_{p^0}^{\infty} d|\vec{p}|\frac{p^{2\lambda}}{(p^2-m_{\phi}^2)^2}\left(\frac{2T_a}{|\vec{p}|}\right)^{n}\left(\frac{2T_b}{|\vec{p}|}\right)^{m}\left(\sum_{r=0}^n\frac{n!}{(n-r)!}\left(\frac{|\vec{p}|\beta_a}{2T_a}\right)^{n-r}\right)\nonumber\\
		&\times \left(\sum_{r=0}^m\frac{m!}{(m-r)!}\left(\frac{|\vec{p}|\beta_b}{2T_b}\right)^{n-r}\right)e^{-\frac{\beta_a|\vec{p}|}{2T_a}}e^{-\frac{\beta_b|\vec{p}|}{2T_b}}.
	\end{align}
	
	In the right panel of figure~\ref{fig:quant_compare}, we show the ratio of the collision term calculated using eq.~\eqref{eq:t_channel_full} after setting $\xi_a=\xi_b=1$ (green) and $\xi_a=\xi_b=-1$ (blue) relative to the Maxwell-Boltzmann collision term (eq.~\eqref{eq:t_channel_mb}). We set $m_a=0$ and $T_b=0.3T_a$ while evaluating the collision terms. The solid lines are plotted assuming constant matrix element, i.e. only $c_{002}$ is non-zero and $m_{\phi}=0$, while the black-dashed line is plotted for electron scattering with millicharged particles using $c_{nm\lambda}$ given in eq.~\eqref{eq:cnml_qed}. One can see that the deviation from Maxwell-Boltzmann becomes notable at $T\gg m$, when all the particles are relativistic.

	\bibliographystyle{utphys}
	\bibliography{references}
\end{document}